\documentclass[
aps,
pra,
twocolumn,
superscriptaddress,
floatfix,
longbibliography
]{revtex4-1}
\usepackage[dvipdfmx]{graphicx}
\usepackage{dcolumn}
\usepackage{bm}
\usepackage{ulem}
\usepackage{amsmath}
\usepackage{amssymb}
\usepackage{txfonts}
\usepackage{hyperref}
\usepackage{color} 
\usepackage{xcolor}
\usepackage{listings}
\usepackage{cancel}

\newcommand{\bs}   {\boldsymbol}
\newcommand{\mb}   {\mathbf}

\newcommand{\e}{{\rm e}}
\newcommand{\imag}{{\rm i}}
\newcommand{\dd}{{\rm d}}

\hypersetup{
  colorlinks=true,
  linkcolor=[rgb]{0.60,0.00,0.00},
  citecolor=[rgb]{0.00,0.00,0.60},
  urlcolor=[rgb]{0.00,0.00,0.60},
  setpagesize=false
}

\definecolor{codegreen}{rgb}{0,0.6,0}
\definecolor{codegray}{rgb}{0.5,0.5,0.5}
\definecolor{codepurple}{rgb}{0.58,0,0.82}
\definecolor{backcolour}{rgb}{0.95,0.95,0.92}

\lstdefinestyle{mystyle}{
    backgroundcolor=\color{backcolour},   
    commentstyle=\color{codegreen},
    keywordstyle=\color{magenta},
    numberstyle=\tiny\color{codegray},
    stringstyle=\color{codepurple},
    basicstyle=\ttfamily\footnotesize,
    breakatwhitespace=false,         
    breaklines=true,                 
    captionpos=b,                    
    keepspaces=true,                 
    numbers=left,                    
    numbersep=5pt,                  
    showspaces=false,                
    showstringspaces=false,
    showtabs=false,                  
    tabsize=2
}

\begin{document}

\title{
  Spatial, spin, and charge symmetry projections for a Fermi-Hubbard model on a quantum computer
}

\author{Kazuhiro~Seki}
\affiliation{Quantum Computational Science Research Team, RIKEN Center for Quantum Computing (RQC), Saitama 351-0198, Japan}

\author{Seiji~Yunoki}
\affiliation{Quantum Computational Science Research Team, RIKEN Center for Quantum Computing (RQC), Saitama 351-0198, Japan}
\affiliation{Computational Quantum Matter Research Team, RIKEN Center for Emergent Matter Science (CEMS), Saitama 351-0198, Japan}
\affiliation{Computational Materials Science Research Team, RIKEN Center for Computational Science (R-CCS),  Hyogo 650-0047,  Japan}
\affiliation{Computational Condensed Matter Physics Laboratory, RIKEN Cluster for Pioneering Research (CPR), Saitama 351-0198, Japan}

\begin{abstract}
  We propose an extended version of the 
  symmetry-adapted variational-quantum-eigensolver (VQE)
  and apply it to a two-component Fermi-Hubbard model
  on a bipartite lattice.
  In the extended symmetry-adapted VQE method, 
  the Rayleigh quotient for the Hamiltonian and a parametrized quantum state
  in a properly chosen subspace is minimized within the subspace 
  and is optimized among the variational parameters implemented
  on a quantum circuit to obtain variationally the ground state and 
  the ground-state energy.  
  The corresponding energy derivative with respect to a variational parameter is 
  expressed as a Hellmann-Feynman-type formula
  of a generalized eigenvalue problem in the subspace,
  which thus allows us to use the parameter-shift rules for its evaluation.  
  The natural-gradient-descent method is also generalized to optimize variational parameters 
  in a quantum-subspace-expansion approach.  
  As a subspace for approximating the ground state of the 
  Hamiltonian, we consider 
  a Krylov subspace generated by the Hamiltonian and a symmetry-projected variational state, 
  and therefore the approximated ground state can restore the 
  Hamiltonian symmetry that is broken 
  in the parametrized variational state
  prepared on a quantum circuit.  
  We show that spatial symmetry operations for fermions in an occupation basis    
  can be expressed as a product of the
  nearest-neighbor fermionic {\sc swap} operations on a quantum circuit.
  We also describe how the spin and charge symmetry operations, i.e., rotations, 
  can be implemented on a quantum circuit. 
  By numerical simulations, we demonstrate that 
  the spatial, spin, and charge symmetry projections can improve 
  the accuracy of the parametrized variational state,
  which can be further improved systematically by expanding the Krylov subspace
  without increasing the number of variational parameters.
\end{abstract}

\date{\today}

\maketitle

\section{Introduction}

Recent technological advances in quantum devices~\cite{Nakamura1999,optical_RMP2007,Ladd2010,RevModPhys.85.623,Chow2014,Barends2014,Riste2015,Kelly2015,Arute2019,Zhong2020,Blais2021}
have suggested that quantum computation of quantum physics and chemistry systems~\cite{Feynman1982}  
is becoming a reality in the not-so-distant future~\cite{arute2020observation,Stanisic2021}.
Currently available quantum computers are, however,
prone to noise and hence the size of a quantum circuit to be reliably executed is limited.
Such quantum computers are called
noisy intermediate-scale quantum (NISQ) computers~\cite{Preskill2018}.
Despite the limitation,
NISQ computers with 50-100 qubits 
are anticipated to show advantage over classical computers with
the best known algorithm for particular tasks
(for example, see Refs.~\cite{Arute2019,Zhong2020,huang2020classical}). 
Thus, in parallel with research to accomplish fault-tolerant quantum computers,  
for which an increasing number of experimental developments towards realization of
logical qubit operations has been reported recently~\cite{Nigg2014,Erhard2021,Satzinger2021,Marques2021,hilder2021faulttolerant}, 
it is of importance to find practical applications of
NISQ computers for further stimulating progress in the field of quantum computing. 
To this and, several quantum-classical hybrid algorithms and
error mitigation schemes have been developed~\cite{Li2017PRX,Endo2018PRX,BonetMonroig2018,Endo2019mitigation}.

The variational-quantum eigensolver (VQE)~\cite{Yung2014,Peruzzo2014,McClean2016,Kandala2017} 
is one of the potentially promising quantum-classical hybrid schemes for
solving eigenvalue problems in quantum chemistry and quantum physics with NISQ computers.
For recent reviews on variational quantum algorithms,
see for example Refs.~\cite{Endo2021,Cerezo2021,tilly2021variational}.
In the VQE, the expectation value of a Hamiltonian of interest
with respect to a variational state represented on a parametrized quantum circuit, i.e., the variational energy,
is evaluated with a quantum computer,
while variational parameters are optimized 
by minimizing the variational energy on a classical computer. 
Depending on the form of variational states
and the assignment of tasks for quantum and classical computers,
several variants of the VQE scheme have been proposed.
For example, VQE-type approaches based on
the quantum-subspace expansion (QSE)~\cite{Colless2018}
performs subspace diagonalization in
an appropriately chosen subspace to 
approximate the target state(s) (e.g. the ground state)
better than the conventional VQE scheme,
at a cost of polynomial numbers of
additional measurements~\cite{nakanishi2018subspacesearch,heya2019subspace}.
Aiming at a systematic construction of the subspace,
a Krylov subspace~\cite{Chatelin} generated by properly chosen initial states with
a real-time evolution operator~\cite{parrish2019quantum,huggins2019nonorthogonal,Stair2020},
an imaginary-time evolution operator~\cite{Motta2019,Yeter-Aydeniz2020}, or
a Hamiltonian~\cite{seki2021} 
is often adopted.

In addition,
to exploit the Hamiltonian symmetry, 
several variants or extensions of the VQE approach   
have been proposed. 
For example, we have proposed the symmetry-adapted VQE (SAVQE) method 
to encompass the Hamiltonian symmetry in the VQE scheme~\cite{Seki2020vqe}. 
In the SAVQE method,
the symmetry that is broken in the variational state generated on a parametrized quantum circuit
is restored with a symmetry-projection operator. 
The nonunitarity of the projection operator is treated classically as a post processing
with increasing the number of measurements, a similar idea 
for treating Jastrow-type correlators reported
earlier in Ref.~\cite{Mazzola2019}.
Moreover, related schemes have been employed to restore symmetry in a continuous group such as 
the SU(2) total-spin conservation~\cite{Tsuchimochi2020,Siwach2021} and 
the U(1) particle-number conservation~\cite{Khamoshi2020,guzman2021accessing},
where the integral over continuous parameters of the group in the projection operator 
is properly discretized.   
Concerning the translational symmetry of periodic systems, 
another VQE-type scheme has been proposed to formulate directly in the reciprocal 
space~\cite{liu2020simulating,manrique2020momentumspace,yoshioka2020variational}.
Furthermore, in a different VQE variant, an appropriate penalty term is introduced into 
a cost function to obtain an eigenstate of the Hamiltonian 
in a desired symmetry sector~\cite{Ryabinkin2019,kuroiwa2020penalty}. 
It should also be noted that the Hamiltonian symmetry can also be 
utilized to mitigate errors due to different noise channels~\cite{BonetMonroig2018}. 
Remarkably, recent experiments 
have demonstrated a significant improvement for mitigating errors in a VQE simulation of a Fermi-Hubbard model 
by an error mitigation technique based on the Hamiltonian symmetry, 
including spin- and particle-number conservations,
time-reversal symmetry, particle-hole symmetry, and spatial symmetry~\cite{Stanisic2021}.

In this paper, we propose a QSE-based VQE method that incorporates the symmetry of Hamiltonian.  
The main idea is based on the SAVQE method~\cite{Seki2020vqe} and the quantum power method (QPM)~\cite{seki2021}, 
both of which have been developed recently by the present authors. 
In the proposed method,
we construct a Krylov subspace by multiplying 
Hamiltonian power onto a quantum state that is 
obtained by applying 
the symmetry-projection operators to a single variational state prepared on a parametrized quantum circuit. 
We then perform a subspace diagonalization by 
minimizing the Rayleigh quotient for the Hamiltonian and the quantum states 
in the Krylov subspace,  
and obtain the lowest eigenvalue of the generalized eigenvalue problem that depends on the variational parameters. 
The variational parameters are optimized so as to minimize the lowest eigenvalue of the generalized eigenvalue problem, 
which gives us the variational ground-state energy with the optimal set of variational parameters that represents the variational 
ground state. 
Considering a two-component Fermi-Hubbard model in a ladder lattice structure, we numerically demonstrate 
the proposed method by showing that 
the estimated ground-state energy as well as the ground-state fidelity can be improved
by the spatial, spin, and charge symmetry projections, and they are further improved systematically 
with expanding the subspace without increasing the number of variational parameters. 

The rest of this paper is organized as follows.
In Sec.~\ref{sec:form},
we briefly summarize a general formalism of the QSE method that is relevant for our purpose.
In Sec.~\ref{sec:opt},
we formulate the natural-gradient-descent (NGD) method
for the QSE scheme, and we describe how the corresponding
energy gradient and the Fubini-Study metric tensor 
can be evaluated with a quantum computer using 
the parameter-shift rule. 
We then describe in Sec.~\ref{sec:ksVQE} the Krylov-extended SAVQE by 
introducing the Krylov subspace generated by the Hamiltonian and a symmetry-projected 
variational state. 
In Sec.~\ref{sec:model}, 
we define the Fermi-Hubbard model on a bipartite lattice, and 
we briefly review its spatial, spin, and charge symmetry.  
We then describe the corresponding symmetry-projection operators.  
In Sec.~\ref{sec:symmetry},
we first show 
how the symmetry operations for a fermion model in general 
can be implemented on a quantum circuit, 
and then we explain the case of the spin and charge symmetry operations.  
In Sec.~\ref{sec:results}, 
we numerically demonstrate the proposed method 
for the Fermi-Hubbard model. 
A conclusion and a discussion are given 
in Sec.~\ref{sec:conclusion}.
Additional details on 
the NGD method and the fermionic symmetry operations are 
provided in Appendixes~\ref{app} and \ref{app:fermi}, respectively.
Matrix representations of typical two-qubit two-level unitary gates
for quantum many-body systems, such as the Givens-rotation gate and the Bogoliubov-transformation gate,  
are provided in Appendix~\ref{app:two_level_unitaries}.
A simple parallelization scheme for classical simulation of
the VQE method is described in Appendix~\ref{app:parallel}.
A remark on the normalization factor of the symmetry-projected quantum state 
is made in Appendix~\ref{app:fidelity}. 
Further numerical results for a different type of variational states
are discussed in Appendix~\ref{app:trotter}. 

\section{Formalism of quantum-subspace expansion}\label{sec:form}

\subsection{Trial state and variational principle}\label{sec:vp}
Consider a subspace
\begin{equation}
  {\cal U}={\rm span}
  \left(|u_0\rangle, |u_1\rangle,\cdots,|u_{d_{\cal U}-1}\rangle \right) 
  \label{eq:u_sub}
\end{equation}
with $\dim {\cal U}\equiv d_{\cal U}$.
The basis states $\{|u_i\rangle\}_{i=0}^{d_{\cal U}-1}$ should be linearly independent
of each other but they are not necessarily orthonormalized.
We assume that $|u_i\rangle$ has a form of
\begin{equation}
  |u_i(\bs{\theta})\rangle =
  \hat{O}_i
  |\psi(\bs{\theta})\rangle,
  \label{Ooperator}
\end{equation}
where $|\psi(\bs{\theta})\rangle$ is a variational state (ansatz) 
parametrized by a set of $N_{\rm v}$ variational parameters
$\bs{\theta}=\{\theta_k\}_{k=1}^{N_{\rm v}}$,
assuming they are real, and
$\hat{O}_i$ is an operator independent of the variational parameters.
$\hat{O}_i$ is not necessarily unitary but it is assumed to be 
given as a linear combination of unitary operators. 
Note that by definition, the parametrized part $|\psi(\bs{\theta})\rangle$ is 
common to all $\{|u_i(\bs{\theta})\rangle\}_{i=0}^{d_{\cal U}-1}$ in Eq.~(\ref{Ooperator}).

We now intend to approximate 
the exact ground state $|\Psi_0\rangle$ of the Hamiltonian $\hat{\cal H}$
within the subspace ${\cal U}$. 
This can be done simply by assuming that  
a trial state $|\Psi_{\cal U}(\bs{\theta})\rangle$ 
for approximating the ground state $|\Psi_0\rangle$ is given by 
\begin{equation}
  |\Psi_{\cal U}(\bs{\theta})\rangle = \sum_{i=0}^{d_{\cal U}-1}
  v_{i}(\bs{\theta}) |u_i(\bs{\theta})\rangle 
  \label{gst}
\end{equation}
in the subspace ${\cal U}$, where $\bs{v}(\bs{\theta})=\{v_i (\bs{\theta})\}_{i=0}^{d_{\cal U}-1}$
are coefficients to be determined.
According to the variational principle,  
the optimal variational parameters $\bs{\theta}_{\rm opt}$
are obtained by minimizing 
the expectation value of the Hamiltonian $\hat{\cal H}$, i.e., the variational energy, 
with respect to both $\bs{\theta}$ and $\bs{v}(\bs{\theta})$: 
\begin{alignat}{1}
  \bs{\theta}_{\rm opt}
  \equiv
  \underset{\bs{\theta}}{\rm arg~min}
  \left\{
  \min_{\bs{v}(\bs{\theta})}
  \left\{
  \frac{\langle \Psi_{\cal U}(\bs{\theta}) |\hat{\cal H} |\Psi_{\cal U}(\bs{\theta})\rangle}
       {\langle \Psi_{\cal U}(\bs{\theta}) |\Psi_{\cal U}(\bs{\theta})\rangle}
       \right\}
       \right\}. 
       \label{eq:theta_opt}
\end{alignat}

\subsection{Subspace diagonalization}\label{sec:subdiag}
Let us consider the optimization with respect to
$\bs{v}(\bs{\theta})$ for a given set of variational parameters $\bs{\theta}$ 
in Eq.~(\ref{eq:theta_opt}) 
and denote the optimal coefficients as 
$\bs{v}_0(\bs{\theta})\equiv\{v_{i,0}(\bs{\theta})\}_{i=0}^{d_{\cal U}-1}$, i.e.,
\begin{equation}
  \bs{v}_0(\bs{\theta})
  =
  \underset{\bs{v} (\bs{\theta})}{\rm arg~min} \left\{ \frac{\langle \Psi_{\cal U}(\bs{\theta}) |\hat{\cal H} |\Psi_{\cal U}(\bs{\theta})\rangle}
           {\langle \Psi_{\cal U}(\bs{\theta}) |\Psi_{\cal U}(\bs{\theta})\rangle} \right\}.
  \label{Esub}
\end{equation}
The optimal coefficients $\bs{v}_{0}(\bs{\theta})$ 
subject to a normalization condition, e.g., 
$\langle \Psi_{\cal U}(\bs{\theta}) |\Psi_{\cal U}(\bs{\theta})\rangle=1$, 
can be determined
by solving the generalized eigenvalue problem~\cite{seki2021} 
\begin{equation}
  \bs{H}(\bs{\theta})\bs{v}_0(\bs{\theta})=E_0(\bs{\theta})
  \bs{S}(\bs{\theta})\bs{v}_0(\bs{\theta}),
  \label{gev}
\end{equation}
where 
$\bs{H}(\bs{\theta})$ is the Hamiltonian matrix with its element 
\begin{equation}
  [\bs{H}(\bs{\theta})]_{ij} =
  \langle u_i (\bs{\theta})|\hat{\cal H} | u_j(\bs{\theta}) \rangle=
  \langle \psi (\bs{\theta})|\hat{O}_i^\dag \hat{\cal H}\hat{O}_j |\psi(\bs{\theta}) \rangle,
  \label{ham}
\end{equation}
$ \bs{S}(\bs{\theta})$ is the overlap matrix with its element 
\begin{equation}
  [\bs{S}(\bs{\theta})]_{ij} =
  \langle u_i (\bs{\theta})| u_j (\bs{\theta})\rangle=
  \langle \psi (\bs{\theta})|\hat{O}_i^\dag \hat{O}_j |\psi(\bs{\theta}) \rangle,  
  \label{ovlp}
\end{equation}
and $E_0(\bs{\theta})$ is the smallest eigenvalue. 
The trial state
\begin{equation}
  |\Psi_{\cal U}^{(0)}(\bs{\theta})\rangle \equiv \sum_{i=0}^{d_{\cal U}-1}
  v_{i,0}(\bs{\theta}) |u_i(\bs{\theta})\rangle
  \label{gs}
\end{equation}
with $[\bs{v}_0]_i=v_{i,0}$ being the $i$th entry of
the corresponding eigenvector in Eq.~(\ref{gev})
is normalized as long as the eigenvector $\bs{v}_{0}(\bs{\theta})$ is normalized 
with respect to $\bs{S}(\bs{\theta})$, i.e.,
\begin{equation}
  \langle \Psi_{\cal U}^{(0)}(\bs{\theta})|\Psi_{\cal U}^{(0)}(\bs{\theta}) \rangle
  = \bs{v}_0^\dag (\bs{\theta})\bs{S}(\bs{\theta}) \bs{v}_0(\bs{\theta}) = 1, 
  \label{normalization}
\end{equation}
where $\bs{v}_0^\dag(\bs{\theta}) = [\bs{v}_0^{\rm T}(\bs{\theta})]^*$.

\subsection{Energy and other expectation values}

The smallest eigenvalue $E_0(\bs{\theta})$ of the generalized eigenvalue problem in Eq.~(\ref{gev}) corresponds 
to the minimum value of the Rayleigh quotient, i.e., 
\begin{equation}
  E_0(\bs{\theta})
  =
  \min_{\bs{v} (\bs{\theta})} \left\{ \frac{\langle \Psi_{\cal U}(\bs{\theta}) |\hat{\cal H} |\Psi_{\cal U}(\bs{\theta})\rangle}
      {\langle \Psi_{\cal U}(\bs{\theta}) |\Psi_{\cal U}(\bs{\theta})\rangle} \right\}
  =
  \bs{v}_0^\dag(\bs{\theta}) \bs{H}(\bs{\theta}) \bs{v}_0 (\bs{\theta}), 
  \label{Esub}
\end{equation}
where the normalization condition for $|\Psi_{\cal U}^{(0)}(\bs{\theta})\rangle$
in Eq.~(\ref{normalization}) is used in the second equality. 
Therefore, the approximated ground-state energy of the Hamiltonian $\hat{\cal H}$ 
is obtained by minimizing $E_0(\bs{\theta})$ with respect to 
the variational parameters $\bs{\theta}$, 
\begin{alignat}{1}
E_0(\bs{\theta}_{\rm opt}) &= \min_{\bs{\theta}} \left\{ E_0(\bs{\theta}) \right\} 
=  \min_{\bs{\theta}} \left\{ \frac{\langle \Psi_{\cal U}^{(0)}(\bs{\theta}) |\hat{\cal H} |\Psi_{\cal U}^{(0)}(\bs{\theta})\rangle}
{\langle \Psi_{\cal U}^{(0)}(\bs{\theta}) |\Psi_{\cal U}^{(0)}(\bs{\theta})\rangle} \right\} \nonumber \\
&=
\min_{\bs{\theta}}
\left\{
\min_{\bs{v} (\bs{\theta})}
\left\{
\frac{\langle \Psi_{\cal U}(\bs{\theta}) |\hat{\cal H} |\Psi_{\cal U}(\bs{\theta})\rangle}
     {\langle \Psi_{\cal U}(\bs{\theta}) |\Psi_{\cal U}(\bs{\theta})\rangle}
     \right\}
     \right\}, 
     \label{eq:Eopt}
\end{alignat}
and the optimized variational parameters $\bs{\theta}_{\rm opt}$ 
give us the approximated ground state $|\Psi_{\cal U}^{(0)}(\bs{\theta}_{\rm opt})\rangle$.  
Once a set of optimal variational parameters $\bs{\theta}_{\rm opt}$ is obtained,
the ground-state expectation value of an observable $\hat{A}$ 
can be approximated as
\begin{equation}
  \langle \Psi_0 |\hat{A}|\Psi_0 \rangle 
  \approx \langle \Psi_{\cal U}^{(0)}(\bs{\theta}_{\rm opt}) |\hat{A} |\Psi_{\cal U}^{(0)}(\bs{\theta}_{\rm opt})\rangle
  =\bs{v}_0^\dag(\bs{\theta}_{\rm opt}) \bs{A}(\bs{\theta}_{\rm opt}) \bs{v}_0 (\bs{\theta}_{\rm opt}), 
\end{equation}
where 
$[\bs{A}(\bs{\theta})]_{ij}
= \langle u_i (\bs{\theta})|\hat{A} | u_j(\bs{\theta}) \rangle
= \langle \psi (\bs{\theta})|\hat{O}_i^\dag \hat{A} \hat{O}_j| \psi(\bs{\theta}) \rangle$ and 
the normalization condition for $|\Psi_{\cal U}^{(0)}(\bs{\theta}_{\rm opt})\rangle$
in Eq.~(\ref{normalization}) is assumed.

\section{Parameter-optimization method}\label{sec:opt}

\subsection{Natural-gradient-descent method}

The NGD method optimizes $\bs{\theta}$ by 
minimizing $E_0(\bs{\theta})$ with the following iteration~\cite{Amari1998}: 
\begin{equation}
  \bs{\theta}^{(x+1)}=
  \bs{\theta}^{(x)} - \tau [\bs{G}(\bs{\theta}^{(x)})]^{-1}
  \nabla E_0(\bs{\theta}^{(x)}),
  \label{iteration}
\end{equation}
where $\bs{\theta}^{(x)}=\{\theta_k^{(x)}\}_{k=1}^{N_{\rm v}}$ are the variational parameters at the $x$th iteration and 
$\tau>0$ is a parameter called learning rate.  
$\bs{G}(\bs{\theta})$ is the Fubini-Study metric tensor~\cite{provost1980,KOLODRUBETZ20171,stokes2020quantum} defined below 
and $\left[ \nabla E_0(\bs{\theta}) \right]_k=\partial E_0(\bs{\theta}) /\partial \theta_k$ is the energy gradient with respect to the 
variational parameter $\theta_k$.  
The gradient descent (GD) method can be obtained by setting $\bs{G}=\bs{I}$ in Eq.~(\ref{iteration}). 
The initial variational parameters $\bs{\theta}^{(1)}$ can be set arbitrary. 
In the following, we shall derive explicit forms of
the energy gradient and the Fubini-Study metric tensor for a QSE-based approach.

From Eqs.~(\ref{gev}) and (\ref{normalization})
and the Hermiticity of $\bs{H}(\bs{\theta})$ and $\bs{S}(\bs{\theta})$,
the energy derivative is given by 
\begin{equation}
  \frac{\partial E_0(\bs{\theta})}{\partial \theta_k}=
  \bs{v}_0^\dag(\bs{\theta}) \left(
  \frac{\partial \bs{H}(\bs{\theta})}{\partial \theta_k}
  -E_0(\bs{\theta})
  \frac{\partial \bs{S}(\bs{\theta})}{\partial \theta_k}
  \right) \bs{v}_0(\bs{\theta}).  
  \label{gradient}
\end{equation}
Equation~(\ref{gradient})
can be considered as a variant of
the Hellmann-Feynman theorem
for a generalized eigenvalue problem in the sense that 
the energy derivative does not require the
derivative of the eigenvector $\bs{v}_0(\bs{\theta})$.
The difference from the Hellmann-Feynman theorem
for a standard eigenvalue problem is that  
Eq.~(\ref{gradient}) involves the derivative of the overlap matrix $\bs{S}(\bs{\theta})$.

The Fubini-Study metric tensor $\bs{G}(\bs{\theta})$ is given by  
\begin{equation}
  [\bs{G}(\bs{\theta})]_{kl}={\rm Re}
  \left[\gamma_{kl}(\bs{\theta})
    -\beta_{k}^*(\bs{\theta})\beta_{l}(\bs{\theta})
    \right],
  \label{Gmat}
\end{equation}
where
\begin{equation}
  \gamma_{kl}(\bs{\theta})=
  \langle \partial_k \Psi_{\cal U}^{(0)}(\bs{\theta})|
  \partial_l \Psi_{\cal U}^{(0)}(\bs{\theta})\rangle
  \label{gamma}
\end{equation}
and
\begin{equation}
  \beta_{k}(\bs{\theta})=\langle \Psi_{\cal U}^{(0)}(\bs{\theta})|
  \partial_k \Psi_{\cal U}^{(0)}(\bs{\theta})\rangle
  \label{beta}
\end{equation}
with $\partial_k \equiv \partial/\partial {\theta_k}$  
(see Appendix~\ref{app} for details).
Note that our definition of $\beta_k$ in Eq.~(\ref{beta})
differs from that in Ref.~\cite{provost1980} by a multiplicative factor $-\imag$. 
The derivative of the normalization condition
$\partial_k \langle \Psi_{\cal U}^{(0)}(\bs{\theta})|\Psi_{\cal U}^{(0)}(\bs{\theta})\rangle=0$
implies that $\beta_k(\bs{\theta})$ is pure imaginary 
and hence $\bs{G}(\bs{\theta})$ is a real symmetric matrix.
By substituting 
\begin{alignat}{1}
  |\partial_k \Psi_{\cal U}^{(0)} (\bs{\theta}) \rangle
  =
  \sum_{i} \partial_{k}  v_{i,0}(\bs{\theta}) |u_i (\bs{\theta}) \rangle   
  +
  \sum_{i} v_{i,0}(\bs{\theta})|{\partial_{k}} u_i (\bs{\theta}) \rangle  
\end{alignat}
into Eqs.~(\ref{gamma}) and (\ref{beta}), we obtain that 
\begin{alignat}{1}
  \gamma_{kl}(\bs{\theta})
  &=
  \sum_{ij} (\partial_{k} v_{i,0}^*) (\partial_{l} v_{j,0}) \langle u_i | u_j \rangle
  +
  \sum_{ij} v_{i,0}^* v_{j,0} \langle \partial_{k} u_i | \partial_{l} u_j \rangle \notag \\
  &+
  \sum_{ij} (\partial_{k} v_{i,0}^*) v_{j,0} \langle u_i | \partial_{l} u_j \rangle 
  +
  \sum_{ij} v_{i,0}^* (\partial_{l} v_{j,0}) \langle \partial_{k} u_i | u_j \rangle \notag \\  
  &=
  (\partial_{k} \bs{v}_0)^\dag \bs{S} (\partial_{l} \bs{v}_0) 
  +
  \bs{v}_0^\dag \bs{S}_{(k,l)} \bs{v}_0 \notag \\
  &+
  (\partial_{k} \bs{v}_0)^\dag \bs{S}_{(,l)} \bs{v}_0 
  +
  \bs{v}_0^\dag [\bs{S}_{(,k)}]^\dag (\partial_l \bs{v}_0)
  \label{gamma2}
\end{alignat}
and 
\begin{alignat}{1}
  \beta_{k}(\bs{\theta})
  &=
  \sum_{ij}
  v_{i,0}^* (\partial_{k} v_{j,0}) \langle u_i|u_j \rangle
  +
  \sum_{ij}
  v_{i,0}^* v_{j,0} \langle u_i| \partial_{k} u_j \rangle \notag \\
  &=
  \bs{v}_0^\dag \bs{S}(\partial_k \bs{v}_0)
  +
  \bs{v}_0^\dag \bs{S}_{(,k)} \bs{v}_0, 
  \label{beta2}
\end{alignat}
where 
$[\bs{S}_{(,k)}(\bs{\theta})]_{ij}\equiv
\langle u_i(\bs{\theta}) | \partial_k u_j(\bs{\theta})\rangle$ 
and
$[\bs{S}_{(k,l)}(\bs{\theta})]_{ij}
\equiv
\langle \partial_k u_i(\bs{\theta})|\partial_l u_j(\bs{\theta}) \rangle$.
Note that the $\bs{\theta}$ dependence of $|u_i\rangle$, $\bs{v}_0$, and $\bs{S}$ 
is implicitly assumed in the right-hand
sides of Eqs.~(\ref{gamma2}) and (\ref{beta2}). 
We should also note that
$[\bs{S}_{(k,l)}(\bs{\theta})]^\dag = \bs{S}_{(l,k)}(\bs{\theta})$ and hence 
not all matrices $\bs{S}_{(k,l)}(\bs{\theta})$ are independent.

To obtain the derivative of the eigenvector $\bs{v}_0 (\bs{\theta})$ with respect to the variational parameters $\bs{\theta}$,
let us first expand it in the subspace ${\cal U}$ as
\begin{equation}
  \frac{\partial \bs{v}_0 (\bs{\theta})}{\partial \theta_k}=
  \sum_{n=0}^{d_{\cal U}-1} c_{k,n}\bs{v}_n(\bs{\theta}),
\end{equation}
where 
$\bs{v}_n$ is the $n$th eigenvector satisfying 
$\bs{H}(\bs{\theta}) \bs{v}_n(\bs{\theta}) =
E_{n}(\bs{\theta}) \bs{S}(\bs{\theta})\bs{v}_n(\bs{\theta})$ with 
$\bs{v}_m^\dag(\bs{\theta}) \bs{S}(\bs{\theta}) \bs{v}_n(\bs{\theta})=\delta_{mn}$ 
for $0 \leqslant m, n \leqslant d_{\cal U}-1$, 
and $\{c_{k,n}\}_{n=0}^{d_{\cal U}-1}$ are
complex numbers to be determined. 
By taking the derivative of
$\bs{H}(\bs{\theta}) \bs{v}_n(\bs{\theta}) =
E_{n}(\bs{\theta}) \bs{S}(\bs{\theta})\bs{v}_n(\bs{\theta})$,
multiplying $\bs{v}_n^\dag(\bs{\theta})$ from the left on it, and
using  $\bs{v}_m^\dag(\bs{\theta}) \bs{S}(\bs{\theta}) \bs{v}_n(\bs{\theta})=\delta_{mn}$
for $n\not= 0$, we find
that $c_{k,n}=  \frac{\bs{v}_n^\dag
  \left(\partial_k \bs{H}- E_0 \partial_k \bs{S} \right)\bs{v}_0}
{E_0-E_{n}} \bs{v}_n$
for $n \not= 0$ and hence 
\begin{equation}
  \frac{\partial \bs{v}_0 (\bs{\theta})}{\partial \theta_k}=
  c_{k,0}\bs{v}_0(\bs{\theta})
  +
  \sum_{n \not = 0 }^{d_{\cal U}-1}
  \frac{\bs{v}_n^\dag(\bs{\theta})
    \left(\tfrac{\partial \bs{H}(\bs{\theta})}{\partial \theta_k}
    - E_0(\bs{\theta}) \tfrac{\partial \bs{S}(\bs{\theta})}{\partial \theta_k}
    \right)
    \bs{v}_0(\bs{\theta})}
       {E_0(\bs{\theta})-E_{n}(\bs{\theta})} \bs{v}_n(\bs{\theta}),
       \label{dvec}
\end{equation}
where $E_0(\bs{\theta}) \not= E_n(\bs{\theta})$ is assumed.
For the remaining coefficient $c_{k,0}$, 
the derivative of the normalization condition 
$\partial_k \langle \Psi_{\cal U}^{(0)}(\bs{\theta})|\Psi_{\cal U}^{(0)}(\bs{\theta}) \rangle= 
\partial_k(\bs{v}_0^\dag(\bs{\theta}) \bs{S}(\bs{\theta}) \bs{v}_0(\bs{\theta}))=0$
implies that 
\begin{alignat}{1}
  &{\rm Re}c_{k,0}=
  -\frac{1}{2}
  \bs{v}_0^{\dag}(\bs{\theta})
  \frac{\partial \bs{S}(\bs{\theta})}{\partial \theta_k} \bs{v}_0(\bs{\theta}), \label{Rec}\\
  &{\rm Im}c_{k,0}\text{:\,undetermined}. \label{Imc}
\end{alignat}
As we shall discuss below,
the coefficient $c_{k,0}$ can be chosen arbitrarily
as far as the Fubini-Study metric tensor $\bs{G}(\bs{\theta})$ is concerned.
For numerical simulations in Sec.~\ref{sec:results},
we set that ${\rm Im}c_{k,0}=0$.

Here are two remarks regarding the coefficient $c_{k,0}$. 
First, ${\rm Re}c_{k,0}$ in Eq.~(\ref{Rec})
ensures that $\beta_k(\bs{\theta})$ is pure imaginary [see the discussion below Eq.~(\ref{beta})],  
which follows from Eq.~(\ref{beta2}) because one can easily show that $\bs{v}_0^\dag \bs{S}(\partial_k \bs{v}_0)=c_{k,0}$ and 
${\rm Re}\left[ \bs{v}_0^\dag \bs{S}_{(,k)} \bs{v}_0\right] = \frac{1}{2}\bs{v}_0^{\dag}(\partial_k \bs{S}) \bs{v}_0$.
This is a natural generalization of the corresponding result 
for real symmetric $\bs{H}(\bs{\theta})$ and $\bs{S}(\bs{\theta})$
in a real inner product space reported in Ref.~\cite{Ruymbeek2019} to a Hermitian case in a complex inner product space.
Second,  
the Fubini-Study metric tensor $\bs{G}(\bs{\theta})$ is independent of $c_{k,0}$. 
To prove this statement, let us consider a shift of the coefficient
$c_{k,0}\mapsto c_{k,0}+\Delta c_{k,0}$, i.e., 
\begin{equation}
  \frac{\partial \bs{v}_0 (\bs{\theta})}{\partial \theta_k} \mapsto
  \frac{\partial \bs{v}_0 (\bs{\theta})}{\partial \theta_k}+\Delta c_{k,0} \bs{v}_0(\bs{\theta}). 
  \label{shift_c0}
\end{equation}
Then, $\gamma_{kl}(\bs{\theta})$ and $\beta_{k}^*(\bs{\theta})\beta_{l}(\bs{\theta})$ are transformed accordingly as
\begin{alignat}{1}
  \gamma_{kl}(\bs{\theta})
  & \mapsto \gamma_{kl}(\bs{\theta})
  + \Delta c_{k,0}^* \Delta c_{l,0}
  + \Delta c_{k,0}^* \beta_{l}(\bs{\theta})
  + \Delta c_{l,0}   \beta_{k}^*,\label{trans_gamma}(\bs{\theta})
 \end{alignat}
  and 
  \begin{alignat}{1}
  \beta_{k}^*(\bs{\theta}) \beta_{l}(\bs{\theta})
  & \mapsto
  (\beta_{k}^*(\bs{\theta})+\Delta c_{k,0}^*)(\beta_{l}(\bs{\theta})+\Delta c_{l,0}) \notag \\
  &=\beta_{k}^*(\bs{\theta}) \beta_{l}(\bs{\theta})
  + \Delta c_{k,0}^* \Delta c_{l,0}
  + \Delta c_{k,0}^* \beta_{l}(\bs{\theta})
  + \Delta c_{l,0}   \beta_{k}^*(\bs{\theta}). \label{trans_beta}
\end{alignat}
Therefore, the Fubini-Study metric tensor $\bs{G}(\bs{\theta})$ is invariant under the shift in Eq.~(\ref{shift_c0}), i.e., 
\begin{equation}
  \bs{G}(\bs{\theta}) \mapsto \bs{G}(\bs{\theta}).
\end{equation}
Notice that the transformations in Eqs.~(\ref{trans_gamma}) and (\ref{trans_beta}) are
similar to those considered in Ref.~\cite{provost1980}, where   
the Fubini-Study metric tensor $\bs{G}(\bs{\theta})$ is constructed so as to be 
invariant under multiplication of a global phase factor to the state
$|\Psi_{\cal U}^{(0)}(\bs{\theta})\rangle$.  
Indeed, the former transformations are exactly the same as the latter ones if $\Delta c_{k,0}$ and $\Delta c_{l,0}$ 
in Eqs.~(\ref{trans_gamma}) and (\ref{trans_beta}) are pure imaginary. 
This also implies that
the undetermined and in principle arbitrarily chosen ${\rm Im}c_{k,0}$ in Eq.~(\ref{Imc}) can be
absorbed into the global phase factor of the state $|\Psi_{\cal U}^{(0)}(\bs{\theta})\rangle$,
which is analogous to
the case of the first-order perturbation theory for a standard (i.e., not a generalized) Hermitian eigenvalue problem
in quantum mechanics, where the imaginary part of the coefficient corresponding to $c_{k,0}$ is also 
undetermined~\cite{Schiff}. 
Since $\partial_k \bs{v}_0(\bs{\theta})$ appears only in the Fubini-Study metric tensor
$\bs{G}(\bs{\theta})$ in the present study,
$\Delta c_{k,0}$ and hence $c_{k,0}$ can be chosen arbitrarily.

\subsection{Derivatives}~\label{shiftrules}

We now derive analytical expressions of the derivatives,
assuming that $|\psi(\bs{\theta})\rangle$ in Eq.~(\ref{Ooperator}) has a particular form of 
\begin{equation}
  |\psi(\bs{\theta})\rangle
  \equiv
  \prod_{k=N_{\rm v}}^{1} 
  \hat{U}_k(\theta_k) \hat{W} |0\rangle^{\otimes N},
  \label{vqe}
\end{equation}
where
$\hat{U}_k(\theta_k)$ is a unitary operator parametrized by $\theta_k$ and 
$\hat{W}$ is a unitary operator independent of the variational parameters.
Since $\{\hat{U}_k(\theta_k)\}_{k=1}^{N_{\rm v}}$ and $\hat{W}$ are unitary, 
$|\psi(\bs{\theta})\rangle$ can be prepared
efficiently on a quantum computer. 
Moreover, we assume that
$\hat{U}_k(\theta_k)$ in Eq.~(\ref{vqe})
is an exponential of an involutory operator $\hat{P}_k$ of the form   
\begin{equation}
\hat{U}_k(\theta_k)
=\e^{-\imag \hat{P}_k \theta_k/2}
=\hat{1}\cos{\tfrac{\theta_k}{2}} - \imag \hat{P}_k \sin{\tfrac{\theta_k}{2}},
\end{equation}
where $\hat{1}$ is the identity operator and the second equality is because of $\hat{P}_k^2=\hat{1}$.

According to the parameter-shift rules~\cite{Li2017,Mitarai2018PRA,Schuld2019}, 
the derivatives appearing in Eqs.~(\ref{gradient}) and (\ref{dvec})
can be evaluated as
\begin{equation}
  \frac{\partial \bs{H}(\bs{\theta})}{\partial \theta_k}
  = \frac{1}{2}
  \left[
    \bs{H}(\bs{\theta}+\tfrac{\pi}{2} \bs{e}_k)
    -\bs{H}(\bs{\theta}-\tfrac{\pi}{2} \bs{e}_k)
    \right]
  \label{dham}
\end{equation}
and
\begin{equation}
  \frac{\partial \bs{S}(\bs{\theta})}{\partial \theta_k}
  = \frac{1}{2}
  \left[
    \bs{S}(\bs{\theta}+\tfrac{\pi}{2} \bs{e}_k)
    -\bs{S}(\bs{\theta}-\tfrac{\pi}{2} \bs{e}_k)
    \right],
  \label{dovlp}
\end{equation}
where 
$[\bs{e}_k]_{k^\prime}=\delta_{kk^\prime}$ 
is the $N_{\rm v}$-dimensional basis vector.
Note that the derivative of the parametrized state $|u_i(\bs{\theta})\rangle$
in Eq.~(\ref{Ooperator}) itself 
is also evaluated simply as  
\begin{equation}
  |\partial_{k} u_i (\bs{\theta}) \rangle = \frac{1}{2}|u_i(\bs{\theta}+\pi \bs{e}_k)\rangle. 
  \label{dpsi}
\end{equation}

Instead of Eqs.~(\ref{dham}) and (\ref{dovlp}), 
the derivatives of $\bs{H}(\bs{\theta})$ and $\bs{S}(\bs{\theta})$ can also be written as 
\begin{equation}
  \frac{\partial \bs{H}(\bs{\theta})}{\partial \theta_k}
  =\bs{H}_{(,k)}(\bs{\theta})+[\bs{H}_{(,k)}(\bs{\theta})]^\dag 
  \label{dham2}
\end{equation}
and
\begin{equation}
  \frac{\partial \bs{S}(\bs{\theta})}{\partial \theta_k}
  =\bs{S}_{(,k)}(\bs{\theta})+[\bs{S}_{(,k)}(\bs{\theta})]^\dag, 
  \label{dovlp2}
\end{equation}
where 
$[\bs{H}_{(,k)}(\bs{\theta})]_{ij}\equiv
\langle u_i(\bs{\theta}) | \hat{\cal H} |\partial_k u_j(\bs{\theta})\rangle$. 
By substituting Eq.~(\ref{dpsi})
into Eqs.~(\ref{dham2}) and (\ref{dovlp2}),
different analytical expressions for the derivatives can be obtained. 
As far as numerical simulations are concerned,
Eqs.~(\ref{dham2}) and (\ref{dovlp2}) are more preferable
than Eqs.~(\ref{dham}) and (\ref{dovlp}). 
This is because the number of the parametrized states required for Eqs.~(\ref{dham2}) and (\ref{dovlp2}) 
is $N_{\rm v}+1$,
i.e., $|\psi(\bs{\theta})\rangle$ and
$\{|\psi(\bs{\theta}+\pi \bs{e}_k)\rangle\}_{k=1}^{N_{\rm v}}$, 
while Eqs.~(\ref{dham}) and (\ref{dovlp}) require $2N_{\rm v}$ states
$\{|\psi(\bs{\theta} \pm \tfrac{\pi}{2} \bs{e}_k)\rangle\}_{k=1}^{N_{\rm v}}$, 
and thus the numerical complexity is approximately half.

\section{Krylov-extended symmetry-adapted VQE}\label{sec:ksVQE}

Equation~(\ref{gs}) implies that 
the selection of the subspace $\cal U$ is crucial for approximating 
the ground state (or any target eigenstate) of a Hamiltonian $\hat{\cal H}$. 
In the Krylov-subspace SAVQE for the ground state calculation, 
the subspace $\cal U$ is constructed as 
\begin{alignat}{1}
  {\cal U}
  &={\cal K}_{d_{\cal U}}(\hat{\cal H},\hat{\cal P} |\psi(\bs{\theta})\rangle) \notag \\
  &={\rm span} \left(
  \hat{\cal P}|\psi(\bs{\theta}) \rangle, 
  \hat{\cal H} \hat{\cal P}|\psi(\bs{\theta}) \rangle, 
  \cdots,
  \hat{\cal H}^{d_{\cal U}-1} \hat{\cal P}|\psi(\bs{\theta}) \rangle
  \right),
  \label{eq:Krylov0}
\end{alignat}
where 
$\hat{\cal P}$ is a symmetry-projection operator and by definition $\hat{\cal H}$ and 
$\hat{\cal P}$ commute to each other, i.e., $[\hat{\cal H},\hat{\cal P}]=0$. 
The explicit form of $\hat{\cal P}$ is determined 
by considering the Hamiltonian symmetry and the details are described in Sec.~\ref{sec:Poperator}.
$|\psi(\bs{\theta})\rangle$ is a quantum circuit ansatz parametrized 
by the variational parameters $\bs{\theta}=\{\theta_k\}_{k=1}^{N_{\rm v}}$ 
as in Eq.~(\ref{vqe}) and 
generally breaks the symmetry of the Hamiltonian $\cal{H}$. 
With the projection operator $\hat{\cal P}$, the state $\hat{\cal P}|\psi(\bs{\theta})\rangle$ 
is projected onto the symmetry sector relevant to the ground state. 
It is obvious
that the Krylov subspace in Eq.~(\ref{eq:Krylov0}) is a special case of the more general subspace 
in Eq.~(\ref{eq:u_sub}) with the operator $\hat{O}_n$ in Eq.~(\ref{Ooperator}) given now by 
\begin{equation}
  \hat{O}_{n}=\hat{\cal H}^{n} \hat{\cal P}.
  \label{eq:ohp}
\end{equation}
In the VQE scheme, each term in the matrix elements of $\bs{H}(\bs{\theta})$ and $\bs{S}(\bs{\theta})$ 
for a given set of 
the variational parameters $\bs{\theta}$ is evaluated on a quantum computer,
while a classical computer is employed to solve 
the generalized eigenvalue problem in Eq.~(\ref{gev}) and to optimize the variational parameters, 
here using the NGD method described in Sec.~\ref{sec:opt}, for which each term in the matrix elements 
of the Fubini-Study tensor $\bs{G}(\bs{\theta})$ is also evaluated on a quantum computer.   
As explained in the next section, the nonunitary projection operator $\hat{\cal P}$ 
can be expressed by a linear combination of unitary operators and hence can be implemented on a quantum computer 
with the appropriate postprocessing~\cite{Seki2020vqe}.

To evaluate the expectation value of $\hat{\cal H}^n$ on a quantum computer, several
direct~\cite{kowalski2020quantum,vallury2020quantum,Claudino2021} 
and approximate~\cite{bespalova2020hamiltonian,gonzalez2021quantum}
methods have been proposed recently.
For a recent review, see for example Ref.~\cite{Aulicino2021}.
Here, we employ the QPM, a scheme introduced in Ref.~\cite{seki2021}, where the Hamiltonian power 
$\hat{\cal H}^n$ is directly treated by approximating it with
a linear combination of unitary time-evolution 
operators at $n+1$ different times. 
Using the second-order symmetric Suzuki-Trotter (ST) decomposition~\cite{Trotter1959,Suzuki1976},
the Hamiltonian power $\hat{\cal H}^n$ is approximated as~\cite{seki2021}
\begin{equation} 
  \hat{\cal H}^{n} =
  \hat{\cal H}^{n}_{\rm ST}(\Delta)
  + O(\Delta^2),
  \label{HST}
\end{equation}
where
\begin{equation} 
  \hat{\cal H}^{n}_{\rm ST}(\Delta) = \frac{\imag^n}{\Delta^n}\sum_{k=0}^{n} (-1)^k\binom{n}{k} \
  \left[\hat{S}_2(\Delta/2)\right]^{n-2k},  
  \label{eq:HST}
\end{equation}
and $\Delta$ is a parameter and should be in general a small real number.  
$\hat{S}_2(\Delta/2)$ is the second-order symmetric ST decomposition
of the time-evolution operator $\e^{-\imag \hat{\cal H} \Delta/2}$ and
$O(\Delta^2)$ in Eq.~(\ref{HST}) is the leading systematic error. 
The explicit form of $\hat{S}_2(\Delta/2)$ depends on the Hamiltonian $\hat{\cal H}$ and 
it is given in Sec.~\ref{sec:std} for the two-component Fermi-Hubbard model in a ladder lattice structure.

As described in detail in Ref.~\cite{seki2021}, the order of the leading systematic error 
can be eliminated systematically with the 
Richardson extrapolation. For instance, the first-order
Richardson extrapolation $\hat{\cal H}_{\rm ST(1)}^n(\Delta)$ of
$\hat{\cal H}_{\rm ST}^n(\Delta)$ can be given by
\begin{equation}
  \hat{\cal H}_{\rm ST(1)}^n (\Delta) =
  \frac{1}{3}\left[ {4 \hat{\cal H}_{\rm ST}^n (\Delta/2) -\hat{\cal H}_{\rm ST}^n (\Delta)} \right],
  \label{HST1}
\end{equation}
which approximates $\hat{\cal H}^n$ to $O(\Delta^4)$, i.e., 
\begin{equation}
\hat{\cal H}^{n} =
\hat{\cal H}^{n}_{\rm ST(1)}(\Delta)
+ O(\Delta^4).
\label{eq:FRE}
\end{equation}
By using Eq.~(\ref{HST1}), we can approximate $\hat{O}_n$ in Eq.~(\ref{eq:ohp}) as 
\begin{equation}
  \hat{O}_n \approx \hat{\cal H}_{\rm ST(1)}^n(\Delta) \hat{\cal P}. 
\end{equation}
Accordingly, the Hamiltonian matrix and the overlap matrix are given respectively as 
\begin{equation}
  [\bs{H}(\bs{\theta})]_{ij} \approx \langle \psi(\bs{\theta})|
  \hat{\cal P}
  \hat{\cal H}_{\rm ST(1)}^i(\Delta)
  \hat{\cal H} 
  \hat{\cal H}_{\rm ST(1)}^j(\Delta)
  \hat{\cal P}
  |\psi(\bs{\theta}) \rangle
  \label{Happrox2}
\end{equation}
and
\begin{equation}
  [\bs{S}(\bs{\theta})]_{ij} \approx \langle \psi(\bs{\theta})|
  \hat{\cal P}
  \hat{\cal H}_{\rm ST(1)}^i(\Delta)
  \hat{\cal H}_{\rm ST(1)}^j(\Delta)
  \hat{\cal P}
  |\psi(\bs{\theta}) \rangle,
  \label{Sapprox2}
\end{equation}
where $\hat{\cal P}^\dag=\hat{\cal P}$ and $[\hat{\cal H}_{\rm ST(1)}^n]^\dag=\hat{\cal H}_{\rm ST(1)}^n$ are used. 
On the other hand, because of $[\hat{\cal H},\hat{\cal P}]=0$, we can also approximate $\hat{O}_n$ in Eq.~(\ref{eq:ohp}) as 
\begin{equation}
 \hat{O}_n \approx \hat{\cal P} \hat{\cal H}_{\rm ST(1)}^n(\Delta). 
 \label{eq:On}
\end{equation}
Correspondingly, the Hamiltonian matrix and the overlap matrix can be given respectively as 
\begin{equation}
  [\bs{H}(\bs{\theta})]_{ij} \approx \langle \psi(\bs{\theta})|
  \hat{\cal H}_{\rm ST(1)}^i(\Delta)
  \hat{\cal H} \hat{\cal P}
  \hat{\cal H}_{\rm ST(1)}^j(\Delta)
  |\psi(\bs{\theta}) \rangle
  \label{Happrox}
\end{equation}
and
\begin{equation}
  [\bs{S}(\bs{\theta})]_{ij} \approx \langle \psi(\bs{\theta})|
  \hat{\cal H}_{\rm ST(1)}^i(\Delta)
  \hat{\cal P}
  \hat{\cal H}_{\rm ST(1)}^j(\Delta)
  |\psi(\bs{\theta}) \rangle,
  \label{Sapprox}
\end{equation}
where $\hat{\cal P}^2=\hat{\cal P}$ is also used. 
Although the apparent forms of $[\bs{H}(\bs{\theta})]_{ij}$ ($[\bs{S}(\bs{\theta})]_{ij}$) 
in Eqs.~(\ref{Happrox2}) and (\ref{Happrox}) [Eqs.~(\ref{Sapprox2}) and (\ref{Sapprox})] are different, 
they are exactly the same up to $O(\Delta^4)$ because of $[\hat{\cal P}, \hat{\cal H}_{\rm ST(1)}^n]=O(\Delta^4)$.

An important difference between Eqs.~(\ref{Happrox2}) and (\ref{Sapprox2})
and Eqs.~(\ref{Happrox}) and (\ref{Sapprox}) is that
the latter formulas contain the projection
operator $\cal{\hat{P}}$ only once for each matrix element.
This is advantageous for quantum-classical hybrid computing 
because the number of measurements required for 
Eqs.~(\ref{Happrox}) and (\ref{Sapprox}) is smaller than that for
Eqs.~(\ref{Happrox2}) and (\ref{Sapprox2}) 
by a factor of the number of terms in $\hat{\cal P}$, 
assuming that $\hat{\cal P}$ is given
by a linear combination of unitary operators.
For this reason, we adopt
Eqs.~(\ref{Happrox}) and (\ref{Sapprox}) for numerical simulations in Sec.~\ref{sec:results}. 
It should be noted that, due to the presence of the projection
operator $\hat{\cal P}$, the number of measurements required for evaluating
the Hamiltonian and overlap matrices in Eqs.~(\ref{Happrox}) and (\ref{Sapprox})
is increased by a factor of the number of terms in $\hat{\cal P}$
from that discussed in Ref.~\cite{seki2021}. 

After optimizing the variational parameters $\bs{\theta}=\{\theta_k\}_{k=1}^{N_{\rm v}}$ in Eq.~(\ref{eq:Eopt}), 
we obtain the approximated ground state $|\Psi_{\cal U}^{(0)}(\bs{\theta}_{\rm opt})\rangle$ in 
the Krylov subspace $\cal U$ defined in Eq.~(\ref{eq:Krylov0}), i.e., 
\begin{equation}
|\Psi_0\rangle \approx |\Psi_{\cal U}^{(0)}(\bs{\theta}_{\rm opt})\rangle = \sum_{i=0}^{d_{\cal U}-1}
  v_{i,0}(\bs{\theta}_{\rm opt}) \hat{\cal P} \hat{\cal H}_{\rm ST(1)}^i(\Delta)  |\psi(\bs{\theta}_{\rm opt}) \rangle. 
  \end{equation}
 The expectation value of any quantity 
$\hat{\cal O}$ with respect to $|\Psi_{\cal U}^{(0)}(\bs{\theta}_{\rm opt})\rangle$ is then evaluated simply as 
\begin{eqnarray}
\langle \Psi_0| \hat{\cal O} |\Psi_0\rangle &\approx&
\frac{\langle \Psi_{\cal U}^{(0)}(\bs{\theta}_{\rm opt}) |\hat{\cal O}| \Psi_{\cal U}^{(0)}(\bs{\theta}_{\rm opt})\rangle}
{ \langle \Psi_{\cal U}^{(0)}(\bs{\theta}_{\rm opt}) | \Psi_{\cal U}^{(0)}(\bs{\theta}_{\rm opt})\rangle} \nonumber \\
&=& \frac{\bs{v}_0^\dag(\bs{\theta}_{\rm opt}) \bs{O}(\bs{\theta}_{\rm opt}) \bs{v}_0(\bs{\theta}_{\rm opt})  }
{  \bs{v}_0^\dag(\bs{\theta}_{\rm opt}) \bs{S}(\bs{\theta}_{\rm opt}) \bs{v}_0(\bs{\theta}_{\rm opt}) },
\label{eq:average}
\end{eqnarray}
where 
\begin{equation}
  [\bs{O}(\bs{\theta})]_{ij} = \langle \psi(\bs{\theta})|
  \hat{\cal H}_{\rm ST(1)}^i(\Delta)
  \hat{\cal O} \hat{\cal P}
  \hat{\cal H}_{\rm ST(1)}^j(\Delta)
  |\psi(\bs{\theta}) \rangle
\end{equation}
and $[\hat{\cal O},\hat{\cal P}]=0$ is assumed. This is easily generalized for a quantity $\hat{\cal O}$ 
with $[\hat{\cal O},\hat{\cal P}]\ne0$.

\section{Model, symmetry, and projection operator}\label{sec:model}

\subsection{Fermi-Hubbard model}
As a demonstration, we apply the extended
SAVQE method to the two-component Fermi-Hubbard model. 
The Hamiltonian $\hat{\cal H}$ of the Fermi-Hubbard model is given by 
\begin{equation}
  \hat{\cal H} =
  -t
  \sum_{\langle i,j \rangle, \sigma} 
  \left(
  \hat{c}_{i\sigma}^\dag \hat{c}_{j\sigma}
  + {\rm H. c.}
  \right)
  +
  U_{\rm H}
  \sum_{i}
  \left(\hat{n}_{i\uparrow} -\frac{1}{2}\right)
  \left(\hat{n}_{i\downarrow} -\frac{1}{2}\right),
   \label{Ham_Hubbard}
\end{equation}
where
$\hat{c}_{i\sigma}^\dag$ ($\hat{c}_{i\sigma}$)
is the creation (annihilation) operator
of a fermion at site $i\,(=1,2,\dots,L)$ with spin $\sigma\,(=\uparrow,\downarrow)$ on a lattice 
having $L$ sites, 
and $\hat{n}_{i\sigma}=\hat{c}_{i\sigma}^\dag \hat{c}_{i\sigma}$
is the fermion density operator. 
$t$ and $U_{\rm H}$ are the hopping and the on-site
interaction parameters, respectively, and 
$\langle i,j \rangle$
runs over all nearest-neighbor pairs of sites $i$ and $j$. 
We assume that $t>0$ and $U_{\rm H}>0$.

We adopt the Jordan-Wigner transformation~\cite{Jordan1928,Rodriguez1959} of the form  
\begin{equation}
  \hat{c}_{i\sigma}^\dag 
  \overset{\rm JWT}{=} 
  \frac{1}{2}(\hat{X}_{i_\sigma} - \imag \hat{Y}_{i_\sigma}) 
  \prod_{k < {i_\sigma}} \hat{Z}_k
  \label{eq:JWT1}
\end{equation}
and
\begin{equation}
  \hat{c}_{i\sigma} 
  \overset{\rm JWT}{=} 
  \frac{1}{2}(\hat{X}_{i_\sigma} + \imag \hat{Y}_{i_\sigma}) 
  \prod_{k < {i_\sigma}} \hat{Z}_k,
  \label{eq:JWT2}
\end{equation}
where ``$\overset{\rm JWT}{=}$'' implies that the 
Jordan-Wigner transformation transforms fermion operators 
on the left-hand side to a qubit representation on the right-hand side, 
and ${\hat X}_{i_\sigma}$, ${\hat Y}_{i_\sigma}$, and ${\hat Z}_{i_\sigma}$ are the Pauli operators at qubit $i_\sigma$. 
We use the convention that the qubits are ordered as $i_\uparrow = i\,(=1,2,\dots,L) $ and $i_\downarrow = i+L\,(=L+1,L+2,\dots,2L)$ 
with the total number of qubits being $N=2L$ in Eqs.~(\ref{eq:JWT1}) and (\ref{eq:JWT2}). 
The fermion density operator is given by
\begin{equation}
  \hat{n}_{i\sigma}
  =\hat{c}_{i\sigma}^\dag \hat{c}_{i\sigma}
  \overset{\rm JWT}{=} 
  \frac{1}{2}(1-\hat{Z}_{i_\sigma}), 
  \label{eq.density}
\end{equation}
and thus it is plausible to say that
the single-particle state $(i,\sigma)$ is occupied (unoccupied) 
if the $i_\sigma$th qubit state is 
$|1\rangle_{i_\sigma}$ ($|0\rangle_{i_\sigma}$) because 
$\hat{n}_{i\sigma}|1\rangle_{i_\sigma}=1 |1\rangle_{i_\sigma}$ 
($\hat{n}_{i\sigma}|0\rangle_{i_\sigma}=0 |0\rangle_{i_\sigma}$), 
where $|1\rangle_{i_\sigma}$ and $|0\rangle_{i_\sigma}$ are the eigenstates of Pauli operator $\hat{Z}_{i_\sigma}$ 
at qubit $i_\sigma$, 
i.e., $\hat{Z}_{i_\sigma}|0\rangle_{i_\sigma}=|0\rangle_{i_\sigma}$ and $\hat{Z}_{i_\sigma}|1\rangle_{i_\sigma}=-|1\rangle_{i_\sigma}$.
The Fermi-Hubbard Hamiltonian in the qubit representation reads 
\begin{equation}
  \hat{\mathcal{H}}
  \overset{\rm JWT}{=}
  -\frac{t}{2} \sum_\sigma \sum_{\langle i_\sigma, j_\sigma \rangle}
  \left(
  \hat{X}_{i_\sigma} \hat{X}_{j_\sigma} +
  \hat{Y}_{i_\sigma} \hat{Y}_{j_\sigma} 
  \right)
  \hat{Z}_{{\rm JW},i_\sigma j_\sigma} 
  + \frac{U_{\rm H}}{4} \sum_{i=1}^{L} \hat{Z}_i \hat{Z}_{i+L}, 
  \label{JWHub}
\end{equation}
where 
$\hat{Z}_{{\rm JW},ij}=\prod_{i \lessgtr k \lessgtr j} \hat{Z}_k$ 
is the Jordan-Wigner string for $i \lessgtr k \lessgtr j$ 
and $\hat{Z}_{{\rm JW},ij}=\hat{1}$ for $i=j\pm 1$.

\begin{center}
  \begin{figure}
    \includegraphics[width=0.35\textwidth]{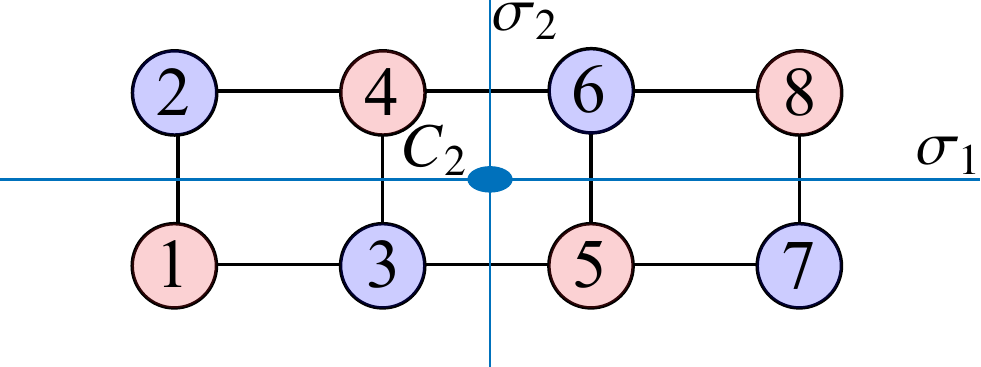}
    \caption{
      Geometry of a $4 \times 2$ lattice in which the Fermi-Hubbard model is defined. 
      Circles represent sites that are labeled by the numbers shown inside the circles. 
      The hopping term is present only between the nearest-neighboring sites connected by black solid lines. 
      The lattice sites belonging to $A$ ($B$) sublattice 
      are denoted with red (blue) circles. 
      The two-fold rotational axis ($C_2$) and the mirror planes ($\sigma_1, \sigma_2$) 
      for the $C_{2v}$ symmetry are also indicated in the figure.
      \label{fig.geometry}
    }
  \end{figure}
\end{center}

In this paper, we consider the Fermi-Hubbard model in a $4 \times 2$ lattice 
under open-boundary conditions (see Fig.~\ref{fig.geometry}) at half filling. 
Therefore, 
to represent the variational state, we use $N=16$ qubits, 
where the first 8 qubits are assigned to the single-particle states  
labeled by site $i$ and spin $\uparrow$, 
and the remaining 8 qubits are 
assigned to the single-particle states  
labeled by site $i$ and spin $\downarrow$ (also see Fig.~\ref{fig:lattice2}).

\subsection{Hamiltonian symmetry}

We now consider the Hamiltonian symmetry. 
From the particular geometry of the cluster 
shown in Fig.~\ref{fig.geometry}, it is obvious that
the model has the $C_{2v}$ point-group symmetry, i.e.,  
\begin{equation}
  \left[\hat{\cal H}, \hat{g}_{m}\right] =0, 
\end{equation}
where $\hat{g}_m$ is a group element of $C_{2v}$.
We note that the $4\times2$ cluster itself possesses
the $D_{2h}$ point-group symmetry that has 8 symmetry operations
(i.e., symmetry elements),
and the $C_{2v}$ group is a subgroup of the $D_{2h}$ group. 
Indeed, the $D_{2h}$ group
can be seen as a direct-product group of the $C_{2v}$ group and the $C_i$ group, 
i.e., $D_{2h}=C_{2v}\times C_i$.
However, in a pure two-dimensional space, the inversion and the two-fold rotation 
around the principal axis are equivalent.
Therefore, among the 8 symmetry operations in the $D_{2h}$ group, 
only the 4 symmetry operations, which are all the symmetry operations of the $C_{2v}$ group, 
are independent. This is the reason why we consider the $C_{2v}$ point-group symmetry here.

It can also be shown that the Fermi-Hubbard model considered here is spin symmetric, i.e., 
\begin{equation}
  \left[\hat{\cal H}, \hat{S}^2 \right] =0, \quad   
  \left[\hat{\cal H}, \hat{S}_z \right] =0, 
  \label{eq:Ssym}
\end{equation}
where $\hat{S}^2=\hat{S}_x^2 + \hat{S}_y^2 + \hat{S}_z^2$,  
$\hat{S}_{x} = \sum_{i}\hat{S}^{x}_i$,   
$\hat{S}_{y} = \sum_{i}\hat{S}^{y}_i$, and    
$\hat{S}_{z} = \sum_{i}\hat{S}^{z}_i$ with   
\begin{alignat}{1}
  \hat{S}_{i}^{x} &=
  \hat{\mb{c}}_i^\dag \mb{s}_x \hat{\mb{c}}_i
  =
  \frac{1}{2}\left( 
  \hat{c}_{i\uparrow}^\dag \hat{c}_{i\downarrow} +
  \hat{c}_{i\downarrow}^\dag \hat{c}_{i\uparrow}
  \right),
  \\
  \hat{S}_{i}^{y} &=
  \hat{\mb{c}}_i^\dag \mb{s}_y \hat{\mb{c}}_i
  =
  \frac{1}{2\imag}\left( 
  \hat{c}_{i\uparrow}^\dag \hat{c}_{i\downarrow} -
  \hat{c}_{i\downarrow}^\dag \hat{c}_{i\uparrow}
  \right),
  \\
  \hat{S}_{i}^{z} &=
  \hat{\mb{c}}_i^\dag \mb{s}_z \hat{\mb{c}}_i
  =
  \frac{1}{2}\left(
  \hat{n}_{i\uparrow} - \hat{n}_{i\downarrow}
  \right).
\end{alignat}
Here, we have introduced that  
$\hat{\mb{c}}_i^\dag =
\begin{bmatrix}
  \hat{c}_{i\uparrow}^\dag & \hat{c}_{i\downarrow}^\dag
\end{bmatrix}$
and 
\begin{equation}
  \mb{s}_x=
  \frac{1}{2}
  \begin{bmatrix}
      0 & 1 \\
      1 & 0
  \end{bmatrix},
  \quad
  \mb{s}_y=
  \frac{1}{2}
  \begin{bmatrix}
      0 & -\imag \\
      \imag & 0
  \end{bmatrix}, 
  \quad
  \mb{s}_z=
  \frac{1}{2}
  \begin{bmatrix}
      1 & 0 \\
      0 & -1
  \end{bmatrix}. 
\end{equation}
Hereafter, we refer to Eq.~(\ref{eq:Ssym}) as the $S$ symmetry.

Due to the bipartite structure of the hopping term (see Fig.~\ref{fig.geometry}), 
the Fermi-Hubbard model has another symmetry~\cite{Yang1989} 
\begin{equation}
  \left[\hat{\cal H}, \hat{\eta}^2 \right] =0, \quad   
  \left[\hat{\cal H}, \hat{\eta}_z \right] =0, 
  \label{eq:Esym}
\end{equation}
where $\hat{\eta}^2=\hat{\eta}_x^2 + \hat{\eta}_y^2 + \hat{\eta}_z^2$,  
$\hat{\eta}_{x} = \sum_{i}\hat{\eta}^{x}_i$,   
$\hat{\eta}_{y} = \sum_{i}\hat{\eta}^{y}_i$, and    
$\hat{\eta}_{z} = \sum_{i}\hat{\eta}^{z}_i$ with   
\begin{alignat}{1}
  \hat{\eta}_{i}^{x} &=
  \hat{\mb{b}}_i^\dag \mb{s}_x \hat{\mb{b}}_i
  =
  \frac{\e^{\imag \phi_i}}{2}
  \left(
  \hat{c}_{i\uparrow}^\dag \hat{c}_{i\downarrow}^\dag
  +
  \hat{c}_{i\downarrow} \hat{c}_{i\uparrow} 
  \right), \label{eq:eta_x}
  \\
  \hat{\eta}_{i}^{y} &=
  \hat{\mb{b}}_i^\dag \mb{s}_y \hat{\mb{b}}_i
  =
  \frac{\e^{\imag \phi_i}}{2\imag}
  \left(
  \hat{c}_{i\uparrow}^\dag \hat{c}_{i\downarrow}^\dag
  -
  \hat{c}_{i\downarrow} \hat{c}_{i\uparrow} 
  \right), \label{eq:eta_y}
  \\
  \hat{\eta}_{i}^{z} &=
  \hat{\mb{b}}_i^\dag \mb{s}_z \hat{\mb{b}}_i
  =
  \frac{1}{2}
  \left(
  \hat{n}_{i\uparrow} + \hat{n}_{i\downarrow} - 1
  \right). \label{eq:eta_z}
\end{alignat}
Here, we have introduced that  
$\hat{\mb{b}}_i^\dag =
\begin{bmatrix}
  \hat{c}_{i\uparrow}^\dag & \e^{\imag \phi_i} \hat{c}_{i\downarrow}
\end{bmatrix}$ and
\begin{equation}
  \e^{\imag \phi_i} =
  \e^{-\imag \phi_i} =
  \begin{cases}
    +1 &  \text{for $i\in A$ sublattice} \\
    -1 &  \text{for $i\in B$ sublattice}
  \end{cases} \label{eq:phii}
\end{equation}
with $A$ and $B$ sublattices being indicated in Fig.~\ref{fig.geometry}. 
Hereafter, we refer to Eq.~(\ref{eq:Esym}) as the $\eta$ symmetry. 
Note that, as in the case of the spin operators ${\hat S}_i^\alpha$ with $\alpha=x,y,z$, 
the $\eta$ operators ${\hat \eta}_i^\alpha$ in Eqs.~(\ref{eq:eta_x})-(\ref{eq:eta_z}) satisfy 
the commutation relations of the angular momentum.  
It is also important to note that $[{\hat S}^\alpha_i,{\hat \eta}^\beta_j]=0$ with $\alpha,\beta=x,y,z$. 
Therefore, any eigenstate of $\hat{\cal H}$ can be simultaneously an eigenstate of ${\hat S}^2$, ${\hat S}_z$, 
${\hat\eta}^2$, and ${\hat\eta}_z$, and thus it is characterized with these eigenvalues.

\subsection{Spin- and $\eta$-singlet ground state}

Via the particle-hole transformation for spin-down fermions 
\begin{alignat}{1}
  \hat{c}_{i\downarrow} &\mapsto \e^{-\imag \phi_i}\hat{c}_{i\downarrow}^\dag
  \label{shiba-ph}
\end{alignat}
with spin-up fermions unaltered, 
the spin and the $\eta$ operators are transformed into each other because  
$\hat{\mb{c}}_{i}^\dag \mapsto \hat{\mb{b}}_{i}^\dag$ and 
$\hat{\mb{b}}_{i}^\dag \mapsto \hat{\mb{c}}_{i}^\dag$, assuming that $\e^{\imag \phi_i}$ is given in Eq.~(\ref{eq:phii}).   
As stated above, this observation also implies that the $\eta$ operators satisfy the same commutation relations 
as the spin operators. 
Note that a unitary operator 
$\hat{U}_i^{\rm ph}=\hat{c}_{i\downarrow}+\e^{-\imag \phi_i} \hat{c}_{i\downarrow}^\dag$~\cite{Tasaki}
satisfies the relation 
$(\hat{U}_i^{\rm ph})^\dag \hat{c}_{i\downarrow} \hat{U}_i^{\rm ph}=\e^{-\imag \phi_i} \hat{c}_{i\downarrow}^\dag$, 
which corresponds to the mapping in Eq.~(\ref{shiba-ph}).

With the same particle-hole transformation, 
the repulsive Fermi-Hubbard model on a bipartite lattice maps to 
the attractive Fermi-Hubbard model on the same lattice~\cite{Shiba1972}.
The particle-hole transformation is useful to gain insight into 
the correspondence between the repulsive and attractive Fermi-Hubbard models 
as well as that between the spin and charge degrees of freedom 
(see for example Refs.~\cite{You2017,Karakuzu2018negU,Fujiuchi2020,Li2020}).
Indeed, the fact~\cite{Lieb1989} that 
the ground state of the attractive Fermi-Hubbard model for even number of fermions 
in the absence of magnetic field has $S=0$ (i.e., spin singlet) 
implies that 
the ground state of the repulsive Fermi-Hubbard model on a bipartite lattice 
at half filling has $\eta=0$ (i.e., $\eta$ singlet). 
Furthermore,
the ground state of the repulsive Fermi-Hubbard model on a bipartite lattice at half filling 
is proved to have $S=\frac{1}{2}| |B| - |A| |$, where $|B|$ ($|A|$) is the 
number of sites in the $B$ ($A$) sublattice~\cite{Lieb1989}. 
Since $|B|=|A|$ in the present case, the ground state has $S=0$.

\subsection{Projection operators}\label{sec:Poperator}

To take into account the spatial, $S$, and $\eta$ symmetry, 
we adopt the projection operator $\hat{\cal P}$ of the form
\begin{equation}
  \hat{\cal P} = \hat{\cal P}^{(\eta)} \hat{\cal P}^{(S)} \hat{\cal P}^{(\alpha)}, 
\end{equation}
where $\hat{\cal P}^{(\alpha)}$, $\hat{\cal P}^{(S)}$, and $\hat{\cal P}^{(\eta)}$ are 
the projection operators of 
the spatial, $S$, and $\eta$ symmetry, respectively. 
In this section, we specify the form of these projection operators. 
Note that these three projection operators commute with each other.

\subsubsection{Spatial symmetry}

The projection operator of the spatial symmetry for 
an irreducible representation $\alpha$ of group $\cal G$ is given by~\cite{Inui}
\begin{equation}
  \hat{\cal P}^{(\alpha)}_{\mu\nu}= \frac{d_{\alpha}}{|\cal G|} 
  \sum_{\hat{g}_m\in{\cal G}} 
  \left[D^{(\alpha)}_{\mu\nu}(\hat{g}_m)\right]^* \hat{g}_m,
  \label{eq:projection0}
\end{equation}
where 
$|{\cal G}|$ is the number of group elements in ${\cal G}$ (i.e., order of group $\cal G$), 
$d_\alpha$ is the dimension of the irreducible representation $\alpha$, 
$D_{\mu\nu}^{(\alpha)}(\hat{g}_m)$ is the $(\mu,\nu)$ entry of 
the representation matrix in the irreducible representation $\alpha$ 
for the symmetry operation $\hat{g}_m$, and 
the sum runs over all the group elements.   
The character of the representation matrix $D^{(\alpha)}_{\mu\nu}(\hat{g}_m)$ is defined by
$\chi^{(\alpha)} (\hat{g}_m) = \sum_{\mu=1}^{d_\alpha} D^{(\alpha)}_{\mu\mu}(\hat{g}_{m})$.  
Note that $D_{\mu\nu}^{(\alpha)}$ satisfies the orthogonality relation
\begin{equation}
  \sum_{\hat{g}_m\in {\cal G}}
  \left[D_{\mu\nu}^{(\alpha)}(\hat{g}_m)\right]^*
  D_{\mu^\prime \nu^\prime}^{(\alpha^\prime)}(\hat{g}_m) = 
  \frac{|{\cal G}|}{d_\alpha} 
  \delta_{\alpha \alpha^\prime}
  \delta_{\mu \mu^\prime}
  \delta_{\nu \nu^\prime}.
\end{equation}
$\hat{\cal P}_{\mu \nu}^{(\alpha)}$ extracts 
a basis state $|\alpha \mu\rangle$ from $|\alpha \nu \rangle$, i.e., 
$\hat{\cal P}_{\mu \nu}^{(\alpha)} |\alpha \nu \rangle = |\alpha \mu \rangle$, 
where $|\alpha \nu \rangle$ is a state such that 
$\hat{g}_m |\alpha \nu \rangle = \sum_{\mu} D_{\mu \nu}^{(\alpha)}(\hat{g}_m)|\alpha\mu\rangle$. 
Strictly speaking, $\hat{\cal P}_{\mu \nu}^{(\alpha)}$ is a projection operator 
only when $\mu=\nu$ because 
$\hat{\cal P}_{\mu \nu}^{(\alpha)}
\hat{\cal P}_{\nu^\prime \mu^\prime}^{(\alpha^\prime)}
=
\delta_{\alpha \alpha^\prime}
\delta_{\nu \nu^\prime}
\hat{\cal P}_{\mu \mu^\prime}^{(\alpha)}
$,
but here we loosely use the term ``projection operator'' for
$\hat{\cal P}_{\mu \nu}^{(\alpha)}$.

For the point group $C_{2v}$,  
$|{\cal G}|=4$, $d_\alpha=1$, and 
$\{\hat{g}_m\}_{m=1}^{|{\cal G}|}
=\{\hat{I}, \hat{C}_2, \hat{\sigma}_1, \hat{\sigma_2}\}$, 
where $\hat{C}_2$ is the $\pi$ rotation about the 
center of the $4 \times 2$ cluster, and $\sigma_1$ and $\sigma_2$ 
are reflections with respect to the corresponding planes 
(see Fig.~\ref{fig.geometry}). 
Omitting the subscript $\mu\nu$ in Eq.~(\ref{eq:projection0}) because of $d_\alpha=1$, we have 
\begin{equation}
  \hat{\cal P}^{(\alpha)} = \frac{1}{4}\sum_{m=1}^{4} \left[\chi^{(\alpha)} (\hat{g}_m)\right]^* \hat{g}_m. 
\end{equation}
The ground state of the Fermi-Hubbard model studied here belongs to the irreducible representation $\alpha=A_1$, in which 
$\chi^{(\alpha)}(\hat{g}_{m})=1$ for all $\hat{g}_{m}$.

\subsubsection{$S$ and $\eta$ symmetry}

Since both $S$ and $\eta$ operators satisfy the same commutation 
relations of the angular momentum, 
we denote by $J$ either $S$ or $\eta$ for convenience. 
According to the theory of rotation group, 
the projection operator of the 
$J$ symmetry ($J=S,\eta$)
is given by~\cite{Ring,Schmid2004,Mizusaki2004,Bally2021}
\begin{equation}
  \hat{\cal P}^{(J)}_{MK}=\frac{2J+1}{\Omega} 
  \int \dd \omega \left[D^{(J)}_{M K}(\omega)\right]^* \hat{R}(\omega),
  \label{eq:Jproj}
\end{equation}
where 
\begin{equation}
  \hat{R}(\omega) = 
  \e^{-\imag \alpha \hat{J}_z }
  \e^{-\imag \beta \hat{J}_y }
  \e^{-\imag \gamma \hat{J}_z }
\end{equation}
is the rotation operator with 
$\omega=(\alpha,\beta,\gamma)$, i.e., the parameters of 
rotation group (Euler angles) specifying the group element, and    
\begin{equation}
\Omega=
\int \dd\omega =
\int_{0}^{2\pi} \dd \alpha 
\int_{0}^{\pi}  \dd \beta \sin{\beta} 
\int_{0}^{2\pi} \dd \gamma 
=8\pi^2
\label{eq:Omega}
\end{equation}  
defines the volume of the parameter $\omega$ space.  

$D_{MK}^{(J)}(\omega)$ in Eq.~(\ref{eq:Jproj}) is the Wigner's $D$ function defined by
\begin{equation}
  D^{(J)}_{MK}(\omega)= \langle J M| \hat{R}(\omega) | J K\rangle 
= \e^{-\imag \alpha M } 
d^{(J)}_{MK}(\beta)
\e^{-\imag \gamma K} 
\end{equation}
with $|JM\rangle$ being an eigenstate of $\hat{J}^2$ and $\hat{J}_z$ such that 
$\hat{J}^2|JM\rangle = J(J+1)|JM\rangle$ and 
$\hat{J}_z|JM\rangle = M|JM\rangle$, and 
\begin{equation}
  d^{(J)}_{MK}(\beta)=
  \langle J M| \e^{-\imag \beta \hat{J}_y} | J K\rangle 
\end{equation} 
is the Wigner's small $d$ function that is real~\cite{Rose,Wigner,Varshalovich}. 
Note that the Wigner's $D$ function $D_{MK}^{(J)}(\omega)$ satisfies the orthogonality relation 
\begin{equation}
  \int \dd\omega 
  \left[D_{MK}^{(J)}(\omega)\right]^*
  D_{M^\prime K^\prime}^{(J^\prime)}(\omega)
  =\frac{\Omega}{2J+1}
  \delta_{JJ^\prime}
  \delta_{MM^\prime}
  \delta_{KK^\prime}.
  \label{eq:Dorthogonal}
\end{equation}
Therefore, 
$\hat{\cal P}_{MK}^{(J)}$ extracts 
a basis state $|J M\rangle$ from $|J K\rangle$, i.e., 
$\hat{\cal P}_{MK}^{(J)} |J K \rangle = |J M \rangle$ 
because 
$\hat{R}(\omega) |J K \rangle = \sum_{M} D_{MK}^{(J)}(\omega)|JM\rangle$. 
In Eqs.~(\ref{eq:Omega}) and (\ref{eq:Dorthogonal}), $J$ is assumed to be integer. 
For a half integer $J$, the range of either $\alpha$ or $\gamma$ integration 
should be $[0, 4\pi]$ and hence $\Omega=16\pi^2$~\cite{Inui,Bally2021,Varshalovich}.
Details on $D_{MK}^{(J)}(\omega)$ and $d_{MK}^{(J)}(\beta)$
can be found, for example, in Ref.~\cite{Varshalovich}.

Since our target state is 
the ground state of the Fermi-Hubbard model 
in the sector of $\eta_z=0$ (i.e., half filling) and
$S_z=0$ (i.e., zero magnetization), 
we can set $M=0$ in Eq.~(\ref{eq:Jproj}) for both $J=\eta$ and $S$. 
In addition, the variational state $|\psi(\bs{\theta})\rangle$,  
whose concrete form is described in Sec.~\ref{varstate},
is constructed to satisfy $\hat{J}_z|\psi(\bs{\theta})\rangle=0$, implying that 
we can set $K=0$ in Eq.~(\ref{eq:Jproj}) for 
$J=\eta$ and $S$. 
For $K=M=0$, the Wigner's $D$ function is given by
\begin{equation}
  D^{J}_{00}(\omega)=d^{J}_{00}(\beta)=P_{J}(\cos{\beta}), 
\end{equation}
where $P_J$ is the $J$th order Legendre polynomial.  
By integrating out $\gamma$, 
the state 
$
\hat{\cal P}^{(J)} |\psi(\bs{\theta})\rangle 
\equiv 
\hat{\cal P}^{(J)}_{00} |\psi(\bs{\theta})\rangle $
is given by~\cite{Tahara2008}
\begin{alignat}{1}
  \hat{\cal P}^{(J)}|\psi(\bs{\theta})\rangle 
  &=
  \frac{2J+1}{4\pi} 
  \int_0^{2\pi} \dd \alpha 
  \int_0^{\pi} \dd \beta  \sin{\beta} P_{J}(\cos{\beta}) 
  \notag \\
  &
  \times
  \e^{-\imag \alpha \hat{J}_z }
  \e^{-\imag \beta \hat{J}_y }
  |\psi(\bs{\theta})\rangle.
  \label{eq:PJ}
\end{alignat}
Furthermore, when a matrix element
$ \langle \psi^\prime |\hat{\cal P}^{(J)}| \psi (\bs{\theta})\rangle$ 
with $\hat{J}_z|\psi^\prime\rangle=0$ is considered, 
one can also integrate out $\alpha$ as~\cite{Tahara2008} 
\begin{alignat}{1}
  \langle \psi^\prime |\hat{\cal P}^{(J)}|\psi(\bs{\theta})\rangle 
  &=
  \frac{2J+1}{2} 
  \int_0^{\pi} \dd \beta  \sin{\beta} P_{J}(\cos{\beta}) 
  \notag \\
  &
  \times
  \langle \psi^\prime | 
  \e^{-\imag \beta \hat{J}_y }
  |\psi(\bs{\theta})\rangle.
  \label{eq:PJ00}
\end{alignat}
Examples of $|\psi^\prime\rangle$ 
particularly relevant for the present study include 
$|\psi(\bs{\theta}) \rangle$,  
$\hat{\cal H}|\psi(\bs{\theta}) \rangle$,
and 
$\hat{J}^2|\psi(\bs{\theta}) \rangle$.
Moreover, 
Eqs.~(\ref{eq:PJ}) and (\ref{eq:PJ00}) still hold even when 
$|\psi(\bs{\theta})\rangle$ in these equations is replaced with
$\hat{\cal H}_{\rm ST(1)}^n|\psi(\bs{\theta})\rangle$
because 
$\hat{J}_z\hat{\cal H}_{\rm ST(1)}^n|\psi(\bs{\theta})\rangle=0$ 
for the ST decomposition employed (see Sec.~\ref{sec:std}).

\subsection{Projection operator as a linear combination of unitaries}
Since the rotation group is a continuous group, $\hat{\cal P}^{(J)}$  
involves integration over continuous variables, i.e., Euler angles, as in Eqs.~(\ref{eq:PJ}) and (\ref{eq:PJ00}).
However, for numerical simulations as well as quantum-classical hybrid calculations, 
a proper discretization of the integration is necessary.
The integration over $\alpha$ and $\beta$ in Eq.~(\ref{eq:PJ}) 
can be discretized with the trapezoidal rule and the Gauss-Legendre quadrature, 
respectively. 
By omitting irrelevant normalization factor, 
$\hat{\cal P}^{(J)}|\psi(\bs{\theta})\rangle$ is now approximated as 
\begin{alignat}{1}
&  \hat{\cal P}^{(J)}|\psi(\bs{\theta})\rangle \nonumber\\
\approx &
  \frac{2J+1}{2N_{J,\rm azimuth}}
  \sum_{i=1}^{N_{J,\rm azimuth}} 
  \sum_{j=1}^{N_{J, \rm polar}}
  w_{j,J}
  P_J(\cos{\beta}_{j,J})
  \e^{-\imag \alpha_{i,J} \hat{J}_z} 
  \e^{-\imag \beta_{j,J} \hat{J}_y} |\psi(\bs{\theta})\rangle,    
\end{alignat}
where $\{\alpha_{i,J}\}_{i=1}^{N_{J,\rm azimuth}}$ are 
$N_{J, \rm azimuth}$ azimuth angles equally spaced in $[0,2\pi]$, 
$\{\beta_{j,J}\}_{j=1}^{N_{J, \rm polar}}$ are polar angles 
such that $P_{N_{J, \rm polar}}(\cos{\beta_{j,J}})=0$,    
and $\{w_{j,J}\}_{j=1}^{N_{J, \rm polar}}$ are the corresponding 
integration weight of the Gauss-Legendre quadrature.

The full symmetry-projected state is now given approximately as 
\begin{alignat}{1}
  &
  \hat{\cal P}^{(\eta)} 
  \hat{\cal P}^{(S)} 
  \hat{\cal P}^{(\alpha)} |\psi(\bs{\theta})\rangle \notag \\    
  \approx &
  \frac{2\eta+1}{2N_{\eta, \rm azimuth}}
  \frac{2S+1}{2N_{S, \rm azimuth}}
  \frac{d_\alpha}{|{\cal G}|}
  \sum_{i=1}^{N_{\eta, \rm azimuth}} 
  \sum_{j=1}^{N_{\eta, \rm polar}}
  \sum_{k=1}^{N_{S, \rm azimuth}} 
  \sum_{l=1}^{N_{S, \rm polar}}
  \sum_{m=1}^{|\cal G|} \notag \\
  \times &
  w_{j,\eta}
  w_{l,S}
  P_{\eta}(\cos{\beta}_{j,\eta})
  P_{S}(\cos{\beta}_{l,S})
  \left[\chi^{(\alpha)}(\hat{g}_m)\right]^* \notag \\
  \times &
  \e^{-\imag \alpha_{i,\eta} \hat{\eta}_z} 
  \e^{-\imag \beta_{j,\eta} \hat{\eta}_y}
  \e^{-\imag \alpha_{k,S}\hat{S}_z} 
  \e^{-\imag \beta_{l,S} \hat{S}_y}
  \hat{g}_m |\psi(\bs{\theta})\rangle. 
  \label{eq:PPP}
\end{alignat}
Note that the operators in the last line are unitary and thus the symmetry-projected state is 
evaluated by applying a linear combination of unitary operators to the state. 
The symmetry-projected state given in Eq.~(\ref{eq:PPP}) is
in general
not normalized even though the state $ |\psi(\bs{\theta})\rangle$ is normalized. 
However, this is not a problem because 
the normalization of states spanning the Krylov subspace $\cal U$ is not required (see Sec~\ref{sec:form} and Sec.~\ref{sec:ksVQE})
but the normalization of the approximated ground state
is guaranteed by Eq.~(\ref{normalization}).

A matrix element similar to that in Eq.~(\ref{eq:PJ00}) 
but for the full projection operators 
can be evaluated simply by setting $N_{J,\rm azimuth}=1$ 
and $\e^{-\imag \alpha_{1,J}\hat{J}_z}=\hat{I}$
in Eq.~(\ref{eq:PPP}) and taking the overlap with $\langle \psi^\prime|$, i.e.,  
\begin{alignat}{1}
  &
  \langle \psi^\prime|
  \hat{\cal P}^{(\eta)} 
  \hat{\cal P}^{(S)} 
  \hat{\cal P}^{(\alpha)} |\psi(\bs{\theta})\rangle\notag \\
  \approx &
  \frac{2\eta+1}{2}
  \frac{2S+1}{2}
  \frac{d_\alpha}{|{\cal G}|}
  \sum_{j=1}^{N_{\eta, \rm polar}}
  \sum_{l=1}^{N_{S, \rm polar}}
  \sum_{m=1}^{|\cal G|} \notag \\
  \times &
  w_{j,\eta}
  w_{l,S}
  P_{\eta}(\cos{\beta}_{j,\eta})
  P_{S}(\cos{\beta}_{l,S})
  \left[\chi^{(\alpha)}(\hat{g}_m)\right]^* \notag \\
  \times &
  \langle \psi^\prime|
  \e^{-\imag \beta_{j,\eta} \hat{\eta}_y}
  \e^{-\imag \beta_{l,S} \hat{S}_y}
  \hat{g}_m |\psi(\bs{\theta})\rangle. 
  \label{eq:PPP2}
\end{alignat}
Concerning quantum-classical hybrid calculations, 
Eq.~(\ref{eq:PPP2}) implies 
that the matrix element on the left-hand side 
can be estimated by evaluating 
the matrix elements 
$
{\Big\{}
{\Big\{}
{\Big\{}
\langle \psi^\prime|
\e^{-\imag \beta_{j,\eta} \hat{\eta}_y}
\e^{-\imag \beta_{l,S} \hat{S}_y}
\hat{g}_m |\psi(\bs{\theta})\rangle
{\Big\}}_{m=1}^{|{\cal G}|}
{\Big\}}_{l=1}^{N_{S, \rm polar}}
{\Big\}}_{j=1}^{N_{\eta, \rm polar}}
$ 
with 
$N_{\eta, \rm polar} N_{S, \rm polar} |\cal G|$ 
different circuit structures 
using quantum computers,  
and then adding all of them with the proper weights 
on classical computers. 
We should recall that 
$d_\alpha=1$ and $|{\cal G}|=4$ for the $C_{2v}$ point group, 
$P_\eta(\cos{\beta}_{j,\eta})=1$ for $\eta=0$,  
$P_S(\cos{\beta}_{l,S})=1$ for $S=0$, and  
$\chi^{(\alpha)}(\hat{g}_m)=1$ for $\alpha=A_1$.

While Eq.~(\ref{eq:PPP2}) suffices for our purpose,
we briefly discuss how many of the integration points
$N_{J, \rm polar}$ and $N_{J, \rm azimuth}$ in Eq.~(\ref{eq:PPP})
are required for the variational state $|\psi(\bs{\theta})\rangle$ 
that will be described in Sec.~\ref{varstate}. 
We find numerically that typically 
$N_{S, \rm polar}=2$ and $N_{S, \rm azimuth}=4$ 
($N_{\eta, \rm polar}=3$ and $N_{\eta, \rm azimuth}=5$) 
are enough to ensure that the calculated value of $S$ ($\eta$)  
with respect to the state $\hat{\cal P}|\psi(\bs{\theta})\rangle$ 
can be regarded as integer within the double-precision arithmetic. 
We also find that typically 
the lowest-order Lebedev quadrature 
$\hat{\cal P}^{(J)} \approx 
  \sum_{i=1}^{6}
  \e^{-\imag \alpha_{i,J} \hat{J}_z} 
  \e^{-\imag \beta_{i,J} \hat{J}_y}$,  
which uses six integration points 
$(\alpha_{i,J}, \beta_{i,J})=
(0,     0),
(0,     \pi/2),
(\pi/2, \pi/2),
(\pi,   \pi/2),
(3\pi/2,\pi/2),$ and 
$(0,     \pi)$ with equal integration weights, 
is adequate at least for the $S$-symmetry projection.  
On the other hand, when we use Eq.~(\ref{eq:PPP2}), where 
$N_{J, \rm azimuth}=1$ by construction, 
we find that typically 
$N_{S, \rm polar}=2$ ($N_{\eta,\rm polar}=3$) 
is sufficient to evaluate $S$ ($\eta$) 
within the double-precision arithmetic. 
However,
to ensure that the symmetry projection is made 
essentially exactly, 
we set $N_{J, \rm polar}=4$ for both $J=S$ and $\eta$ projections 
in our numerical simulations.

\section{Symmetry operation on quantum circuit}\label{sec:symmetry}

In this section, we describe how 
the spatial symmetry operators $\hat{g}_m$ and 
the rotation operators 
$\e^{-\imag \alpha \hat{J}_y}$ and 
$ \e^{-\imag \beta \hat{J}_z}$ with $J=S,\eta$
for fermions can be implemented on a quantum circuit,  
assuming the qubit representation of fermion operators by the Jordan-Wigner transformation 
in Eqs.~(\ref{eq:JWT1}) and (\ref{eq:JWT2}).

\subsection{Quantum circuit for spatial symmetry operations}

Let $\hat{g}_{m}$ be a spatial-symmetry operation that transfers 
the local state at the $i$th qubit to
the $m(i)$th qubit. 
Such operation can be represented as a permutation 
\begin{equation}
  {\sigma}_{m}\equiv
  \underbrace{
  \left(
  \begin{matrix}
    N & N-1 & \cdots & 2 & 1 \\
    m(N) & m(N-1) & \cdots & m(2) & m(1) \\ 
  \end{matrix}
  \right)
  }_{N \text{\ columns}},
  \label{permS}
\end{equation}
where $N$ is the number of qubits. 
We first briefly review an implementation of the 
spatial-symmetry operations for spins, and then proceed to
the spatial-symmetry operations for fermions.

\subsubsection{Spatial symmetry operations for spins}

\begin{center}
  \begin{figure}
    \includegraphics[width=1\columnwidth]{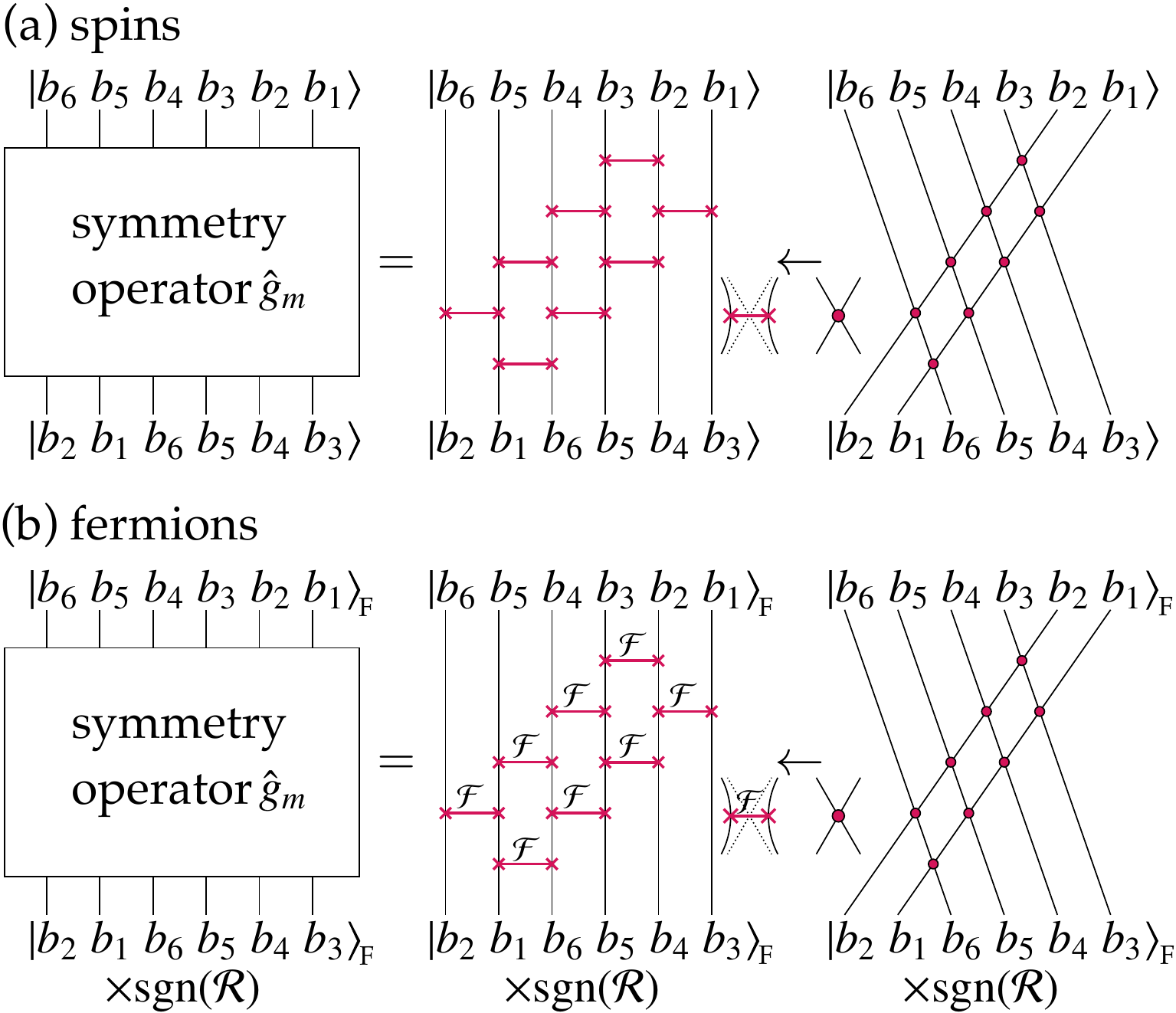}
    \caption{
      Amida-lottery construction of
      a symmetry operator $\hat{g}_m$
      on a quantum circuit for
      (a) spin models in the computational basis~\cite{Seki2020vqe} 
      and (b) fermion models in an occupation basis
      [see Eq.~(\ref{fermion_basis})].
      The figure refers to the case of $N=6$ qubits.
      The vertical lines represent qubits and
      ${\rm sgn}(\cal R)$ in (b) denotes an appropriate sign factor
      for fermions [see Eq.~(\ref{fermion_bin}) and also Eq.~(\ref{reorder})].
      In the middle panel of (a), {\rm swap} gates $\hat{\cal S}_{i\delta(i)}$
      are highlighted with thick red lines.
      In (b),
      a fermionic-{\rm swap} gate $\hat{\cal F}_{i\delta(i)}$ is
      represented by a two-qubit gate with symbol $\cal F$.
      Assuming that the circuit evolves from top to bottom, 
      the middle panels in (a) and (b) show that 
      the symmetry operation $\hat{g}_m$
      in this example can be expressed as
      $\hat{g}_m=
      \hat{\cal S}_{45}
      \hat{\cal S}_{56}\hat{\cal S}_{34}
      \hat{\cal S}_{45}\hat{\cal S}_{23}
      \hat{\cal S}_{34}\hat{\cal S}_{12}
      \hat{\cal S}_{23}$ for spin models
      and
      $\hat{g}_m=
      \hat{\cal F}_{45}
      \hat{\cal F}_{56}\hat{\cal F}_{34}
      \hat{\cal F}_{45}\hat{\cal F}_{23}
      \hat{\cal F}_{34}\hat{\cal F}_{12}
      \hat{\cal F}_{23}$ for fermion models, respectively. 
    }\label{fig:amida}
  \end{figure}
\end{center}

For a system composed of spins, each having spin 1/2, the spatial-symmetry operator $\hat{g}_m$ can be expressed as a product
of nearest-neighbor {\sc swap} operators $\hat{\cal S}_{i\delta(i)}$ 
\begin{equation}
  \hat{g}_{m} = \prod_{{\sigma}_m} \hat{\cal S}_{i\delta(i)},
  \label{amida}
\end{equation}
because any permutation can be expressed as a product of
transpositions (i.e., nearest-neighbor {\sc swap} operations). 
Here, $\delta(i)(=i\pm 1)$ denotes a neighboring qubit of qubit $i$,
and the product $\prod_{{\sigma}_m}$
should contain 
sequences of the {\sc swap} operators of the form
$\hat{\cal S}_{\delta(\cdots \delta(\delta(k))), m(k)}
\cdots
\hat{\cal S}_{\delta(k), \delta(\delta(k))}  
\hat{\cal S}_{k, \delta(k)}$
for every $k$. 
Such a product in Eq.~(\ref{amida})
can be constructed accordingly to the
``Amida lottery'' construction~\cite{Seki2020vqe},
as shown in Fig.~\ref{fig:amida}(a) for the case of 
\begin{equation}
  {\sigma}_{m}=
  \left(
  \begin{matrix}
    6 & 5 & 4 & 3 & 2 & 1 \\
    4 & 3 & 2 & 1 & 6 & 5 \\
  \end{matrix}
  \right) 
  \label{perm6}
\end{equation}
with $|b_6\, b_5\, b_4\, b_3\, b_2\, b_1\rangle=|b_6\rangle_6 |b_5\rangle_5 |b_4\rangle_4 |b_3\rangle_3 |b_2\rangle_2 |b_1\rangle_1$ 
being transferred to 
$|b_2\, b_1\, b_6\, b_5\, b_4\, b_3\rangle=|b_2\rangle_6 |b_1\rangle_5 |b_6\rangle_4 |b_5\rangle_3 |b_4\rangle_2 |b_3\rangle_1$,  
where $|b\rangle_i$ is the local state $b$ ($=0$ or $1$) at the $i$th qubit.

We note that {\sc swap} operator $\hat{\cal S}_{i j}$ acts on qubits $i$ and $j$ as 
\begin{alignat}{1}
  &\hat{\cal S}_{ij}|0\rangle_i |0\rangle_{j} =  |0\rangle_i |0\rangle_{j}, \label{eq:swap1} \\
  &\hat{\cal S}_{ij}|0\rangle_i |1\rangle_{j} =  |1\rangle_i |0\rangle_{j}, \label{eq:swap2} \\
  &\hat{\cal S}_{ij}|1\rangle_i |0\rangle_{j} =  |0\rangle_i |1\rangle_{j}, \label{eq:swap3} \\
  &\hat{\cal S}_{ij}|1\rangle_i |1\rangle_{j} =  |1\rangle_i |1\rangle_{j}.  \label{eq:swap4}
\end{alignat}
The matrix representation of $\hat{\cal S}_{ij}$
in these four basis states 
$|0\rangle_i |0\rangle_{j}$,
$|0\rangle_i |1\rangle_{j}$,
$|1\rangle_i |0\rangle_{j}$, and
$|1\rangle_i |1\rangle_{j}$ 
is thus given by 
\begin{equation}
  \hat{\cal S}_{ij}
  \overset{\cdot}{=}
  \begin{bmatrix}
  1& 0& 0& 0 \\
  0& 0& 1& 0 \\ 
  0& 1& 0& 0 \\ 
  0& 0& 0& 1
  \end{bmatrix}, 
\end{equation}
where $\overset{\cdot}{=}$ implies that the operator on the
left-hand side can be represented in the matrix form on the right-hand side
with the basis states given above.

\subsubsection{Spatial symmetry operations for fermions}

For a system composed of fermions, $\hat{g}_m$ has to take into account 
the anticommutation relations of fermion operators,
in addition to the permutation of local states. 
This implies that symmetry operators can no longer be
given as a product of the {\sc swap} operators
for fermions in general.
As it is explained in Appendix~\ref{app:fermi}, 
the symmetry operation for fermions can be implemented simply
by replacing the {\sc swap} operators $\hat{\cal S}_{i\delta(i)}$ in Eq.~(\ref{amida})
with the fermionic-{\sc swap} operators $\hat{\cal F}_{i\delta(i)}$:
\begin{equation}
  \hat{g}_{m}=
  \prod_{{\sigma}_{m}} \hat{\cal F}_{i \delta(i)}.
  \label{amida2}
\end{equation}
Here, 
$\hat{\cal F}_{i\delta(i)}$ transforms a fermion creation operator $\hat{c}_i^\dag$ in such a way that 
$\hat{\cal F}_{i\delta(i)} \hat{c}_i^\dag \hat{\cal F}_{i\delta(i)}^{-1}=\hat{c}_{\delta(i)}^\dag$,
implying that 
$\hat{g}_m \hat{c}_i^\dag \hat{g}_m^{-1}=\hat{c}_{m(i)}^\dag$.
Note that the subscript $i\ (=1,2,\cdots,N)$ here labels 
all the single-particle local states 
including sites and spins.

Before discussing quantum circuits for
fermionic symmetry operations, 
let us first explain how a state of the form
\begin{alignat}{1}
  |b_N b_{N-1} \cdots b_{1} \rangle_{\rm F}
  &\equiv \prod_{i=N}^{1} (\hat{c}_i^\dag)^{b_i}|0\rangle_{\rm F} \notag \\
  &=
  (\hat{c}_N^\dag)^{b_N}
  (\hat{c}_{N-1}^\dag)^{b_{N-1}}
  \cdots
  (\hat{c}_1^\dag)^{b_1} |0\rangle_{\rm F}
  \label{fermion_basis}
\end{alignat}
is transformed by $\hat{g}_m$,
where
$b_{i}=0$ or $1$ with $(\hat{c}_i^\dag)^0\equiv\hat{1}$ and 
$|0\rangle_{\rm F}$ denotes the fermion vacuum
such that $\hat{c}_i|0\rangle_{\rm F}=0$ for any $i$.
We call a state of the form in Eq.~(\ref{fermion_basis}) an occupation-basis state.
Notice that here we have adopted the convention for the order of 
the single-particle local states $i\ (=1,2,\cdots,N)$ such that the smaller $i$ resides on the further right, 
indicating the product $\prod_{i=N}^{1}\cdots$, 
although any convention for the order can be used. 
Then, the occupation-basis state is transformed by $\hat{g}_m$ as 
\begin{alignat}{1}
  &\hat{g}_m |b_N b_{N-1} \cdots b_{1} \rangle_{\rm F} \notag \\
  =&\hat{g}_m\prod_{i=N}^{1} (\hat{c}_i^\dag)^{b_i}|0\rangle_{\rm F} \notag \\
  =&\prod_{i=N}^{1} (\hat{c}_{m(i)}^\dag)^{b_i}|0\rangle_{\rm F} \notag  \\
  =&\prod_{m=N}^{1} (\hat{c}_{m(i)}^\dag)^{b_i}|0\rangle_{\rm F}\times {\rm sgn}({\cal R}) \notag \\
  =&\prod_{i=N}^{1} (\hat{c}_{i}^\dag)^{b_{m^{-1}(i)}}|0\rangle_{\rm F}\times {\rm sgn}({\cal R}) \notag \\
  =&\,\, |b_{m^{-1}(N)} b_{m^{-1}(N-1)} \cdots b_{m^{-1}(1)} \rangle_{\rm F} \times {\rm sgn}({\cal R}), \label{fermion_bin} 
\end{alignat}
where ${\rm sgn}(\cal R)$ denotes an appropriate sign factor
due to the reordering of the fermion operators, 
and $m^{-1}(i)$ denotes the number obtained by
applying the inverse permutation $\sigma_m^{-1}$ to $i$.
More details on fermionic symmetry operations 
are briefly reviewed in Appendix~\ref{app:fermi}.
An example shown in Fig.~\ref{fig:amida}(b)
corresponds to a symmetry operation
\begin{alignat}{1}
  \hat{g}_m & |b_6\, b_5\, b_4\, b_3\, b_2\, b_1 \rangle_{\rm F} \notag \\
  = & \hat{g}_m 
  (\hat{c}_6^\dag)^{b_6}
  (\hat{c}_5^\dag)^{b_5}
  (\hat{c}_4^\dag)^{b_4}
  (\hat{c}_3^\dag)^{b_3}
  (\hat{c}_2^\dag)^{b_2}
  (\hat{c}_1^\dag)^{b_1}
  |0\rangle_{\rm F}
  \notag \\
  =&
  (\hat{c}_4^\dag)^{b_6}
  (\hat{c}_3^\dag)^{b_5}
  (\hat{c}_2^\dag)^{b_4}
  (\hat{c}_1^\dag)^{b_3}
  (\hat{c}_6^\dag)^{b_2}
  (\hat{c}_5^\dag)^{b_1}
  |0\rangle_{\rm F}
  \notag \\
  =&
  (\hat{c}_6^\dag)^{b_2}
  (\hat{c}_5^\dag)^{b_1}
  (\hat{c}_4^\dag)^{b_6}
  (\hat{c}_3^\dag)^{b_5}
  (\hat{c}_2^\dag)^{b_4}
  (\hat{c}_1^\dag)^{b_3}
  |0\rangle_{\rm F}
  \times {\rm sgn}({\cal R}) \notag \\
  =& |b_2\, b_1\, b_6\, b_5\, b_4\, b_3 \rangle_{\rm F} \times {\rm sgn}({\cal R}) 
  \label{eq:example} 
\end{alignat}
with the permutation $\sigma_m$ given in Eq.~(\ref{perm6}). 
We should note that the ${\rm sgn}({\cal R})$ is not
the parity of the permutation $\sigma_m$
but depends on how fermions occupy the single-particle local states.

Let us now discuss quantum circuits for fermionic symmetry operations.
Following the convention of the Jordan-Wigner transformation 
in Eqs.~(\ref{eq:JWT1}) and (\ref{eq:JWT2}), 
here we assume the correspondence that 
the $i$th fermionic single-particle state is occupied
(unoccupied) if the state of the $i$th
qubit is $|1\rangle_{i}$ ($|0\rangle_{i}$).
  The fermion vacuum in the occupation basis can then be represented in the computational basis as
  \begin{equation}
    |0\rangle_{\rm F} \overset{\rm JWT}{=}
    |0\rangle_N|0\rangle_{N-1}\cdots|0\rangle_{1}.
  \end{equation}
  Note however that, in general, the occupation-basis state
  $|b_N b_{N-1} \cdots b_{1} \rangle_{\rm F}$
  under the Jordan-Wigner transformation
  is not identical to the computational basis state
  $|b_N b_{N-1} \cdots b_{1} \rangle = |b_N\rangle_N |b_{N-1}\rangle_{N-1} \cdots |b_1\rangle_1$, i.e., 
  \begin{equation}
    |b_N b_{N-1} \cdots b_{1} \rangle_{\rm F}
    \overset{\rm JWT}{\not =}
    |b_N b_{N-1} \cdots b_{1} \rangle
  \end{equation}
  because of the sign factor due to the anticommutation relation of fermions, e.g., 
$|11\rangle_{\rm F}=\hat{c}_2^\dag \hat{c}_1^\dag |0\rangle_{\rm F}
  \overset{\rm JWT}{=}-|1\rangle_{2}|1\rangle_{1}
  = - |11\rangle$.
The sign factor due to the anticommutation relations of fermions
on a quantum state can be tracked by
using the fermionic-{\sc swap} operator~\cite{Bravyi2002,Essler2005,Verstraete2009,Barthel2009,Wecker2015,Kivlichan2018}
on a quantum circuit, 
which does not depend on the convention adopted for the occupation-basis state.
The explicit form of the fermionic-{\sc swap}
operator in terms of the fermion operators is given in Appendix~\ref{app:fermi}.

\begin{center}
  \begin{figure}
    \includegraphics[width=1\columnwidth]{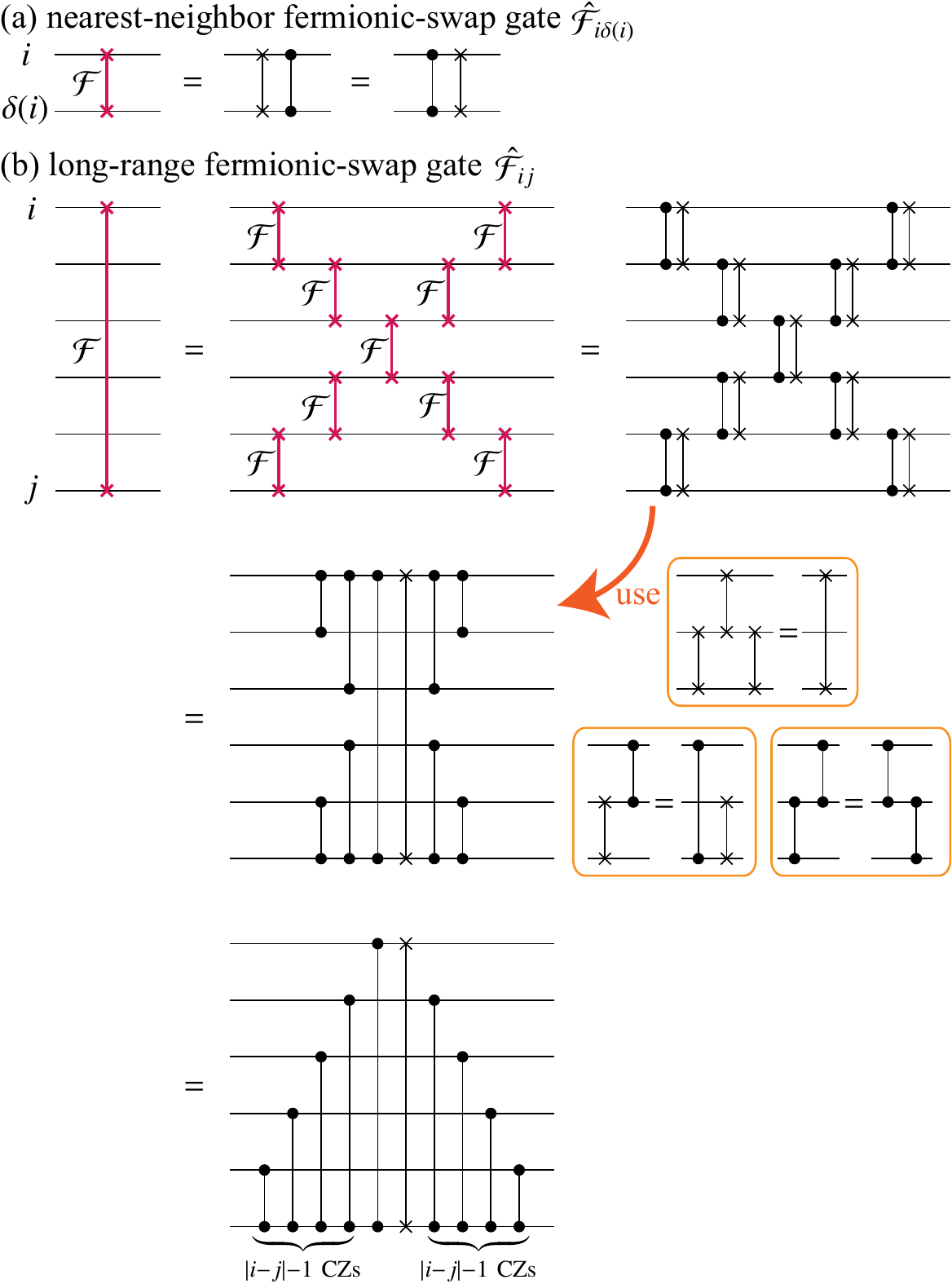}
    \caption{
      Fermionic-{\sc swap} gate
      acting on a pair of 
      (a) nearest-neighbor qubits $i$ and $\delta(i)=i\pm 1$ and 
      (b) distant qubits $i$ and $j$
      (the figure refers to the case of $|i-j|=5$) 
      is decomposed into a product of the {\sc swap} and CZ gates . 
      In (b), the third and fourth equalities follow from 
      the gate identities shown in the boxes. 
    }
    \label{fig:fswap}
  \end{figure}
\end{center}

The nearest-neighbor fermionic-{\sc swap} operator $\hat{\cal F}_{i \delta(i)}$ acting on neighboring
qubits $i$ and $\delta(i)$ in
the computational basis is defined as 
\begin{alignat}{1}
  &\hat{\cal F}_{i\delta(i)}|0\rangle_i |0\rangle_{\delta(i)} =  |0\rangle_i |0\rangle_{\delta(i)},
  \label{fswap00}  \\
  &\hat{\cal F}_{i\delta(i)}|0\rangle_i |1\rangle_{\delta(i)} =  |1\rangle_i |0\rangle_{\delta(i)},
  \label{fswap01}  \\
  &\hat{\cal F}_{i\delta(i)}|1\rangle_i |0\rangle_{\delta(i)} =  |0\rangle_i |1\rangle_{\delta(i)},
  \label{fswap10}  \\
  &\hat{\cal F}_{i\delta(i)}|1\rangle_i |1\rangle_{\delta(i)} = -|1\rangle_i |1\rangle_{\delta(i)}.
  \label{fswap11}  
\end{alignat}
The matrix representation of $\hat{\cal F}_{i\delta(i)}$
in the above four basis states is thus given by 
\begin{equation}
  \hat{\cal F}_{i\delta(i)}
  \overset{\cdot}{=}
  \begin{bmatrix}
  1& 0& 0& 0 \\
  0& 0& 1& 0 \\ 
  0& 1& 0& 0 \\ 
  0& 0& 0& -1
  \end{bmatrix}.
  \label{eq:fsmatrix}
\end{equation}
Since the controlled-$Z$ (CZ) operator acting on qubits $i$ and $j$ is represented as 
\begin{equation}
  \widehat{\rm CZ}_{ij}
  \overset{\cdot}{=}
  \begin{bmatrix}
  1& 0& 0& 0 \\
  0& 1& 0& 0 \\ 
  0& 0& 1& 0 \\ 
  0& 0& 0& -1
  \end{bmatrix}, 
\end{equation}
the nearest-neighbor fermionic-{\sc swap} operator can be written as
[also see Fig.~\ref{fig:fswap}(a)]
\begin{equation}
  \hat{\cal F}_{i\delta(i)}
  \overset{\rm JWT}{=}
  \hat{\cal S}_{i\delta(i)} \widehat{\rm CZ}_{i\delta(i)}
  = \widehat{\rm CZ}_{i\delta(i)} \hat{\cal S}_{i\delta(i)}.
  \label{eq:nnfswap}
\end{equation}

The decomposition of the fermionic-swap operator as in Eq.~(\ref{eq:nnfswap})
is valid only when the qubits are nearest neighbored, i.e., 
$\delta(i)=i \pm 1$.
As shown in Fig.~\ref{fig:fswap}(b),
the long-range fermionic-{\sc swap} operator 
$\hat{\cal F}_{i j}$ for $j\not= \delta(i)$ is inherently nonlocal  
and can be implemented as a product of nearest-neighbor
fermionic-{\sc swap} operators [also see Fig.~\ref{fig:amida}(b)].
Furthermore, using the identities shown
in Fig.~\ref{fig:fswap}(b), $\hat{\cal F}_{i j}$
can be expressed by
$\hat{\cal S}_{ij}\widehat{\rm CZ}_{ij}$ sandwiched between
two sequences of $|i-j|-1$ CZ gates.
The example shown in Fig.~\ref{fig:fswap}(b) can be written as  
\begin{alignat}{1}
  \hat{\cal F}_{ij}
  &
  \overset{\rm JWT}{=}
  \left[\prod_{i \lessgtr k \lessgtr j} \widehat{\rm CZ}_{jk}\right]
  \hat{f}_{ij}
  \left[\prod_{i \lessgtr k \lessgtr j} \widehat{\rm CZ}_{jk}\right],
  \label{longfswap}
\end{alignat}
where
\begin{equation}
  \hat{f}_{ij}
  \equiv\hat{\cal S}_{ij}\widehat{\rm CZ}_{ij}
  =\widehat{\rm CZ}_{ij}\hat{\cal S}_{ij}
  \overset{\cdot}{=}
  \begin{bmatrix}
  1& 0& 0& 0 \\
  0& 0& 1& 0 \\ 
  0& 1& 0& 0 \\ 
  0& 0& 0& -1
  \end{bmatrix}.
  \label{eq:fmatrix}
\end{equation}
Note that this matrix is exactly the same as that in Eq.~(\ref{eq:fsmatrix}), 
but now qubits $i$ and $j$ are not necessarily nearest neighbored.

\begin{center}
  \begin{figure}
    \includegraphics[width=1\columnwidth]{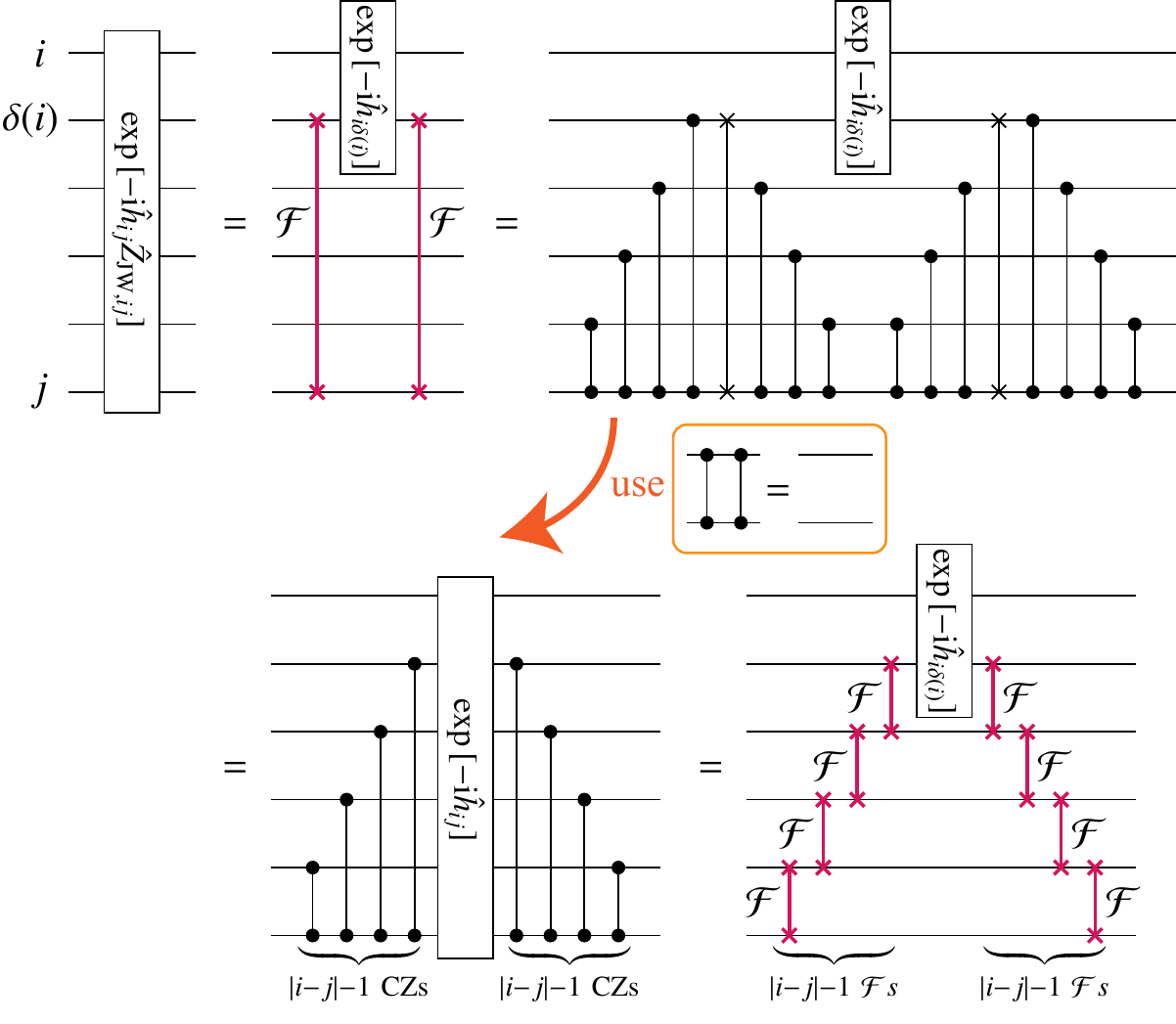}
    \caption{
      Implementation of the ($|i-j|+1$)-qubit unitary gate 
      $\exp[-\imag \hat{h}_{ij} \hat{Z}_{{\rm JW},ij}]$ (top left)
      by a product of two-qubit unitary gates (bottom).
      The figure refers to the case of $|i-j|=5$. 
      The second equality follows from the decomposition
      of the long-range fermionic-{\sc swap} gate shown in Fig.~\ref{fig:fswap}(b),
      and the third equality follows from 
      the identity $\widehat{\rm CZ}_{kl}^2=\hat{I}$.
      The last equality makes use of 
      the decomposition of 
      the long-range fermionic-{\sc swap} gate into a product of the nearest-neighbor fermionic-{\sc swap} gates  
      [for a variant of this decomposition, see the first equality in Fig.~\ref{fig:fswap}(b)].
    }
    \label{fig:fswap_exp}
  \end{figure}
\end{center}

Let us now consider how 
one can implement a unitary operator of the form $\exp{\left[-\imag\hat{q}_{ij}\right]}
\overset{\rm JWT}{=}\exp{\left[-\imag\hat{h}_{ij}\hat{Z}_{{\rm JW},ij}\right]}$ in a quantum circuit, 
where $\hat{q}_{ij}\overset{\rm JWT}{\equiv}\hat{h}_{ij}\hat{Z}_{{\rm JW},ij}$ is Hermitian and quadratic 
in terms of fermion operators, satisfying that $\hat{\cal F}_{kj} \hat{q}_{ij} \hat{\cal F}_{kj} = \hat{q}_{ik}$, and 
$\hat{h}_{ij}$ is a Hermitian operator in the qubit representation acting on qubits $i$ and $j$, i.e.,
$[\hat{h}_{ij},\hat{Z}_{{\rm JW},ij}]=0$. 
Examples of $\hat{q}_{ij}$ include
$\hat{q}_{ij} = \theta (\hat{c}_{i}^\dag\hat{c}_j + {\rm H.c.})$
and
$\hat{q}_{ij} = \imag \theta (\hat{c}_{i}^\dag\hat{c}_j^\dag - {\rm H.c.})$ with real $\theta$.
First,
the equalities 
$\hat{\cal F}_{kj} \exp[-\imag \hat{q}_{ij}] \hat{\cal F}_{kj}
=\exp[-\imag \hat{q}_{ik}] \overset{\rm JWT}{=} \exp[-\imag \hat{h}_{ik} \hat{Z}_{{\rm JW},ik}]$
and $\hat{Z}_{{\rm JW},i\delta(i)}=\hat{I}$
imply that,
by using two long-range fermionic-{\sc swap} gates ${\cal F}_{\delta(i)j}$,
where qubit $\delta(i)$ is located between qubits $i$ and $j$,
we can remove the Jordan-Wigner string $\hat{Z}_{{\rm JW},ij}$ from
the exponent (first equality in Fig.~\ref{fig:fswap_exp}): 
\begin{equation}
  \exp\left[-\imag \hat{h}_{ij} \hat{Z}_{{\rm JW},ij}\right]
  =\hat{\cal F}_{\delta(i)j}
  \exp\left[-\imag \hat{h}_{i\delta(i)}\right]
  \hat{\cal F}_{\delta(i)j}.
\end{equation}
Then, by representing the
long-range fermionic-{\sc swap} gates $\hat{\cal F}_{\delta(i)j}$
as a product of the CZ and the {\sc swap} gates (second equality in Fig.~\ref{fig:fswap_exp})
and using the identity $\widehat{\rm CZ}_{kl}^2=\hat{I}$, we
can cancel the redundant $\widehat{\rm CZ}_{kl}$ gates and
obtain (third equality in Fig.~\ref{fig:fswap_exp}) 
\begin{equation}
  \exp\left[-\imag \hat{h}_{ij} \hat{Z}_{{\rm JW},ij}\right]
  =\left[
    \prod_{i \lessgtr k \lessgtr j} \widehat{\rm CZ}_{jk}
    \right]
  \exp\left[-\imag \hat{h}_{i j}\right]
  \left[
    \prod_{i \lessgtr k \lessgtr j}
    \widehat{\rm CZ}_{jk}
    \right].
  \label{eq:fswap_exp}
\end{equation}
A similar strategy
for eliminating the redundancy in 
consecutive Jordan-Wigner strings
has been reported in Ref.~\cite{Hastings2015}.

Two remarks are in order. First, by substituting $\hat{h}_{ij} \propto \hat{X}_i\hat{X}_j+\hat{Y}_i\hat{Y}_j$
in Eq.~(\ref{eq:fswap_exp}), we can reproduce
the quantum circuit for the   
exponentiated hopping term of fermions reported previously in Ref.~\cite{Reiner2019},
i.e., an exchange-type gate~\cite{McKay2016} 
sandwiched between two sequences of $|i-j|-1$ CNOT gates.
Second, the ($|i-j|+1$)-qubit  unitary gate  
$\exp\left[-\imag \hat{h}_{ij} \hat{Z}_{{\rm JW},ij}\right]$
can also be implemented 
by a product of neighboring two-qubit gates 
at a cost of additional $2(|i-j|-1)$ fermionic-{\sc swap} gates (last equality in Fig.~\ref{fig:fswap_exp}).
For example,  assuming that $i < \delta(i)=i+1 < j$, $\exp\left[-\imag \hat{h}_{ij} \hat{Z}_{{\rm JW},ij}\right]$ can also be implemented as  
\begin{equation}
  \exp\left[-\imag \hat{h}_{ij} \hat{Z}_{{\rm JW},ij}\right]
  =\left[
    \prod_{k=\delta(i)+1}^{j} \hat{\cal F}_{k,k-1} 
  \right]
    \exp\left[-\imag \hat{h}_{i \delta(i)}\right]
    \left[
      \prod_{k=j}^{\delta(i)+1} \hat{\cal F}_{k,k-1} 
    \right].
  \label{eq:fswap_exp2}
\end{equation}
The resulting quantum circuit first 
moves the local state at the $j$th qubit to the qubit next to qubit $i$ by applying the nearest-neighbor 
fermionic-{\sc swap} gates successively, thus 
keeping track of the antisymmetric nature of fermion exchange, 
then it applies the gate $\exp[-\imag \hat{h}_{i\delta(i)}]$ on qubits $i$ and $\delta(i)$,
and finally it moves the local state at the $\delta(i)$th qubit to the original location, i.e., the $j$th qubit, by undoing 
the fermionic-{\sc swap} operations successively~\cite{Stoudenmire2010}.    
We should note here that such a way of exploiting the nearest-neighbor
fermionic-{\sc swap} gates has been discussed extensively  for simulating fermions 
under the Jordan-Wigner transformation 
in Refs.~\cite{Kivlichan2018,Cai2020,Cade2020}.
While Eq.~(\ref{eq:fswap_exp2}) might be preferable for a real
quantum device, depending on the connectivity of qubits
and its native gate set,  
we employ Eq.~(\ref{eq:fswap_exp}) 
for our numerical simulations with classical computers 
because of the smaller number of operations.

\subsection{Quantum circuit for spin rotation}

\begin{center}
  \begin{figure*}
    \includegraphics[width=1\textwidth]{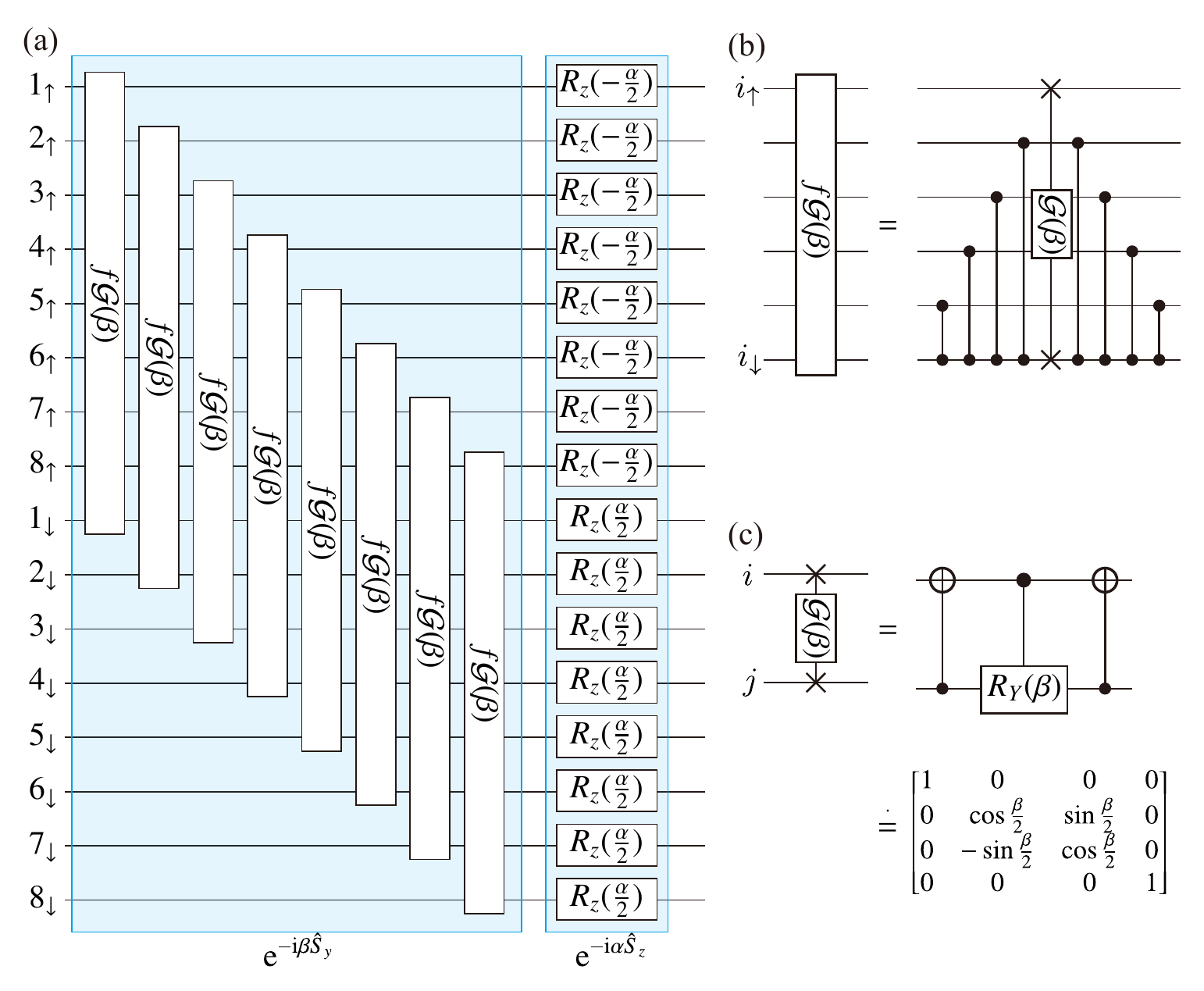}
    \caption{      
        (a) A quantum circuit to implement spin rotation 
        $\e^{-\imag \alpha \hat{S}_z} \e^{-\imag\beta \hat{S}_y}$ 
        for the $4 \times 2$ site Fermi-Hubbard model. 
        $f{\cal G}(\beta)$ denotes a fermionic Givens-rotation gate 
        for distant qubits $i_\uparrow$ and $i_\downarrow$ defined in Eq.~(\ref{eq:def_G}).
        For evaluation of Eq.~(\ref{eq:PPP2}),
        $\e^{-\imag \alpha \hat{S}_z}$ is not required and hence
        the single-qubit rotation $R_{z}(\theta)$ should be replaced with identity.
        (b) A decomposition of a fermionic Givens-rotation gate $f{\cal G}(\beta)$ 
        into a Givens-rotation gate ${\cal G}(\beta)$ 
        sandwiched with CZ gates, as in Eq.~(\ref{CZGCZ}). 
        (c) A decomposition of the Givens-rotation gate ${\cal G}(\beta)$ defined in Eq.~(\ref{Givens}). 
        Here, $\hat{R}_{Y}(\theta)=\e^{-\imag  \theta \hat{Y}_i/2}$ acting on qubit $i$. 
        The matrix representation of ${\cal G}(\beta)$ with the basis states 
        $\{ |0\rangle_i|0\rangle_j, |0\rangle_i|1\rangle_j, |1\rangle_i|0\rangle_j, |1\rangle_i|1\rangle_j \}$ is also shown. 
    }  \label{fig.spincircuit}
  \end{figure*}
\end{center}

With the Jordan-Wigner transformation, the local spin operators 
are represented as  
\begin{alignat}{1}
  \hat{S}_{i}^{x} &
  \overset{\rm JWT}{=}
  \frac{1}{4} \left(
  \hat{X}_{i_\uparrow} \hat{X}_{i_\downarrow} +
  \hat{Y}_{i_\uparrow} \hat{Y}_{i_\downarrow} \right)
  \hat{Z}_{{\rm JW}, i_\uparrow i_\downarrow},
  \\
  \hat{S}_{i}^{y} &
  \overset{\rm JWT}{=}
  \frac{1}{4} \left(
  \hat{X}_{i_\uparrow} \hat{Y}_{i_\downarrow} -
  \hat{Y}_{i_\uparrow} \hat{X}_{i_\downarrow} \right)   
  \hat{Z}_{{\rm JW}, i_\uparrow i_\downarrow},
  \\
  \hat{S}_{i}^{z} &
  \overset{\rm JWT}{=}
  \frac{1}{4} \left(
  \hat{Z}_{i_\downarrow}-\hat{Z}_{i_\uparrow} 
    \right).
\end{alignat}
Therefore, the rotation operator 
$  \e^{-\imag \beta \hat{S}_y} =
  \prod_{i=1}^{L}
  \e^{-\imag \beta \hat{S}^y_i}$
can be given by a product of 
\begin{alignat}{1}
  \e^{-\imag \beta \hat{S}^y_i}&
  \overset{\rm JWT}{=}
  \exp\left[
    -
    \imag\frac{\beta}{4} \left(
    \hat{X}_{i_\uparrow} \hat{Y}_{i_\downarrow} -
    \hat{Y}_{i_\uparrow} \hat{X}_{i_\downarrow} 
    \right)\hat{Z}_{{\rm JW},i_\uparrow,i_\downarrow} \right]
  \label{XYYX}\\
  &=
  \left[\prod_{i_\uparrow <k < i_\downarrow} \widehat{{\rm CZ}}_{i_\downarrow k}\right]
  \hat{\cal G}_{i_\uparrow i_\downarrow}\left(\beta \right)
  \left[\prod_{i_\uparrow <k < i_\downarrow} \widehat{{\rm CZ}}_{i_\downarrow k}\right]
  \label{CZGCZ} \\
  &\equiv \widehat{f{\cal G}}_{i_\uparrow i_\downarrow}(\beta), \label{eq:def_G} 
\end{alignat}
where the 
${\rm CZ}$ gates in Eq.~(\ref{CZGCZ})
account for the Jordan-Wigner string in Eq.~(\ref{XYYX}),
as shown for the more general case in Eq.~(\ref{eq:fswap_exp}), and 
\begin{equation}
  \hat{\cal G}_{i j}(\theta)=
  \exp\left[
    {-}
    \imag\frac{\theta}{4} \left(
    \hat{X}_{i} \hat{Y}_{j} -
    \hat{Y}_{i} \hat{X}_{j} 
    \right) \right]
  \label{Givens}
\end{equation}
is the Givens-rotation gate for $i\ne j$~\cite{Wecker2015} 
whose matrix representation in the computational basis is given by 
\begin{equation}
  \hat{\cal G}_{i j}(\theta)\overset{\cdot}{=}
  \begin{bmatrix}
    1 & 0 & 0 & 0 \\
    0 &  \cos{\tfrac{\theta}{2}} &  \sin{\tfrac{\theta}{2}} & 0 \\
    0 & -\sin{\tfrac{\theta}{2}} & \cos{\tfrac{\theta}{2}} & 0 \\
    0 & 0 & 0 & 1 
  \end{bmatrix}.
  \label{eq:mgivens}
\end{equation}
For deriving this matrix representation, it is useful to notice that 
$(\hat{X}_i\hat{Y}_j-\hat{Y}_i\hat{X}_j)^2=2(1-\hat{Z}_i\hat{Z}_j)$ and 
$(1-\hat{Z}_i\hat{Z}_j) (\hat{X}_i\hat{Y}_j-\hat{Y}_i\hat{X}_j) = 2(\hat{X}_i\hat{Y}_j-\hat{Y}_i\hat{X}_j)$ 
when $i\ne j$. A more detailed description for a general single-particle fermion operator is found in Appendix~\ref{app:Givens}.
In Eq.~(\ref{eq:def_G}), we have defined the fermionic Givens-rotation gate $\widehat{f{\cal G}}_{i_\uparrow i_\downarrow}(\beta)$, 
which is an extension of the Givens-rotation gate for fermions. 

Similarly, 
the rotation around the $z$ direction by $\alpha$ can be given by 
\begin{equation}
  \e^{-\imag \alpha \hat{S}^z_i}
  \overset{\rm JWT}{=}
  \hat{R}_{Z,i_\uparrow}\left(-\frac{\alpha}{2}\right)
  \hat{R}_{Z,i_\downarrow}\left(\frac{\alpha}{2}\right),
\end{equation}
where
\begin{equation}
  \hat{R}_{Z,i}(\theta)=\e^{-\imag  \theta \hat{Z}_i/2}.
\end{equation}
Figure~\ref{fig.spincircuit} shows a quantum circuit 
corresponding to the product of rotations 
$\e^{-\imag \alpha \hat{S}_z}\e^{-\imag \beta \hat{S}_y}
=
\prod_{i} \e^{-\imag \alpha \hat{S}_i^z}
\prod_{i} \e^{-\imag \beta \hat{S}_i^y}$. 
For the Givens rotation, we adopt the 
gate decomposition given in Ref.~\cite{Jiang2018}.
One can also find another way of implementing the rotation 
$\e^{-\imag \beta \hat{S}_y}$ in Refs.~\cite{Wecker2015,Tsuchimochi2020}

\subsection{Quantum circuit for $\eta$ rotation}

With the Jordan-Wigner transformation, the local $\eta$ operators 
are represented as 
\begin{alignat}{1}
  \hat{\eta}_{i}^{x} &
  \overset{\rm JWT}{=}
  \frac{\e^{\imag \phi_i}}{4} \left(
  \hat{X}_{i_\uparrow} \hat{X}_{i_\downarrow} -
  \hat{Y}_{i_\uparrow} \hat{Y}_{i_\downarrow}
  \right)
  \hat{Z}_{{\rm JW}, i_\uparrow i_\downarrow},
  \\
  \hat{\eta}_{i}^{y} &
  \overset{\rm JWT}{=}
  -
  \frac{\e^{\imag \phi_i}}{4} \left(
  \hat{X}_{i_\uparrow} \hat{Y}_{i_\downarrow} +
  \hat{Y}_{i_\uparrow} \hat{X}_{i_\downarrow}
  \right)
    \hat{Z}_{{\rm JW}, i_\uparrow i_\downarrow},
  \\
  \hat{\eta}_{i}^{z} &
  \overset{\rm JWT}{=}
  -\frac{1}{4} \left(\hat{Z}_{i_\uparrow} + \hat{Z}_{i_\downarrow}
  \right).
\end{alignat}
Therefore, the rotation operator 
$  \e^{-\imag \beta \hat{\eta}_y} =
  \prod_{i=1}^{N}
  \e^{-\imag \beta \hat{\eta}^y_i}$
can be given by a product of 
\begin{alignat}{1}
  \e^{-\imag \beta \hat{\eta}^y_i}&\overset{\rm JWT}{=}
  \exp\left[    
    \imag\frac{\e^{\imag \phi_i}\beta}{4} \left(
    \hat{X}_{i} \hat{Y}_{j} +
    \hat{Y}_{i} \hat{X}_{j} 
    \right)\hat{Z}_{{\rm JW},i_\uparrow,i_\downarrow} \right]
  \label{XYYXB}\\
  &=
  \left[\prod_{i_\uparrow <k < i_\downarrow} \widehat{{\rm CZ}}_{i_\downarrow k}\right]
  \hat{\cal B}_{i_\uparrow i_\downarrow}\left(-\e^{\imag \phi_i}\beta \right)
  \left[\prod_{i_\uparrow <k < i_\downarrow} \widehat{{\rm CZ}}_{i_\downarrow k}\right]
  \label{CZBCZ} \\
  & \equiv \widehat{f{\cal B}}_{i_\uparrow i_\downarrow}(-\e^{\imag \phi_i}\beta),
  \label{eq:def_B}
\end{alignat}
where the 
${\rm CZ}$ gates in Eq.~(\ref{CZBCZ})
account for the Jordan-Wigner string in Eq.~(\ref{XYYXB}),
as shown for the more general case in Eq.~(\ref{eq:fswap_exp}), and 
\begin{equation}
  \hat{\cal B}_{i j}(\theta)=
  \exp\left[
    {-}
    \imag\frac{\theta}{4} \left(
    \hat{X}_{i} \hat{Y}_{j} +
    \hat{Y}_{i} \hat{X}_{j} 
    \right) \right].
    \label{eq:BTG}
\end{equation}
is the Bogoliubov-transformation gate~\cite{Wecker2015,Jiang2018} 
whose matrix representation in the computational basis is given by 
\begin{equation}
  \hat{\cal B}_{i j}(\theta)\overset{\cdot}{=}
  \begin{bmatrix}  
    \cos{\tfrac{\theta}{2}} & 0 & 0 &  - \sin{\tfrac{\theta}{2}}  \\
    0 & 1 & 0 & 0 \\
    0 & 0 & 1 & 0 \\
    \sin{\tfrac{\theta}{2}} & 0 & 0 &   \cos{\tfrac{\theta}{2}}  
  \end{bmatrix}.
  \label{eq:BT}
\end{equation}
For deriving this matrix representation, it is useful to notice that 
$(\hat{X}_i\hat{Y}_j+\hat{Y}_i\hat{X}_j)^2=2(1+\hat{Z}_i\hat{Z}_j)$ and 
$(1+\hat{Z}_i\hat{Z}_j) (\hat{X}_i\hat{Y}_j+\hat{Y}_i\hat{X}_j) = 2(\hat{X}_i\hat{Y}_j+\hat{Y}_i\hat{X}_j)$ 
when $i\ne j$. A more detailed description for a general anomalous single-particle fermion operator 
is found in Appendix~\ref{app:Bogoliubov}.
In Eq.~(\ref{eq:def_B}), we have defined the fermionic Bogoliubov-transformation gate 
$\widehat{f{\cal B}}_{i_\uparrow i_\downarrow}(\beta)$, 
which is an extension of the Bogoliubov-transformation gate for fermions.

\begin{center}
  \begin{figure*}
    \includegraphics[width=1\textwidth]{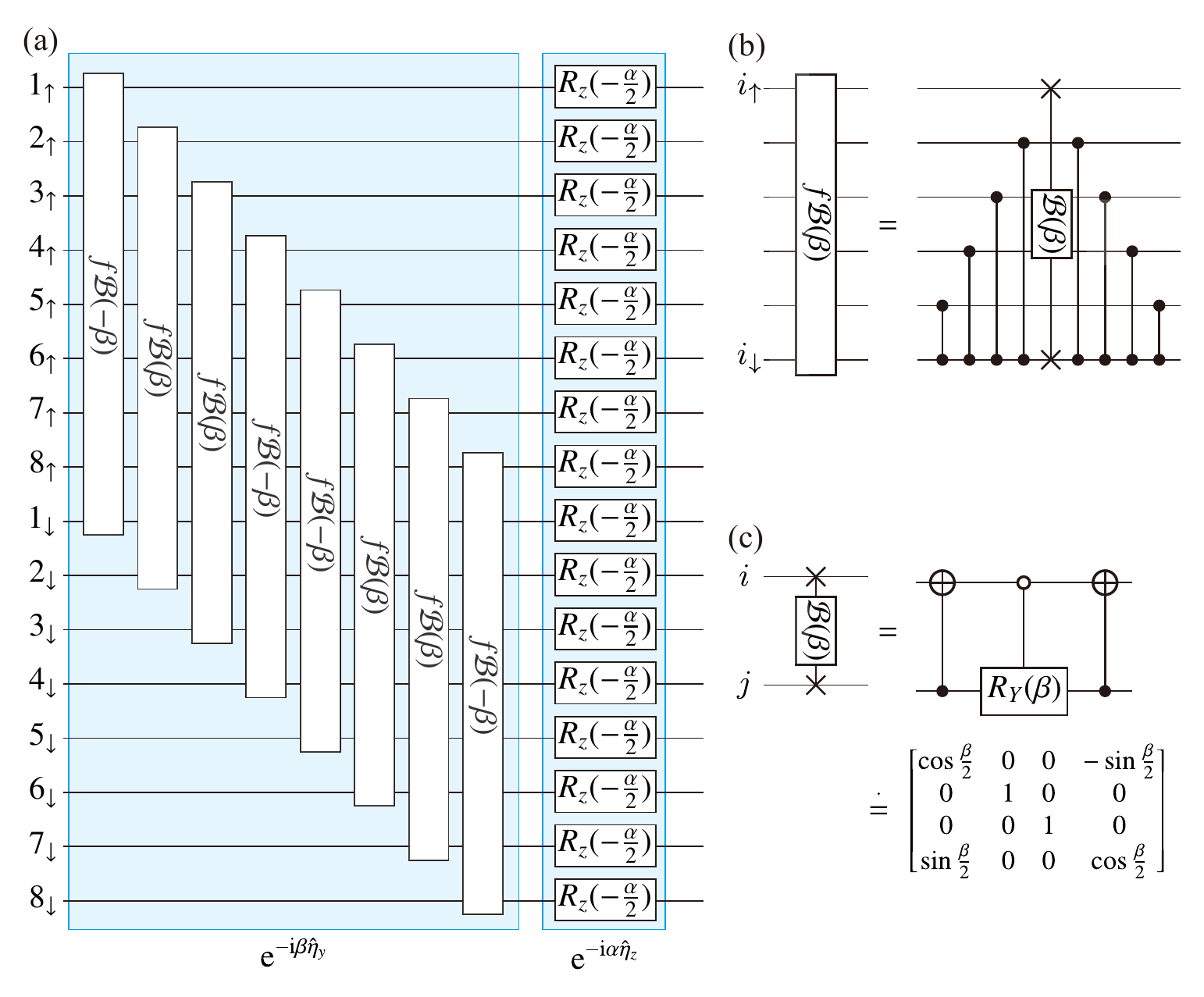}
    \caption{
      (a) A quantum circuit to implement $\eta$ rotation 
      $\e^{-\imag \beta \hat{\eta}_y} \e^{-\imag\alpha \hat{\eta}_z} $ 
      for the $4 \times 2$ site Fermi-Hubbard model. 
      $f{\cal B}(\beta)$ denotes a fermionic Bogoliubov-transformation gate
      for distant qubits $i_\uparrow$ and $i_\downarrow$ defined in Eq.~(\ref{eq:def_B}).  
      For evaluation of Eq.~(\ref{eq:PPP2}),
      $\e^{-\imag \alpha \hat{\eta}_z}$ is not required and hence
      the single-qubit rotation $R_z(\theta)$ should be replaced with identity.
      (b) A decomposition of a fermionic Bogoliubov-transformation gate $f{\cal B}(\beta)$ 
      into a Bogoliubov-transformation gate ${\cal B}(\beta)$ 
      sandwiched with CZ gates, as in Eq.~(\ref{CZBCZ}),
      assuming that site $i$ represented by qubits $i_\uparrow$ and 
      $i_\downarrow$ belongs to $B$ sublattice.
      When site $i$ represented by qubits $i_\uparrow$ and 
      $i_\downarrow$ belongs to $A$ sublattice,
      $f{\cal B}(\beta)$ and ${\cal B}(\beta)$ should be replaced with 
      $f{\cal B}(-\beta)$ and ${\cal B}(-\beta)$, respectively. 
      (c) A decomposition of the Bogoliubov-transformation gate $f{\cal B}(\beta)$ defined in Eq.~(\ref{eq:BTG}). 
      Here, $\hat{R}_{Y}(\theta)=\e^{-\imag  \theta \hat{Y}_i/2}$ acting on qubit $i$.
      The matrix representation of ${\cal B}(\beta)$ with the basis states 
      $\{ |0\rangle_i|0\rangle_j, |0\rangle_i|1\rangle_j, |1\rangle_i|0\rangle_j, |1\rangle_i|1\rangle_j \}$ is also shown. 
    }
    \label{fig.etacircuit}
  \end{figure*}
\end{center}

Similarly, 
the rotation around the $z$ direction by $\alpha$ can be given as 
\begin{equation}
  \e^{-\imag \alpha \hat{\eta}^z_i}
  \overset{\rm JWT}{=}
  \hat{R}_{Z,i_\uparrow}\left(-\frac{\alpha}{2}\right)
  \hat{R}_{Z,i_\downarrow}\left(-\frac{\alpha}{2}\right).
\end{equation}
Figure~\ref{fig.etacircuit} shows a quantum circuit 
corresponding to the product of rotations 
$\e^{-\imag \alpha \hat{\eta}_z}\e^{-\imag \beta \hat{\eta}_y}
=
\prod_{i} \e^{-\imag \alpha \hat{\eta}_i^z}
\prod_{i} \e^{-\imag \beta \hat{\eta}_i^y}$. 
For the Bogoliubov transformation, we adopt the 
gate decomposition given in Ref.~\cite{Jiang2018}.
One can also find another way of implementing the rotation 
$\e^{-\imag \beta \hat{\eta}_y}$ in Ref.~\cite{Wecker2015}.

\section{Numerical simulations}\label{sec:results}

In this section, we demonstrate 
the Krylov-extended SAVQE method 
by numerically simulating the two-component Fermi-Hubbard model on the $4\times2$ 
cluster under open boundary conditions at half filling, i.e., one fermion per site.
To speed up the calculations, a simple strategy of parallelizing numerical simulations is 
employed (see Appendix~\ref{app:parallel}).

\subsection{Variational state}\label{varstate}

\begin{center}
  \begin{figure*}
    \includegraphics[width=1\textwidth]{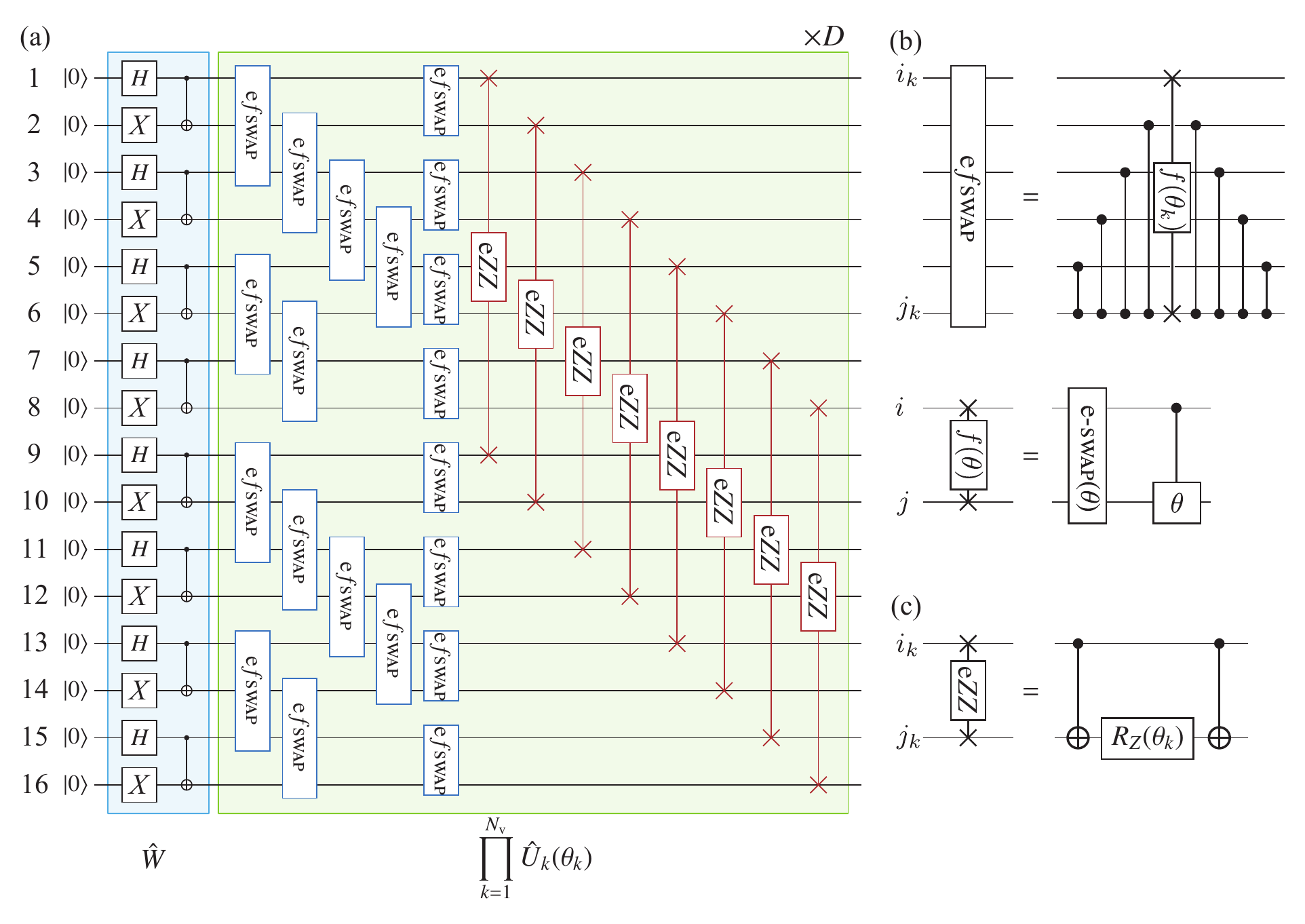}
    \caption{
      (a) A quantum circuit for preparing
      the variational state $|\psi(\bs{\theta}) \rangle=\prod_{k=N_{\rm v}}^1 \hat{U}_k(\theta_k)
      \hat{W} |0\rangle^{\otimes N}$.
      The parametrized gates corresponding to 
      $\hat{U}_k(\theta_k)=\exp(-\imag \hat{\cal F}_{i_k j_k} \theta_{k}/2)$
      and
      $\hat{U}_k(\theta_k)=\exp(-\imag \hat{Z}_{i_k} \hat{Z}_{j_k} \theta_{k}/2)$
      are denoted as e$f${\sc swap} and e$ZZ$, respectively. 
      The qubit numbers indicated in the left most side correspond to the numbering of qubits 
      in Fig.~\ref{fig:lattice2} for the two-component Fermi-Hubbard model on the $4\times2$ cluster. 
      (b) A decomposition of the e$f${\sc swap} gate
      $\hat{U}_k(\theta_k)=\exp(-\imag \hat{\cal F}_{i_k j_k} \theta_{k}/2)$, as given in Eq.~(\ref{efswap}).
      The lower part of the panel shows that 
      the parametrized gate $\hat{f}_{ij}(\theta) = \exp(-\imag\hat{f}_{ij}\theta/2)$ 
      can be expressed as a product of the exponential-{\sc swap} gate $\exp(-\imag\hat{\cal S}_{ij}\theta/2)$ 
      and the controlled-phase gate ${\rm CPHASE}_{ij}(\theta)$ [also see Eq.~(\ref{matF})]. 
      (c) A decomposition of the e$ZZ$ gate
      $\hat{U}_k(\theta_k)=\exp(-\imag \hat{Z}_{i_k} \hat{Z}_{j_k} \theta_{k}/2)$.
    } \label{fig.circuit}
  \end{figure*}
\end{center}

As a variational state $|\psi(\bs{\theta})\rangle$ in Eq.~(\ref{vqe}) 
for the ground state of the Fermi-Hubbard model,
we first construct a product state
of two-site bonding orbitals for each spin of fermions 
by applying 
Hadamard, Pauli $X$, and CNOT gates to $|0\rangle^{\otimes N}$
in an appropriate manner [see Fig.~\ref{fig.circuit}(a)].  
Note that the bonding state is equivalent to one of the Bell states,
$\tfrac{1}{\sqrt{2}}(|0\rangle |1\rangle + |1\rangle |0\rangle)$ 
(for its preparation, see Ref.~\cite{NielsenChuang}).
The preparation of the product state of bonding orbitals 
is independent of the variational parameters and 
hence corresponds to $\hat{W}|0\rangle^{\otimes N}$ in Eq.~(\ref{vqe}), 
also indicated in Fig.~\ref{fig.circuit}(a). 
Note that the total number of fermions for the state represented by this first part of the quantum circuit is 
$N/2=L$ (i.e., half filling) with the same number of up and down fermions, assuming that $L$ is even. 
This implies that the expectation values of $\hat{S}_z$ and $\hat{\eta}_z$ are both zero. 
As described below, we will construct a parametrized part of the quantum circuit 
for the variational state $|\psi(\bs{\theta})\rangle$ that preserves these features, i.e., 
$\langle\psi(\bs{\theta})| \hat{S}_z |\psi(\bs{\theta})\rangle = \langle\psi(\bs{\theta})| \hat{\eta}_z |\psi(\bs{\theta})\rangle = 0$ 
for an arbitrary set of variational parameters $\bs{\theta}$. 
However, this does not necessarily imply that the expectation values of ${\hat S}^2$ and ${\hat\eta}^2$ are zero. 
Instead, these expectation values are generally nonzero, as shown later in Figs.~\ref{fig.energyd1n1}(c) and \ref{fig.energyd1n1}(d), 
for example.

In order to construct a parametrized part of the quantum circuit for the variational state $|\psi(\bs{\theta})\rangle$ in Eq.~(\ref{vqe}),  
let $N_{\rm hop}$ ($N_{\rm int}$) be the number of hopping (interaction) terms 
in the Hamiltonian $\hat{\cal H}$ in Eq.~(\ref{Ham_Hubbard}). 
For $1 \leqslant k \leqslant N_{\rm hop}$ in Eq.~(\ref{vqe}), 
we apply 
\begin{alignat}{1}
  \hat{U}_k(\theta_k)&
  =\exp{(-\imag \hat{\cal F}_{i_k j_k} \theta_{k}/2)} \notag \\
  &
  =
  \hat{I} \cos\frac{\theta_k}{2} - \imag \hat{\cal F}_{i_k j_k} \sin{\frac{\theta_k}{2}} \notag \\
  &
  \overset{\rm JWT}{=}
  \left[\prod_{i_k \lessgtr m \lessgtr j_k} \widehat{{\rm CZ}}_{i_k m}\right]
  \hat{f}_{i_k j_k}\left(\theta_k \right)
  \left[\prod_{i_k \lessgtr m \lessgtr j_k} \widehat{{\rm CZ}}_{i_k m}\right]  
  \label{efswap}
\end{alignat}
to every pair of qubits $i_k$ and $j_k$ between which the hopping ($t$) term is
present in the Hamiltonian $\hat{\cal H}$.   
Since the long-range fermionic-{\sc swap} operator  
$\hat{\cal F}_{i_k j_k}$ is nonlocal, the operator $\hat{U}_k(\theta_k)$ above operates also onto 
all qubits between qubits $i_k$ and $j_k$ 
[see Fig.~\ref{fig.circuit}(b)]. 
We refer to this gate as an e$f${\sc swap} gate.
In Eq.~(\ref{efswap}), the parametrized two-qubit gate 
$\hat{f}_{i j}\left(\theta \right)$ is defined as 
\begin{equation}
  \hat{f}_{i j}
    \left(\theta \right) \equiv \exp(-\imag\hat{f}_{ij}\theta/2) 
    \label{eq:expf}
\end{equation}
and $\hat{f}_{ij}$ is given in Eq.~(\ref{eq:fmatrix}). 
Note that the last equality in Eq.~(\ref{efswap}) can be proved simply from Eq.~(\ref{longfswap}) and 
$\hat{f}_{ij}^2=\hat{I}$.
Since $\hat{f}_{ij} =
\hat{\cal S}_{ij} \widehat{{\rm CZ}}_{ij}=
\hat{\cal S}_{ij} + (\widehat{{\rm CZ}}_{ij} - \hat{I})$
and $\hat{\cal S}_{ij}$ commutes with $\widehat{{\rm CZ}}_{ij}$,  
the parametrized two-qubit gate 
$\hat{f}_{i j}\left(\theta \right)$ in Eq.~(\ref{eq:expf}) can be 
given by a product of 
the exponential-{\sc swap} gate
$\exp(-\imag\hat{\cal S}_{ij}\theta/2)$
and 
the controlled-phase gate
${\rm CPHASE}_{ij}(\theta) \equiv \exp{[-\imag (\widehat{{\rm CZ}}_{ij} - \hat{I})\theta/2]}$, i.e.,  
\begin{alignat}{1}
  \hat{f}_{i j}
  \left(\theta \right)
  =&
  \exp(-\imag\hat{\cal S}_{ij}\theta/2)\ {\rm CPHASE}_{ij}(\theta)\notag \\ 
  \overset{\cdot}{=}&
  \begin{bmatrix}
    \e^{-\imag \theta/2} & 0 & 0 & 0 \\
    0 &  \cos{\tfrac{\theta}{2}} &  -\imag\sin{\tfrac{\theta}{2}} & 0 \\
    0 & -\imag\sin{\tfrac{\theta}{2}} &  \cos{\tfrac{\theta}{2}} & 0 \\
    0 & 0 & 0 & \e^{-\imag \theta/2} 
  \end{bmatrix} 
  \begin{bmatrix}
    1 & 0 & 0 & 0 \\
    0 & 1 & 0 & 0 \\
    0 & 0 & 1 & 0 \\
    0 & 0 & 0 & \e^{\imag \theta} 
  \end{bmatrix}\notag \\
 =&
  \begin{bmatrix}
    \e^{-\imag \theta/2} & 0 & 0 & 0 \\
    0 &  \cos{\tfrac{\theta}{2}} &  -\imag\sin{\tfrac{\theta}{2}} & 0 \\
    0 & -\imag\sin{\tfrac{\theta}{2}} &  \cos{\tfrac{\theta}{2}} & 0 \\
    0 & 0 & 0 & \e^{+\imag \theta/2} 
  \end{bmatrix}
  \label{matF}
\end{alignat}
in the computational basis [also see Fig.~\ref{fig.circuit}(b)]. 
From this matrix representation, we can indeed confirm directly that 
$\hat{f}_{i j}\left(\theta \right) = \hat{I} \cos\frac{\theta}{2} - \imag \hat{f}_{i j} \sin{\frac{\theta}{2}}$.
For $N_{\rm hop}+1 \leqslant k \leqslant N_{\rm hop}+N_{\rm int}$, we apply
\begin{equation}
  \hat{U}_k(\theta_k)=\exp{(-\imag \hat{Z}_{i_k} \hat{Z}_{j_k} \theta_{k}/2)}
  \label{eZZ}
\end{equation}
to every pair of qubits $i_k$ and $j_k$ between which the interaction ($U_{\rm H}$) term is
present in the Hamiltonian $\hat{\cal H}$. As shown in Fig.~\ref{fig.circuit}(c), this gate (eZZ gate) is easily implemented 
in the circuit. 

We define the depth $D$ of the entire quantum circuit for the variational state $|\psi(\bs{\theta})\rangle$ 
in such a way that
a single layer of gates contains the sequences
of $N_{\rm hop} + N_{\rm int}$ gates,
$\prod_{k=1}^{N_{\rm hop}+N_{\rm int}} \hat{U}_{k}(\theta_k)$,
and hence the total number $N_{\rm v}$ 
of the variational parameters is $N_{\rm v}=D(N_{\rm hop}+N_{\rm int})$
[see Fig.~\ref{fig.circuit}(a)]. 
Because 
$\hat{\cal F}_{ij}^2 = \hat{1}$ and
$(\hat{Z}_{i}\hat{Z}_j)^2 = \hat{1}$ are satisfied, 
this parametrized variational state allows us to use the parameter-shift rules 
for derivatives 
described in Sec.~\ref{shiftrules}.

\subsection{Suzuki-Trotter decomposition}\label{sec:std}

\begin{center}
  \begin{figure}
    \includegraphics[width=0.35\textwidth]{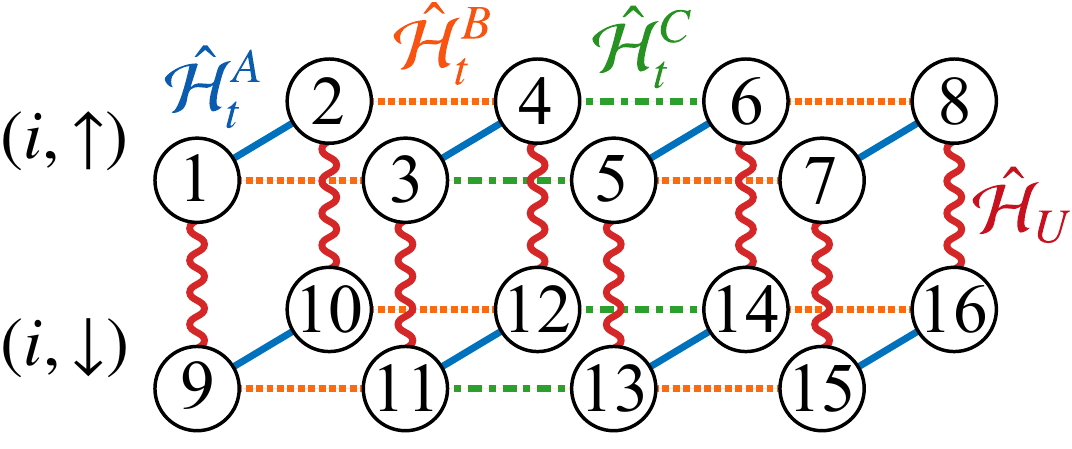}
    \caption{
    In the ST decomposition, the Fermi-Hubbard Hamiltonian $\hat{\cal H}$ on the $4\times 2$ cluster 
    is divided into four parts $\hat{\cal H}_t^{A}$, $\hat{\cal H}_t^{B}$, 
    $\hat{\cal H}_t^{C}$, and $\hat{\cal H}_U$, where $\hat{\cal H}_t^{A}$, $\hat{\cal H}_t^{B}$, and $\hat{\cal H}_t^{C}$ are 
    the hopping ($t$) terms between sites connected by different types of bonds (indicated by blue, orange, and green lines, 
    respectively), and $\hat{\cal H}_U$ is 
    the interaction ($U_{\rm H}$) term (indicated by red curvy lines). 
    Circles represent qubits that are numbered from 1 to 8 for single-particle states at site $i\,(=1,2,\dots,8)$ with spin up and 
    from 9 to16 for single-particle states at site $i\,(=1,2,\dots,8)$ with spin down. 
    This numbering of qubits follows the spin-uniform labeling (see Fig.~\ref{fig.labelings}). 
    }\label{fig:lattice2}
  \end{figure}
\end{center}

As described in Sec.~\ref{sec:ksVQE}, we employ the QPM to generate the Hamiltonian power 
$\hat{\cal H}^n$ for the Krylov subspace $\cal U$ in Eq.~(\ref{eq:Krylov0}). 
In the QPM, the Hamiltonian power $\hat{\cal H}^n$ is approximated with controlled accuracy by a 
linear combination of ST decomposed time-evolution operators. 
In particular, as shown in Eqs.~(\ref{HST}) and (\ref{eq:HST}),  
we employ the second-order symmetric ST decomposition of the time-evolution operator $\e^{-\imag \hat{\cal H} \Delta}$:
\begin{equation}
\e^{-\imag \hat{\cal H} \Delta} = \hat{S}_2(\Delta) + O(\Delta^3) 
\end{equation}
with 
\begin{eqnarray}
  \hat{S}_2(\Delta) &=&
  \e^{-\imag \hat{\cal H}_t^A \Delta/2}
  \e^{-\imag \hat{\cal H}_t^B \Delta/2}
  \e^{-\imag \hat{\cal H}_t^C \Delta/2}  \nonumber \\
  &&\quad\quad \times
  \e^{-\imag \hat{\cal H}_U \Delta}
  \e^{-\imag \hat{\cal H}_t^C \Delta/2}
  \e^{-\imag \hat{\cal H}_t^B \Delta/2}
  \e^{-\imag \hat{\cal H}_t^A \Delta/2},  
\end{eqnarray}
where $\hat{\cal H}_t^{A}$, $\hat{\cal H}_t^{B}$, and $\hat{\cal H}_t^{C}$ are the hopping ($t$) terms 
between sites connected by different types of bonds and $\hat{\cal H}_U$ is the 
interaction ($U_{\rm H}$) term in the Fermi-Hubbard Hamiltonian $\hat{\cal H}$, i.e., 
$\hat{\cal H} = \hat{\cal H}_t^A + \hat{\cal H}_t^B + \hat{\cal H}_t^C + \hat{\cal H}_U$  
(see Fig.~\ref{fig:lattice2}). This decomposition scheme is used for $\hat{H}_{\rm ST}^n(\Delta)$ in Eq.~(\ref{eq:HST}) and 
thus also for $\hat{H}_{\rm ST(1)}^n(\Delta)$ in Eq.~(\ref{HST1}).

\subsection{Numerical results}\label{sec:numerical}

In order to assess the accuracy of the Krylov-extended SAVQE, we evaluate 
the variational energy 
\begin{equation}
  E_0(\bs{\theta}^{(x)}) = \langle \Psi_{\cal U}(\bs{\theta}^{(x)}) |\hat{\cal{H}} |\Psi_{\cal U}(\bs{\theta}^{(x)})\rangle, 
\end{equation}
the fidelity of the ground state 
\begin{equation}
  F(\bs{\theta}^{(x)})=\left|\langle \Psi_0 |\Psi_{\cal U}(\bs{\theta}^{(x)}) \rangle \right|^2,  
\end{equation}
the expectation value of the total spin squared 
\begin{equation}
  \langle \hat{S}^2 \rangle_{\bs{\theta}^{(x)}} = \langle \Psi_{\cal U}(\bs{\theta}^{(x)}) |\hat{S}^2 |\Psi_{\cal U}(\bs{\theta}^{(x)})\rangle,
\end{equation}
and 
the expectation value of the total $\eta$ squared 
\begin{equation}
  \langle \hat{\eta}^2 \rangle_{\bs{\theta}^{(x)}} = \langle \Psi_{\cal U}(\bs{\theta}^{(x)}) |\hat{\eta}^2 |\Psi_{\cal U}(\bs{\theta}^{(x)})\rangle,
\end{equation}
as a function of the number of the optimization iteration $x$ in Eq.~(\ref{iteration}). 
Here, $|\Psi_0\rangle$ is the exact ground state obtained by the Lanczos exact diagonalization method, 
\begin{equation}
 |\Psi_{\cal U}(\bs{\theta}^{(x)}) \rangle 
 = \frac{ |\Psi_{\cal U}^{(0)}(\bs{\theta}^{(x)}) \rangle}{ \sqrt{ \langle \Psi_{\cal U}^{(0)}(\bs{\theta}^{(x)})   |\Psi_{\cal U}^{(0)}(\bs{\theta}^{(x)}) \rangle} }
\end{equation}
is the approximated ground state with the variational parameters $\bs{\theta}^{(x)}$ at the $x$th optimization iteration, and 
$|\Psi_{\cal U}^{(0)}(\bs{\theta}^{(x)}) \rangle$ is given in Eq.~(\ref{gs}). 
All these quantities except for the fidelity $F(\bs{\theta}^{(x)})$ can be calculated as in Eq.~(\ref{eq:average}) 
along with Eq.~(\ref{eq:PPP2}). 
For the fidelity calculation, a careful treatment for the normalization factor 
of the projection operator is required (see Appendix~\ref{app:fidelity}). 
Note also that the expectation values of $\hat{S}_z$ and $\hat{\eta}_z$ are both zero by construction, i.e., 
$\langle \Psi_{\cal U}(\bs{\theta}^{(x)}) |\hat{S}_z |\Psi_{\cal U}(\bs{\theta}^{(x)})\rangle 
= \langle \Psi_{\cal U}(\bs{\theta}^{(x)}) |\hat{\eta}_z |\Psi_{\cal U}(\bs{\theta}^{(x)})\rangle =0$, 
independently of $\bs{\theta}^{(x)}$.

The numerical results shown in this section are all for the Hubbard interaction $U_{\rm H}/t=4$. 
Although the learning rate $\tau$ of the parameter optimization in Eq.~(\ref{iteration}) can be
varied during the optimization iteration in general,
here we fix $\tau=0.025/t$ throughout the optimization iteration. 
The parameter $\Delta$ appearing in Eq.~(\ref{eq:HST}) for the QPM is set to be $\Delta=0.05/t$. 
When the first-order Richardson extrapolation is employed, the expected systematic error for approximating the Hamiltonian power 
$\hat{\cal H}^n$ is $O(\Delta^4)$ [see Eq.~(\ref{eq:FRE})]. 
Indeed, we find that the errors in $\bs{H}$ and $\bs{S}$ are negligible for this $\Delta$ value.

\begin{center}
  \begin{figure}
    \includegraphics[width=\columnwidth]{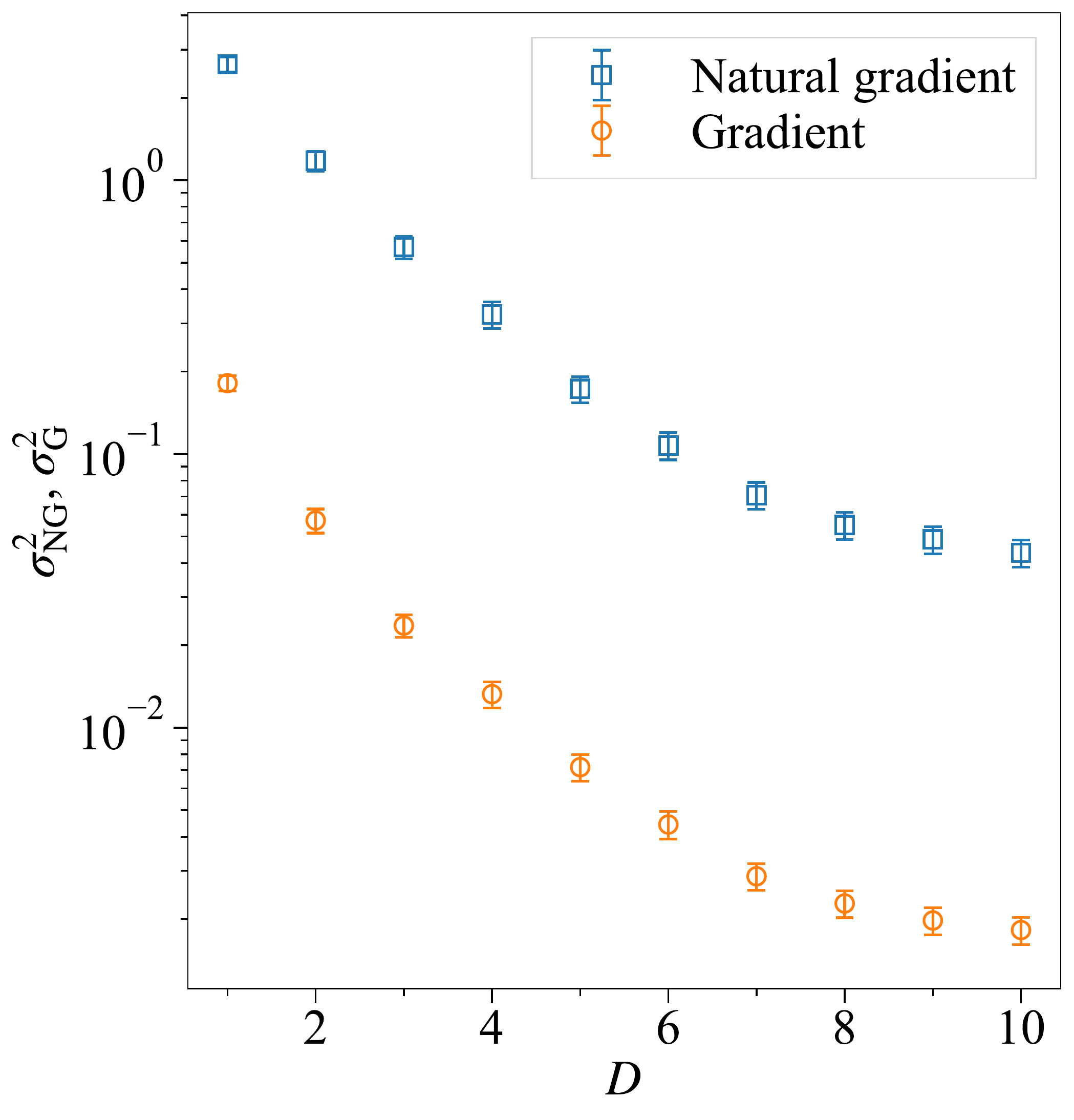}
    \caption{
      Variance of the natural gradient $\sigma_{\rm NG}^2$ (squares) and
      that of the gradient $\sigma_{\rm G}^2$ (circles) as a function of the number $D$ of layers in a quantum circuit 
      shown in Fig.~\ref{fig.circuit} for the parametrized 
      variational state $|\psi(\bs{\theta})\rangle$ without any symmetry projections. The Fermi-Hubbard model on the $4\times2$ cluster 
      is considered and hence there are $N=16$ qubits. The number $N_\nu$ of variational parameters increase linearly in $D$, 
      i.e., $N_\nu=28D$ in this case. 
    }\label{fig:vargrad}
  \end{figure}
\end{center}

Let us first examine
the vanishing-gradient or the barren-plateau problem~\cite{McClean2018} for our particular study.
While the barren-plateau problem
in the context of variational quantum algorithms
has been discussed often with its dependence on the number of
qubits, our focus here is on the comparison between
the natural gradient and the gradient of the cost function, i.e., the expectation value of energy $E_0(\bs{\theta})$, 
for a fixed number of qubits.
Figure~\ref{fig:vargrad} shows the variance of the natural gradient
defined as
\begin{equation}
  \sigma_{\rm NG}^2 \equiv
  \frac{1}{R}\sum_{r=1}^{R}\left[
    \frac{1}{N_{\rm v}}\sum_{k=1}^{N_{\rm v}}
         [         \tilde{\nabla} E_0(\bs{\theta}_r)]_k^2\right],
  \label{eq:sigmaNG}
\end{equation}
where $\bs{\theta}_r=\{\theta_{r,k}\}_{k=1}^{N_{\rm v}}$ is a set of 
the randomly generated variational parameters,
$R$ is the number of the random instances, and 
\begin{equation}
  \tilde{\nabla}E_0(\bs{\theta}_r) \equiv
        [\bs{G}(\bs{\theta}_r)]^{-1}\nabla E_0(\bs{\theta}_r) 
\end{equation}
is the natural gradient~\cite{Amari1998} of the expectation value of energy 
$E_0(\bs{\theta}_r)$ [see Eq.~(\ref{iteration})].
In the calculations, we set $d_{\cal U}=1$ and the projection operator of
the form
$\hat{\cal{P}}=\hat{I}$ (i.e., without any symmetry-projection operators), 
and we estimate $\sigma_{\rm NG}^2$ from $R=64$ random instances. 
The error bar in Fig.~\ref{fig:vargrad}
indicates the standard error of the mean. 
For comparison, we also calculate the variance of the gradient 
\begin{equation}
  \sigma_{\rm G}^2 \equiv
  \frac{1}{R}\sum_{r=1}^{R}\left[
    \frac{1}{N_{\rm v}}\sum_{k=1}^{N_{\rm v}}
         [\nabla E_0(\bs{\theta}_r)]_k^2\right],
  \label{eq:sigmaG}
\end{equation}
in the same manner with the same sets of the randomly generated variational parameters.
In Eqs.~(\ref{eq:sigmaNG}) and (\ref{eq:sigmaG}), 
it is assumed that
the averages of the natural gradient and the gradient over the random
instances vanish.

In Euclidean space, the natural gradient should coincide with the
gradient. However, as shown in Fig.~\ref{fig:vargrad}, 
this is not the case here. Namely, 
the variational-parameter space is certainly non Euclidean.
Indeed, the steepest-descent direction of $E_0(\bs{\theta})$
in the variational-parameter space, which can be
seen as a Riemannian manifold where
the Fubini-Study metric tensor $\bs{G}(\bs{\theta})$ is attributed at each point $\bs{\theta}$,  
is in general not along $-\nabla E_0(\bs{\theta})$
but along $-\tilde{\nabla} E_0(\bs{\theta})=-
[\bs{G}(\bs{\theta})]^{-1}\nabla E_0(\bs{\theta})$
(see Appendix~\ref{app}).
Even when the natural gradient is used,
the variance decreases exponentially in $D$ 
until it is saturated for large $D$, 
which is similar to the results reported in Ref.~\cite{McClean2018}.  
This indicates that the natural gradient cannot solve the
barren-plateau problem. 
However, it is remarkable to notice in Fig.~\ref{fig:vargrad} that the variance of the
natural gradient is more than one order of magnitude larger than
that of the gradient.
The larger variance of the natural gradient
than that of the gradient suggests that 
the natural gradient can alleviate the barren-plateau problem
by capturing the correct steepest-descent direction 
at each point $\bs{\theta}$ in the variational-parameter space, 
at the expense of computing the Fubini-Study metric
tensor $\bs{G}(\bs{\theta})$.

\begin{center}
  \begin{figure}
    \includegraphics[width=\columnwidth]{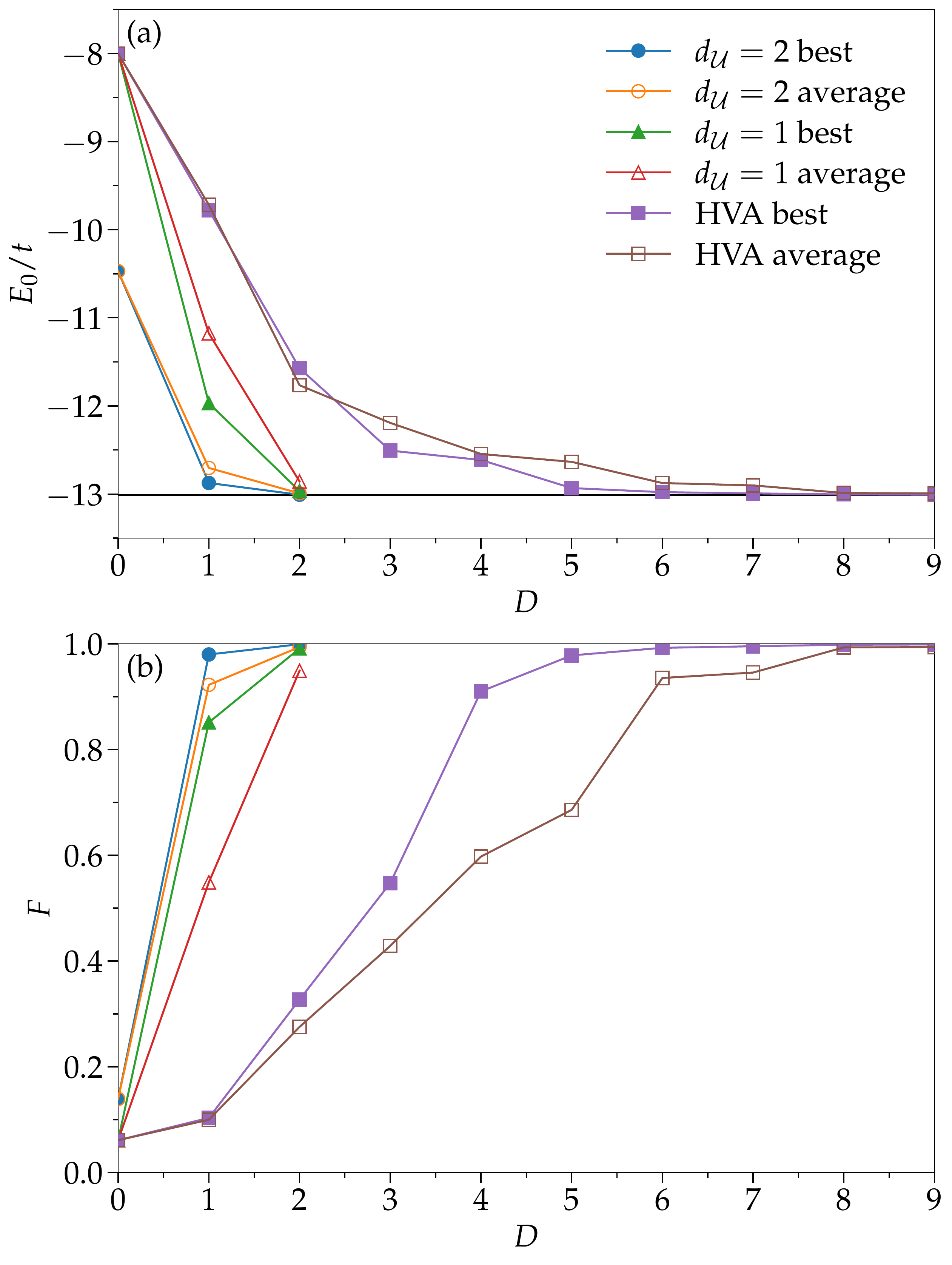}
    \caption{
      (a) The ground-state energy $E_0$ and (b) the ground-state fidelity $F$
      as a function of the number $D$ of layers
      for various variational states with the Krylov subspace dimension $d_{\mathcal{U}}=1$ and 2.
      The horizontal line in (a) indicates
      the exact ground-state energy $E_{\rm exact}(=-13.01250315t)$
      obtained by the Lanczos exact-diagonalization method. 
      For comparison, the results obtained for a HVA are also shown. 
      Here, we perform 64 different independent calculations with different sets of random initial parameters $\bs{\theta}^{(1)}$ 
      and show the results averaged over these 64 independent calculations (open symbols) 
      and the best results in terms of the fidelity among these 64 independent 
      calculations (solid symbols). 
    }  \label{fig.mainresults}
  \end{figure}
\end{center}

Figure~\ref{fig.mainresults} shows the
ground-state energy and the ground-state fidelity as a
function of the number $D$ of the layers of the parametrized
gates (see Fig.~\ref{fig.circuit}) calculated for 
the variational states with the full projection operator 
$\hat{\cal P}=
\hat{\cal P}^{(\eta)}
\hat{\cal P}^{(S)}
\hat{\cal P}^{(\alpha)}$.
We perform 64 independent calculations 
with different sets of random initial parameters $\bs{\theta}^{(1)}$,
and we evaluate these quantities for the well optimized variational
parameters $\bs{\theta}^{(1000)}$.
Figure~\ref{fig.mainresults} shows the results averaged over the 64 independent calculations
and also the best results in terms of the fidelity among the 64 independent calculations. 
For comparison, we also show
the results obtained for a Hamiltonian variational ansatz (HVA)~\cite{Wecker2015vqe,Wiersema2020},
starting also with the initial state $\hat{W}|0\rangle^{\otimes N}$,   
which also preserves the spatial, $S$, and $\eta$ symmetry
without using the projection operators.
Each layer of the HVA consists of $O(L^2)$
two qubit gates due to the Jordan-Wigner string, as in $|\psi(\bs{\theta})\rangle$. 
Further details of the HVA are found in Appendix~\ref{app:trotter}.

As shown in Fig.~\ref{fig.mainresults}(b), the fidelity of $F\approx 0.949$ is achieved
at $D=2$ and $d_{\cal U}=1$ on average,
and  $F\approx 0.991$ is achieved at best.
By increasing the dimension of the subspace to $d_{\cal U}=2$,
$F\approx 0.922$ on average and $F\approx 0.980$ at best at $D=1$, and
$F\approx 0.994$ on average and $F\approx 0.999$ at best at $D=2$ are achieved.
Accordingly, as shown in Fig.~\ref{fig.mainresults}(a), the ground-state energy $E_0$ 
is also systematically improved with increasing the number $D$ of layers or the Krylov subspace dimension 
$d_{\mathcal U}$ to attain almost the exact energy already at $D=1$ and $d_{\mathcal U}=2$ at best and 
the exact energy at $D=2$ on average.  
On the other hand, the HVA, which also respects all the Hamiltonian symmetry,   
requires six layers on average and four layers at best 
to achieve high fidelity of 0.9.
These results are consistent with
the empirical observation that 
the more symmetry is broken to restore it by projection, 
the better quality a variational wave function acquires~\cite{Shi2014}. 

\begin{center}
  \begin{figure}
    \includegraphics[width=\columnwidth]{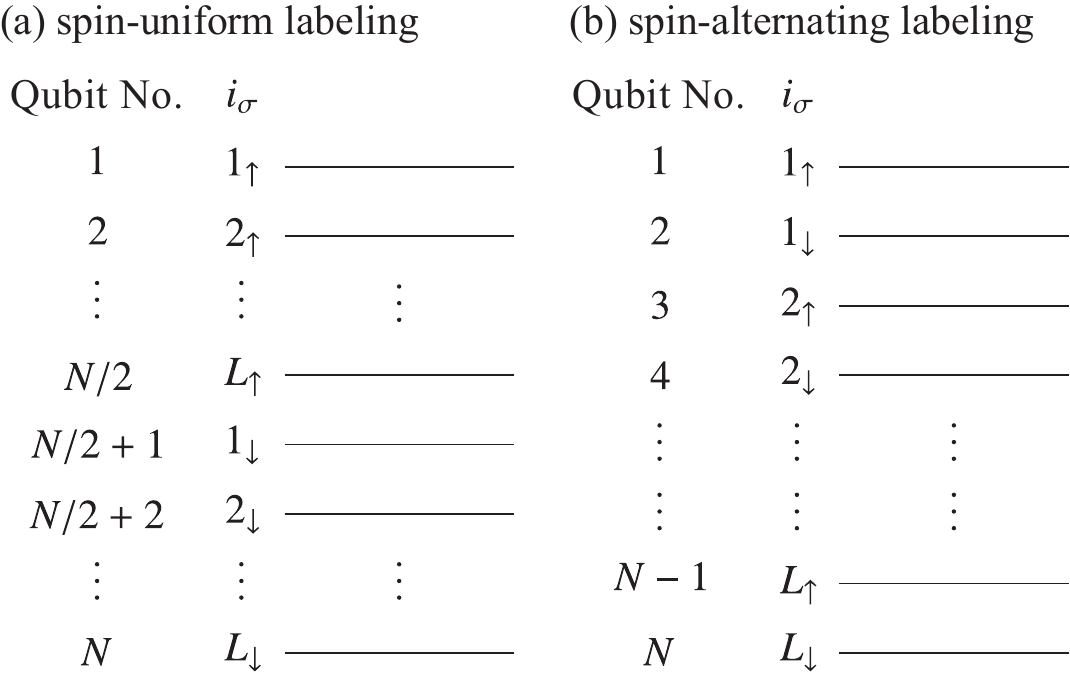}
    \caption{
      Labeling schemes for qubits.  
      (a) spin-uniform labeling and (b) spin-alternating labeling.
      Here, $L=N/2$ is the number of sites. 
    }  \label{fig.labelings}
  \end{figure}
\end{center}

It should be noted that
applying the projection operators and the Hamiltonian powers in the Krylov-extended SAVQE 
requires more gates than the HVA, as summarized in Table~\ref{Table}.  
Moreover, the number of CZ gates required for 
the Jordan-Wigner string depends on the way of mapping
fermion indexes (i.e., local single-particle states) to qubits.
Figure~\ref{fig.labelings} shows two
particular labeling schemes, which we refer to as 
spin-uniform and spin-alternating labelings.
Generally, the full projection operator
requires $O(L^2)$ more two-qubit gates, and
the Hamiltonian power requires
$O(L)$ two-qubit gates (assuming the Fermi-Hubbard model on the ladder lattice as 
in Fig.~\ref{fig:lattice2}),
which are comparable to one to two more
layers in the HVA (see Appendix~\ref{app:trotter}). 
As shown in Table~\ref{Table},
CZ gates for the spin and $\eta$ rotations 
can be eliminated by using the spin-alternating labeling,
while the number of CZ gates for the spatial-symmetry operation 
is double.
Furthermore, the number of CZ gates 
for the time-evolution operator by the hopping term scales similarly but with different prefactors in the two 
different labelings for the Fermi-Hubbard model on the ladder lattice. 
Overall, the two labeling schemes are comparable 
in terms of the number of CZ gates required for the Fermi-Hubbard model 
on the ladder lattice. However, the spin-alternating labeling would be preferred 
for the Fermi-Hubbard model on a two-dimensional square lattice.

\begin{table*}
  \caption{
    Order estimation and counting of the number of
    two-qubit gates required for
    the symmetry and the time-evolution operations 
    of the two-component Fermi-Hubbard model
    in a ladder lattice structure under open-boundary conditions such as the one shown in Fig.~\ref{fig.geometry}. 
    Two labeling schemes for qubits,
    spin-uniform and spin-alternating labelings,
    are considered (see Fig.~\ref{fig.labelings}).
    ``$\times 2$'' (``$\times 4$'') in the right-most column
    indicates that the number of the CZ gates required
    for the Jordan-Wigner (JW) string in the spin-alternating labeling 
    is approximately doubled (quadrupled) as compared with that in the spin-uniform labeling.
    ``Givens", ``Bogoliubov", and ``Exchange'' are three different types of two-qubit gates 
    defined in Eqs.~(\ref{Givens}), (\ref{eq:BTG}), and (\ref{eq:kinetic}), respectively.
    $L$ is the number of sites and
    $\hat{\cal{H}}_t\equiv \hat{\cal H}-\hat{\cal H}_{U}$
    is the hopping term of the Hamiltonian. 
  }
  \label{Table}
  \begin{tabular}{llllll}
    \hline
    \hline
    unitary operators &
    \multicolumn{5}{c}{two-qubit gates}\\
    \hline
    & \multicolumn{2}{c}{spin-uniform labeling}
    & &\multicolumn{2}{c}{spin-alternating labeling}
    \\
    \cline{2-3}
    \cline{5-6}
    spatial symmetry $\hat{g}$ & CZ and {\sc swap} & CZ for {\rm JW} string & & CZ and {\sc swap} & CZ for {\rm JW} string \\
    & $O(L)$ & $O(L^2)$    & & $O(L)$ & $O(L^2) \times 2$\\
    \\
    spin rotation $\e^{-\imag \hat{S}_y \beta}$ & Givens & CZ for {\rm JW} string & & Givens & CZ for {\rm JW} string \\
    & $L$ & $2L(L-1)$ & & $L$ & $0$\\
    \\
    $\eta$ rotation $\e^{-\imag \hat{\eta}_y \beta}$ & Bogoliubov & CZ for {\rm JW} string & & Bogoliubov & CZ for {\rm JW} string \\
    & $L$ & $2L(L-1)$    & & $L$ & $0$\\
    \\
    time evolution by hopping
    $\e^{-\imag \hat{\cal H}_t \Delta}$~\footnote{The same counting of the number of the CZ gates is applied for the product of exponentiated fermionic-{\sc swap} operators [defined in Eq.~(\ref{efswap})] in the variational state $|\psi(\bs{\theta})\rangle$ in Eq.~(\ref{vqe}).}
    & Exchange & CZ for {\rm JW} string & & Exchange & CZ for {\rm JW} string \\
    & $O(L)$ & $O(L)$~\footnote{
    For the Fermi-Hubbard model on a one-dimensional (two-dimensional square) lattice under open-boundary conditions, the number of the CZ gates required is 0 [$O(L^{3/2})$].}    & & $O(L)$ & $O(L)\times 4$~\footnote{
    For the Fermi-Hubbard model on a one-dimensional (two-dimensional square) lattice under open boundary conditions, the number of the CZ gates required is $O(L)$ [$O(L^{3/2})$].}\\
    \\
    time evolution by interaction
    $\e^{-\imag \hat{\cal H}_U \Delta}$
    & CNOT & CZ for {\rm JW} string & & CNOT & CZ for {\rm JW} string \\
    & $2L$ & 0    & & $2L$ & 0\\
    \hline
    \hline
  \end{tabular}
\end{table*}

\begin{center}
  \begin{figure*}
    \includegraphics[width=2\columnwidth]{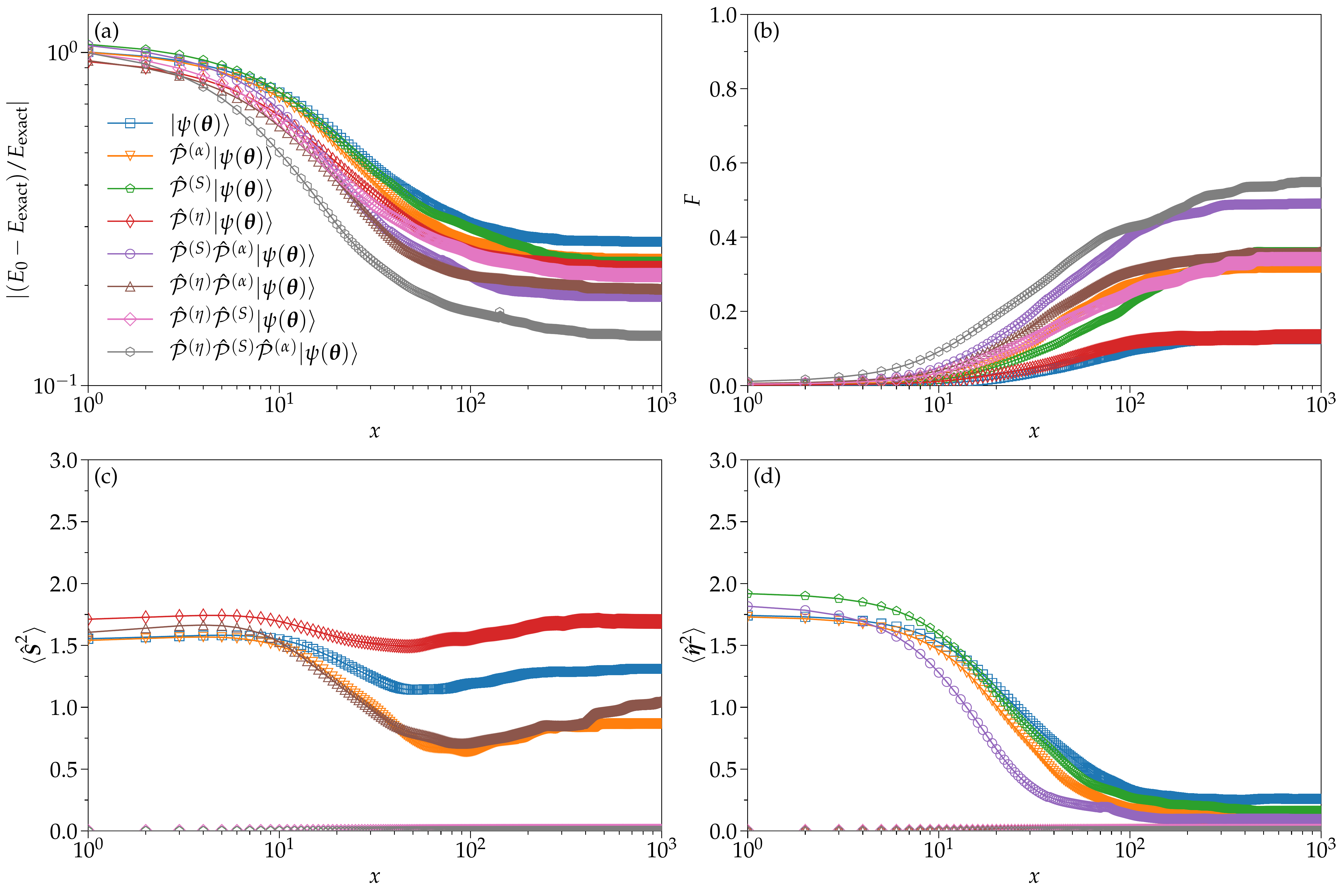}
    \caption{
      (a) The ground-state energy $E_0(\bs{\theta}^{(x)})$,
      (b) the ground-state fidelity $F(\bs{\theta}^{(x)})$,
      (c) the expectation value of total spin squared $ \langle \hat{S}^2 \rangle_{\bs{\theta}^{(x)}}$, and
      (d) the expectation value of total $\eta$ squared $\langle \hat{\eta}^2 \rangle_{\bs{\theta}^{(x)}}$
      as a function of the optimization iteration $x$ 
      with various combinations of projection operators in $|\Psi_{\cal U}(\bs{\theta})\rangle$ for   
      the Fermi-Hubbard model on a $4\times 2$ cluster with open boundary conditions 
      at $U_{\rm H}/J=4$. $E_{\rm exact}$ in (a) is the exact ground-state energy. 
      The quantum circuit of depth $D=1$ is used for 
      $|\psi(\bs{\theta})\rangle$ and hence the number $N_{\rm v}$ of variational parameters is 28. 
      The dimension of the Krylov subspace is $d_{\cal U}=1$. 
      Each result for various combinations of projection operators in $|\Psi_{\cal U}(\bs{\theta})\rangle$ is 
      averaged over 64 independent calculations started with 64 different sets of random initial parameters $\bs{\theta}^{(1)}$, 
      which are common to all calculations. 
    }  \label{fig.energyd1n1}
  \end{figure*}
\end{center}

Now we show the results 
not only for the variational states with the full projection operator 
$\hat{\cal P}=
\hat{\cal P}^{(\eta)}
\hat{\cal P}^{(S)}
\hat{\cal P}^{(\alpha)}$, 
but also with various combinations of the projection operators, 
$\hat{\cal P}^{(\alpha)}$,
$\hat{\cal P}^{(S)}$, and
$\hat{\cal P}^{(\eta)}$,  
to resolve the efficiency of each projection operator.
Figure~\ref{fig.energyd1n1} shows the results for the  quantum 
circuit of depth $D=1$ and the Krylov-subspace dimension $d_{\cal U}=1$. 
Each of the results is average over 64 independent calculations 
with different sets of random initial parameters $\bs{\theta}^{(1)}$
distributed in $[-0.05,0.05]$. 
As shown in Fig.~\ref{fig.energyd1n1}(b),
the fidelity at $x=1$ is essentially zero for all cases 
with these 64 different sets of 
initial parameters $\bs{\theta}^{(1)}$. 
As compared to the results for the bare variational state $|\psi(\bs{\theta})\rangle$ without any projection operators, 
a slight improvement in the ground-state energy and 
a substantial improvement in the ground-state fidelity are found for
$\hat{\cal P}^{(S)}|\psi(\bs{\theta})\rangle$ and also for $\hat{\cal P}^{(\alpha)}|\psi(\bs{\theta})\rangle$. 
On the other hand, 
a slight improvement in the energy and no improvement in the fidelity are found for 
$\hat{\cal P}^{(\eta)}|\psi(\bs{\theta})\rangle$. 
Among the variational states with double projection operators, 
those with the spatial-symmetry projection, 
$\hat{\cal P}^{(S)}\hat{\cal P}^{(\alpha)}|\psi(\bs{\theta})\rangle$ and 
$\hat{\cal P}^{(\eta)}\hat{\cal P}^{(\alpha)}|\psi(\bs{\theta})\rangle$,  
give the better energy than that without the spatial-symmetry projection, 
$\hat{\cal P}^{(\eta)}\hat{\cal P}^{(S)}|\psi(\bs{\theta})\rangle$.
However, the fidelity for   
$\hat{\cal P}^{(S)}\hat{\cal P}^{(\alpha)}|\psi(\bs{\theta})\rangle$ is substantially larger 
than that for 
$\hat{\cal P}^{(\eta)}\hat{\cal P}^{(\alpha)}|\psi(\bs{\theta})\rangle$.  
It is not surprising that the variational state with the full projection operator, i.e., 
$\hat{\cal P}^{(\eta)}\hat{P}^{(S)}\hat{\cal P}^{(\alpha)}|\psi(\bs{\theta})\rangle$, 
achieves the best ground-state energy and fidelity.

As expected, $\langle \hat{S}^2 \rangle_{\bs{\theta}^{(x)}}$ and $\langle \hat{\eta}^2 \rangle_{\bs{\theta}^{(x)}}$ are 
exactly zero, regardless of the values of variational parameters $\bs{\theta}^{(x)}$, for 
the variational states containing the spin-symmetry projection $\hat{\cal P}^{(S)}$ 
and the $\eta$-symmetry projection $\hat{\cal P}^{(\eta)}$, respectively, shown in Figs.~\ref{fig.energyd1n1}(c) 
and \ref{fig.energyd1n1}(d). 
It is also found that while 
$\langle \hat{\eta}^2 \rangle_{\bs{\theta}^{(x)}}$ for the variational states without $\hat{\cal P}^{(\eta)}$
tends to decrease toward zero with increasing the optimization iteration $x$, 
$\langle \hat{S}^2 \rangle_{\bs{\theta}^{(x)}}$ for the variational states without $\hat{\cal P}^{(S)}$ remains finite. 
Moreover, $\langle \hat{S}^2 \rangle_{\bs{\theta}^{(x)}}$ for the variational state $\hat{\cal P}^{(\eta)}|\psi(\bs{\theta})\rangle$ 
converges to a larger value 
than that for the bare variational state $|\psi(\bs{\theta})\rangle$. This can explain the less improved fidelity for 
$\hat{\cal P}^{(\eta)}|\psi(\bs{\theta})\rangle$ in  Fig.~\ref{fig.energyd1n1}(b), 
although the energy is improved better than that for the bare variational state 
$|\psi(\bs{\theta})\rangle$ [see Fig.~\ref{fig.energyd1n1}(a)], in accordance with the variational principle.

\begin{center}
  \begin{figure*}
    \includegraphics[width=2\columnwidth]{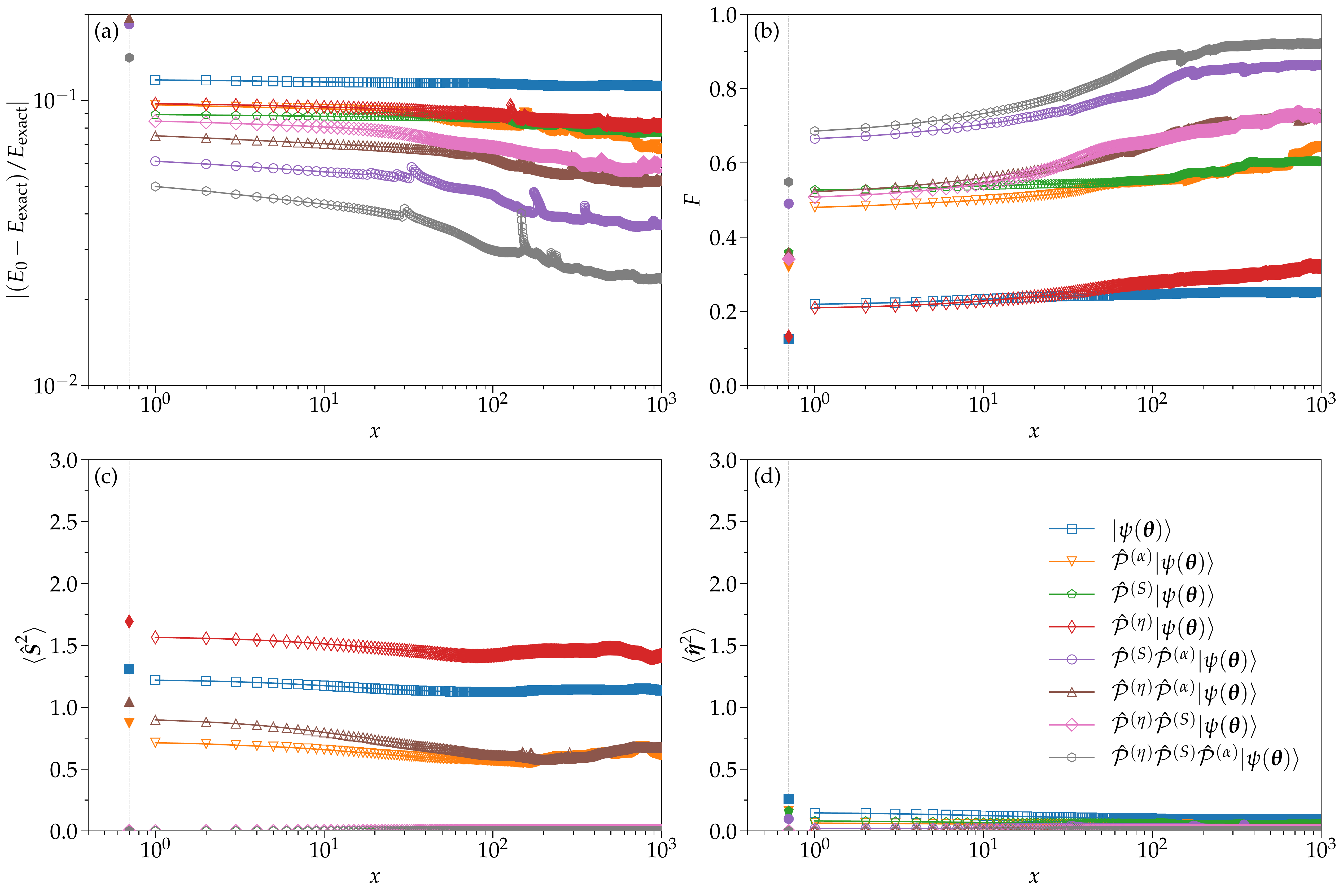}
    \caption{
      Same as Fig.~\ref{fig.energyd1n1} but with the Krylov-subspace dimension $d_{\cal U}=2$. 
      The 64 different sets of optimized variational parameters $\bs{\theta}^{(x)}$ at the last optimization iteration $x=10^3$ obtained 
      for each variational state with $d_{\cal U}=1$ in Fig.~\ref{fig.energyd1n1} are 
      used as the initial sets of variational parameters $\bs{\theta}^{(1)}$ for the corresponding variational states here. 
      For comparison, the results obtained at the last optimization iteration $x=10^3$ in Fig.~\ref{fig.energyd1n1} are 
      also plotted by solid symbols at $x=0.7$. 
      The spiky behavior is occasionally observed in the ground state energy $E_0(\bs{\theta}^{(x)})$ 
      during the parameter optimization iteration. This occurs for particular sets of calculations when the learning rate $\lambda$ 
      in Eq.~(\ref{iteration}) is too large to guarantee the monotonic decrease of $E_0(\bs{\theta}^{(x)})$ with the optimization iteration $x$ 
      (for more details, see Appendix~\ref{ap:ngd}).
    }\label{fig.energyd1n2}
  \end{figure*}
\end{center}

Figure~\ref{fig.energyd1n2} shows the 
same results as in Fig.~\ref{fig.energyd1n1} but with the 
subspace dimension $d_{\cal U}=2$. 
Here, the initial variational parameters are 
set to be the optimal values obtained with $d_{\cal U}=1$ in Fig.~\ref{fig.energyd1n1}. 
As shown in Figs.~\ref{fig.energyd1n2}(a) and 
\ref{fig.energyd1n2}(b), 
by expanding the Krylov subspace, 
both the ground-state energy and the ground-state fidelity are
significantly improved from $d_{\cal U}=1$, 
without increasing the number of the variational parameters. 
Note that the most noticeable improvement of the energy and the fidelity is achieved already at the moment 
when the Krylov subspace is expanded without further optimizing 
the variational parameters. 
The further improvement of these quantities is also observed with increasing the optimization 
iteration $x$. The relative accuracy among the eight different variational states follows essentially the same trend as in the case 
with $d_{\cal U}=1$ shown in Fig.~\ref{fig.energyd1n1}. 
However, we should note that the great improvement of $\langle \hat{S}^2 \rangle_{\bs{\theta}^{(x)}}$ and 
$\langle \hat{\eta}^2 \rangle_{\bs{\theta}^{(x)}}$ is not found even when the Krylov subspace is expanded to $d_{\cal U}=2$, 
as shown in Figs.~\ref{fig.energyd1n2}(c) and \ref{fig.energyd1n2}(d).

\begin{center}
  \begin{figure*}
    \includegraphics[width=2\columnwidth]{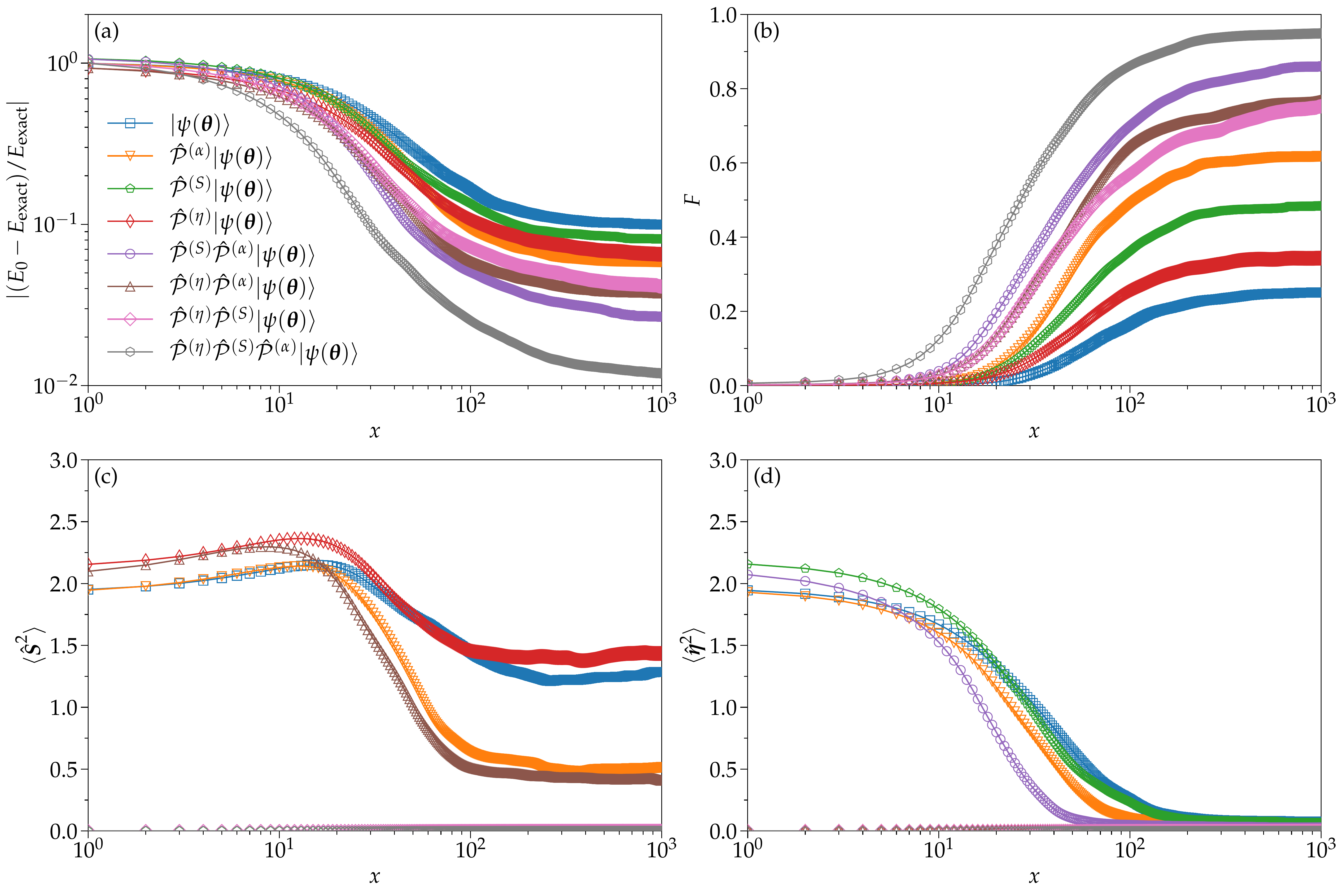}
    \caption{
      {
        Same as Fig.~\ref{fig.energyd1n1} but with the quantum circuit of depth $D=2$ and hence the number $N_{\rm v}$ of 
        variational parameters is 56. 
      }
      \label{fig.energyd2n1}
    }
  \end{figure*}
\end{center}

Figure~\ref{fig.energyd2n1} shows the 
same results as in Fig.~\ref{fig.energyd1n1} but with 
the circuit of depth $D=2$. 
As shown in Fig.~\ref{fig.energyd2n1}(a), the relative error of the ground-state energy 
is decreased by an order of magnitude for all the variational states as compared with 
that for the variational states with $D=1$ in Fig.~\ref{fig.energyd1n1}(a). 
Similarly, the ground-state fidelity in Fig.~\ref{fig.energyd2n1}(b) is increased almost 
two times larger that that for the variational states with $D=1$ in Fig.~\ref{fig.energyd1n1}(b). 
We can also notice in Fig.~\ref{fig.energyd2n1}(c) that no improvement of $\langle \hat{S}^2 \rangle_{\bs{\theta}^{(x)}}$ is made 
for the variational state $\hat{\cal P}^{(\eta)}|\psi(\bs{\theta})\rangle$ 
against the bare variational state $|\psi(\bs{\theta})\rangle$, as in the case of $D=1$ in Fig.~\ref{fig.energyd1n1}(c), 
even though the ground-state energy is indeed improved. 
This is consistent with the moderate improvement of the ground-state fidelity for $\hat{\cal P}^{(\eta)}|\psi(\bs{\theta})\rangle$. 
As shown in Fig.~\ref{fig.energyd2n1}(d), $\langle \hat{\eta}^2 \rangle_{\bs{\theta}^{(x)}}$ converges essentially to zero for all the 
variational states even without containing the $\eta$-symmetry projection $\hat{\cal P}^{(\eta)}$. 
The spatial-symmetry-projected state $\hat{\cal P}^{(\alpha)}|\psi(\bs{\theta})\rangle$ is the best among the variational states 
with a single projection operator, while the variational state with the spatial-symmetry and the spin-symmetry projections, 
$\hat{\cal P}^{(S)}\hat{\cal P}^{(\alpha)}|\psi(\bs{\theta})\rangle$, is the best among 
those with double projection operators.

\begin{center}
  \begin{figure*}
    \includegraphics[width=2\columnwidth]{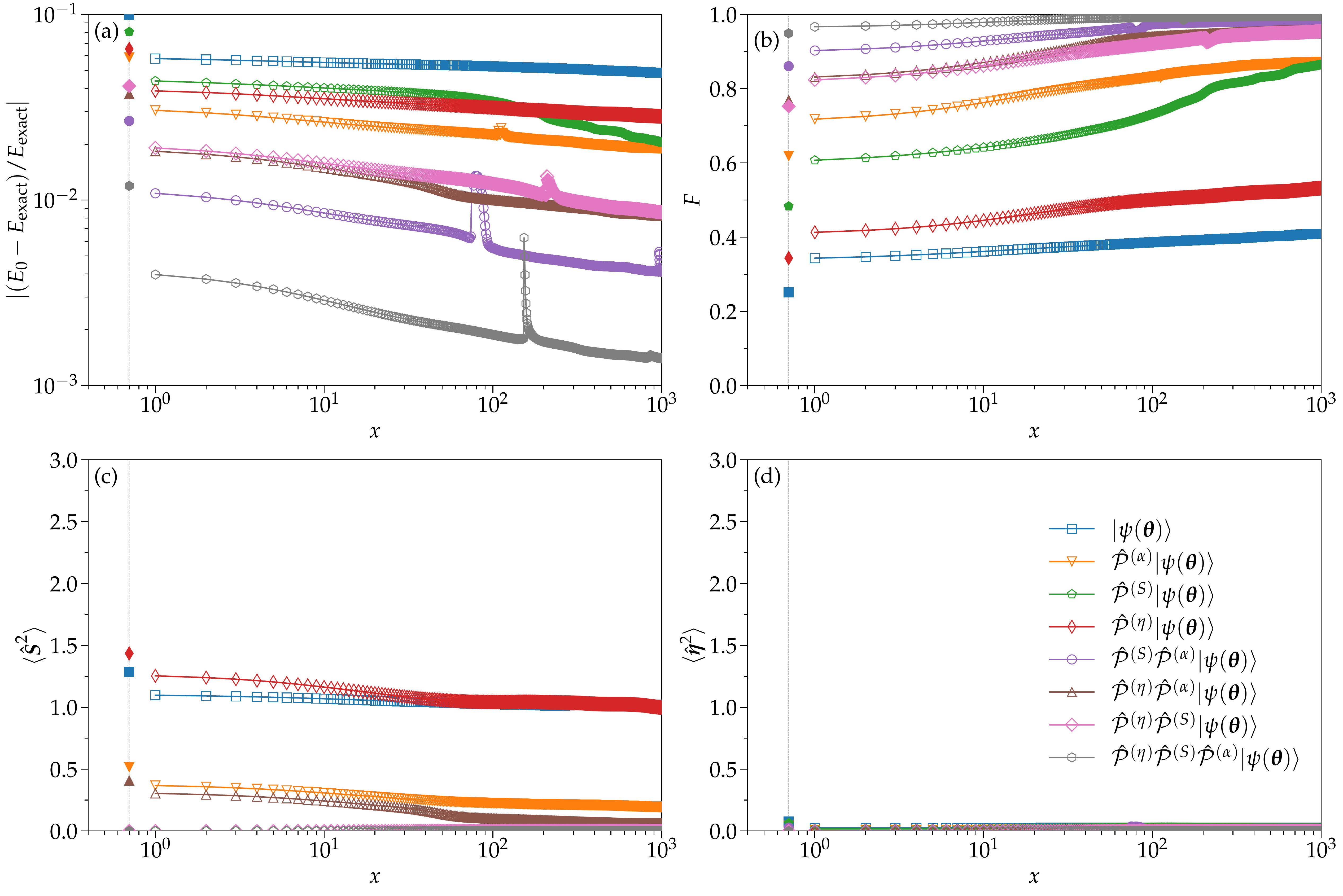}
    \caption{
      Same as Fig.~\ref{fig.energyd2n1} but with the Krylov-subspace dimension $d_{\cal U}=2$. 
      The 64 different sets of optimized variational parameters $\bs{\theta}^{(x)}$ at the last optimization iteration $x=10^3$ obtained 
      for each variational state with $d_{\cal U}=1$ in Fig.~\ref{fig.energyd2n1} are 
      used as the initial sets of variational parameters $\bs{\theta}^{(1)}$ for the corresponding variational states here. 
      For comparison, the results obtained at the last optimization iteration $x=10^3$ in Fig.~\ref{fig.energyd2n1} are 
      also plotted by solid symbols at $x=0.7$.
      The spiky behavior is occasionally observed in the ground state energy $E_0(\bs{\theta}^{(x)})$ 
      during the parameter optimization iteration. This is due to the same reason described in the caption of Fig.~\ref{fig.energyd1n2}.
    }
    \label{fig.energyd2n2}
  \end{figure*}
\end{center}

Figure~\ref{fig.energyd2n2} shows the 
same results as in Fig.~\ref{fig.energyd2n1} but with 
the subspace dimension $d_{\cal U}=2$. 
Here, the initial variational parameters are 
set to be the optimal values obtained with 
$d_{\cal U}=1$ and $D=2$ in Fig.~\ref{fig.energyd2n1}. 
As in the case with $D=1$, 
the ground-state energy and the ground-state fidelity 
are substantially improved over the results for $d_{\cal U}=1$ and $D=2$ shown in Fig.~\ref{fig.energyd2n1}. 
The most remarkable improvement is indeed achieved when the Krylov subspace is expanded without further optimizing the 
variational parameters.  
These quantities are systematically improved by further optimizing the variational parameters. 
As expected, the best variational state is the one with full projected operators, i.e., 
$\hat{\cal P}^{(\eta)}\hat{P}^{(S)}\hat{\cal P}^{(\alpha)}|\psi(\bs{\theta})\rangle$, which 
exhibits, for example, 
the ground-state fidelity as large as 0.9962.
The competitive second best is the variational state with 
the spatial-symmetry and spin-symmetry projections, i.e., $\hat{\cal P}^{(S)}\hat{\cal P}^{(\alpha)}|\psi(\bs{\theta})\rangle$. 
We should also emphasize that the bare variational state $|\psi(\bs{\theta})\rangle$ without any projection operators and 
the variational state with containing only the $\eta$ symmetry projection, $\hat{\cal P}^{(\eta)}|\psi(\bs{\theta})\rangle$, are 
particularly not satisfactory in terms of the ground-sate energy and fidelity as well as $\langle \hat{S}^2 \rangle_{\bs{\theta}^{(x)}}$. 
Our numerical results thus clearly demonstrate that the variational states can be improved by imposing the Hamiltonian symmetry 
on the states without increasing the number of variational parameters.  

\section{Conclusion and Discussion}\label{sec:conclusion}

To conclude, we have proposed a QSE-based VQE scheme that allows us 
to restore the Hamiltonian symmetry of the variational ground state 
in a Krylov subspace generated by
the Hamiltonian and a symmetry-projected state. 
We have described a systematic way to
implement the spatial symmetry operations for fermions 
with the fermionic-{\sc swap} gates, 
and we have also shown how to implement the spin and $\eta$ rotations 
required for the spin- and $\eta$-symmetry projections,  
assuming a fermion-to-qubit mapping with the Jordan-Wigner transformation. 
Moreover, we have generalized the NGD method for the parameter optimization in a QSE-based VQE scheme.  
We have numerically demonstrated the proposed method for the two-component Fermi-Hubbard model on the $4\times2$ cluster, 
and we have found
that the symmetry projections onto the appropriate symmetry sector improve substantially the accuracy of the 
ground state, which is further improved by extending the Krylov-subspace dimension 
without increasing the number of variational parameters 
in the parametrized quantum circuit.

The proposed method is variational and can be regarded as an extension of the SAVQE scheme
and the Krylov-subspace diagonalization method. This is  
because the proposed method is improved 
(i) from the SAVQE scheme by increasing the subspace dimension $d_{\cal U}$ 
without increasing the number of variational parameters, and  
(ii) from the Krylov-subspace diagonalization method by optimizing 
the variational parameters $\bs{\theta}$, but without increasing the subspace dimension $d_{\cal U}$.
The improvement (i) requires more quantum resources than the VQE method, while the improvement (ii) 
requires more classical resources than the Krylov-subspace diagonalization method with the QPM~\cite{seki2021}. 
Therefore, the proposed method allows us to flexibly choose how to use quantum and classical resources 
to improve the results, depending on the performance of available quantum and classical computers.

Finally, we make a few remarks on the $\eta$ symmetry, 
which might not be as familiar as  
the spatial and spin symmetry. 
Recently, the $\eta$ symmetry or the $\eta$-pairing state, which was originally introduced Yang~\cite{Yang1989} 
and was subsequently used to solve the one-dimensional Fermi-Hubbard model analytically~\cite{Essler1991}, 
attracted renewed interest in the context of 
photo excitations~\cite{Kitamura2016,Kaneko2019,Fujiuchi2020,Ejima2020}, 
scar states~\cite{Mark2020,Moudgalya2020},
and a nonequilibrium steady state~\cite{nakagawa2021eta}
in quantum many-body systems.  
The $\eta$ symmetry also exists in other systems such as 
Kondo-lattice systems~\cite{Tsunetsugu1997,Shirakawa2020} and  
spin-orbit-coupled systems~\cite{Li2020,Moudgalya2020} 
under certain constraints of the Hamiltonian.  
Moreover,
$N$-particle generalizations of the
$\eta$-pairing states as eigenstates of 
an extended SU($N$) Fermi-Hubbard model have been reported~\cite{Yoshida2021}.
Therefore, the $\eta$ symmetry and in particular its symmetry projection can also help to prepare the ground states of 
these quantum many-body systems on a quantum computer.

\acknowledgements
The authors would like to thank Tomonori Shirakawa for helpful comments.
A part of the numerical simulations has been done
using the HOKUSAI supercomputer at RIKEN
(Project ID: Q21532) and also supercomputer Fugaku installed in RIKEN R-CCS.
This work is supported by Grant-in-Aid for Research Activity start-up (No.~JP19K23433), 
Grant-in-Aid for Scientific Research (B) (No.~JP18H01183), and 
Grant-in-Aid for Scientific Research (A) (No.~JP21H04446) from MEXT, Japan. 
This work is also supported in part by the COE research grant in computational science from 
Hyogo Prefecture and Kobe City through the Foundation for Computational Science. 

\appendix

\section{Details on the natural-gradient-descent method}\label{app}

We provide additional details on the NGD method. 
We first review how the Fubini-Study metric tensor $\bs{G}(\bs{\theta})$ in Eq.~(\ref{Gmat}) arises. 
Then we derive the NGD iteration for the
energy optimization in Eq.~(\ref{iteration}). 
Finally, we show the positive semidefiniteness of the
matrices related to $\bs{G}(\bs{\theta})$.

\subsection{Fubini-Study metric}

Consider a state $|\Psi \rangle$ and 
its differentiation $|\dd \Psi \rangle$.   
The projection of 
$|\dd \Psi \rangle$ orthogonal to $|\Psi \rangle$ is given by  
\begin{equation}
  |\dd \Psi_\perp\rangle \equiv |\dd \Psi \rangle - |\Psi\rangle \langle \Psi |\dd \Psi \rangle, 
\end{equation}
assuming that $\langle \Psi | \Psi \rangle = 1$. 
In terms of $|\dd \Psi_\perp \rangle$, 
the Fubini-Study metric~\cite{Brody_2001},
denoted as $\dd s^2$, can be written as~\cite{Braunstein1994}
\begin{equation}
  \dd s^2 \equiv \langle \dd \Psi_\perp | \dd \Psi_\perp \rangle
  = \langle \dd \Psi|\dd \Psi \rangle -
  \langle \dd \Psi|\Psi \rangle 
  \langle \Psi|\dd \Psi \rangle.
  \label{eq:ds}
\end{equation}
Assume that $|\Psi \rangle$ is parametrized with 
real variational parameters $\bs{\theta}=\{\theta_k\}_{k=1}^{N_{\rm v}}$.
By expanding
$|\dd \Psi \rangle$ with respect to $\dd \bs{\theta}$, we obtain that 
\begin{equation}
  \dd s^2 = \sum_{k=1}^{N_{\rm v}} \sum_{l=1}^{N_{\rm v}} G_{kl}(\bs{\theta}) \dd \theta_k \dd \theta_l
  = \dd \bs{\theta}^T \bs{G}(\bs{\theta}) \dd \bs{\theta},
  \label{metric}
\end{equation}
where $\dd \bs{\theta}^T = [\dd\theta_1\ \dd\theta_2\ \dots\ \dd\theta_{N_{\rm v}}]$ and 
\begin{equation}
G_{kl}(\bs{\theta}) =
  {\rm Re}
  \langle \partial_k \Psi(\bs{\theta})|\partial_l \Psi(\bs{\theta}) \rangle
  -
  \langle \partial_k \Psi(\bs{\theta})|\Psi(\bs{\theta}) \rangle
  \langle \Psi(\bs{\theta})|\partial_l \Psi(\bs{\theta}) \rangle,
  \label{eq:apG}
\end{equation}
indicating that $\bs{G}(\bs{\theta})$ defines 
the Fubini-Study metric in the variational-parameter space.

\subsection{Natural-gradient-descent iteration}\label{ap:ngd}

To minimize the variational energy $E_0(\bs{\theta})$,
one may wish to find a direction $\delta \bs{\theta}$
in the variational-parameter space 
along which the energy $E_0(\bs{\theta}+\delta \bs{\theta})$ is most decreased, 
when the parameters are displaced by 
a given constant distance, e.g., $||\delta \bs{\theta}||=d$.  
Here, $||\delta \bs{\theta}||$ is the distance in the neighborhood of $\bs{\theta}$
defined by
$||\delta \bs{\theta}||^2= \delta \bs{\theta}^T \bs{G}(\bs{\theta}) \delta \bs{\theta}$.
Such a direction $\delta \bs{\theta}$ can be found by minimizing the function
\begin{equation}
  f(\bs{\theta},\delta \bs{\theta}) =
  E_0(\bs{\theta}+\delta \bs{\theta})
  + \frac{\lambda}{2} \left(\delta \bs{\theta}^T \bs{G}(\bs{\theta}) \delta \bs{\theta}
  - d^2 \right)
  \label{functionalF}
\end{equation}
with respect to $\delta \bs{\theta}$, 
where $\lambda$ is a Lagrange multiplier for 
the equidistant constraint~\cite{Amari1998,Yunoki2006}.
By substituting the linear approximation of the energy 
$E_0(\bs{\theta}+\delta \bs{\theta}) \approx E_0(\bs{\theta})
+ \nabla E_0(\bs{\theta})^T \delta \bs{\theta}$ 
to Eq.~(\ref{functionalF}),  
the steepest direction $\delta \bs{\theta}^*$ such that 
$\nabla_{\delta \bs{\theta}} 
f(\bs{\theta},\delta \bs{\theta})|_{\delta \bs{\theta}=\delta \bs{\theta}^*}=0$  
is obtained as 
\begin{equation}
  \delta \bs{\theta}^* = - \frac{1}{\lambda}
  [\bs{G}(\bs{\theta})]^{-1}\nabla E_0(\bs{\theta}).
  \label{NGD}
\end{equation}
By discretizing Eq.~(\ref{NGD}) as 
$\delta \bs{\theta}^* \to \bs{\theta}^{(x+1)}-\bs{\theta}^{(x)}$ and
$\bs{\theta}\to \bs{\theta}^{(x)}$
and denoting $\tau=1/\lambda$, we obtain Eq.~(\ref{iteration}). 
Although 
$\tau$ can be chosen adaptively depending on $x$, 
the value of $\tau$ 
is fixed throughout the variational-parameter optimization 
in the numerical simulations shown in Sec.~\ref{sec:numerical} and Appendix~\ref{app:trotter}.

From Eq.~(\ref{iteration}),
the change in energy per iteration,
$\delta E_0(\bs{\theta}^{(x)})
=E_0(\bs{\theta}^{(x+1)}) - E_0(\bs{\theta}^{(x)})$, 
can be approximated as 
\begin{alignat}{1}
  \delta E_0(\bs{\theta}^{(x)})
  &\approx
  \nabla E_0(\bs{\theta}^{(x)})^T \delta \bs{\theta}^{(x)} \notag \\
  &=-\tau
  \nabla E_0(\bs{\theta}^{(x)})^T [\bs{G}(\bs{\theta}^{(x)})]^{-1} \nabla E_0(\bs{\theta}^{(x)}), \\
  &=-\frac{1}{\tau}
  \delta \bs{\theta}^{(x)T} \bs{G}(\bs{\theta}^{(x)}) \delta \bs{\theta}^{(x)}, 
  \label{dE2}
\end{alignat}
where $\delta \bs{\theta}^{(x)}=\bs{\theta}^{(x+1)}-\bs{\theta}^{(x)}$. 
Since $\bs{G}(\bs{\theta}^{(x)})$ is positive semidefinite, 
the quadratic forms in the last lines 
are greater than or equal to zero. 
Therefore, if $\tau$ is chosen positive and not too large
so that the above approximation holds, 
the energy is guaranteed to decrease after every iteration, i.e., 
$E_0(\bs{\theta}^{(x+1)}) \leqslant E_0(\bs{\theta}^{(x)})$. 
In our numerical simulations in Figs.~\ref{fig.energyd1n2}(a), \ref{fig.energyd2n2}(a), and \ref{fig:lattice_hva}(a), 
an abrupt increase of $E_0(\bs{\theta}^{(x)})$ occasionally occurs. This is because for particular sets of calculations, 
the value of $\lambda$ set during the optimization iteration is too large to satisfy the condition assumed above.

We note that if the energy in Eq.~(\ref{functionalF})
is expanded up to the second order in $\delta \bs{\theta}$ as
$E_0(\bs{\theta}+\delta{\bs{\theta}}) \approx
E_0(\bs{\theta}) + \nabla E_0(\bs{\theta})^{T} \delta\bs{\theta}
+ \frac{1}{2} \delta \bs{\theta}^T \bs{h}(\bs{\theta}) \delta \bs{\theta}$,
where $\bs{h}(\bs{\theta})$ is the Hessian,
Eq.~(\ref{NGD}) is replaced with 
\begin{equation}
  \delta \bs{\theta}^* = -
  \left[\lambda\bs{G}(\bs{\theta}) + \bs{h}(\bs{\theta})\right]
  ^{-1}
  \nabla E_0(\bs{\theta}).
  \label{NGDNewton}
\end{equation}
Equation~(\ref{NGDNewton}) leads to
a Levenberg-Marquardt-type method,
and reduces to the Newton method if $\lambda=0$. 
Note that the Hessian $\bs{h}(\bs{\theta})$ is not positive semidefinite in general.

\subsection{Positive semidefinite matrices}

Although we already know that 
the Fubini-Study metric tensor $\bs{G}(\bs{\theta})$ is positive semidefinite by definition [see Eq.~(\ref{metric})],
it is instructive to explore some inequalities for $\bs{G}(\bs{\theta})$. 
First, it is convenient to 
define a ${\cal D} \times N_{\rm v}$ Jacobian matrix $\bs{{\cal J}}(\bs{\theta})$ with its component 
\begin{equation}
  {\cal J}_{mk}(\bs{\theta})
  \equiv \frac{\partial \Psi_m(\bs{\theta}) }{\partial \theta_k},   
  \label{eq.Jacobi}
\end{equation}
where $\langle e_m | \Psi(\bs{\theta}) \rangle =\Psi_{m}(\bs{\theta})$
with $\{|e_m\rangle\}_{m=1}^{\cal D}$ being an arbitrary complete orthonormal set, 
and hence $\langle \Psi(\bs{\theta})|\Psi(\bs{\theta})\rangle = \sum_{m=1}^{\cal D}|\Psi_m(\bs{\theta})|^2=1$.  
Inserting the resolution of the identity 
$1=\sum_{m=1}^{\cal D} |e_m \rangle \langle e_m|$ 
into Eqs.~(\ref{gamma}) and (\ref{beta}) yields
\begin{eqnarray}
  \gamma_{kl}(\bs{\theta}) &=& \sum_{m=1}^{\cal D} {\cal J}_{mk}^*(\bs{\theta}) {\cal J}_{ml}(\bs{\theta})
  \label{eq.gammaJ}   
\end{eqnarray}
and 
\begin{eqnarray}
  \beta_k^* (\bs{\theta})
  &=& \sum_{m=1}^{\cal D} {\cal J}_{mk}^*(\bs{\theta}) \Psi_{m}(\bs{\theta}). 
  \label{eq.betaJ}
\end{eqnarray}
Equation~(\ref{eq.gammaJ}) implies that 
$\bs{\gamma}(\bs{\theta})=\bs{{\cal J}}^\dag(\bs{\theta}) \bs{{\cal J}}(\bs{\theta})$ is a Gram matrix and
hence positive semidefinite, i.e., $\bs{\gamma}(\bs{\theta}) \geqslant 0$.

We then consider a real matrix
$[\bs{B}(\bs{\theta})]_{kl} \equiv \beta_k^*(\bs{\theta}) \beta_l(\bs{\theta})$,
which is also a Gram matrix and positive semidefinite, i.e., $\bs{B}(\bs{\theta}) \geqslant 0$. 
For any vector $\bs{x}$, it follows that 
\begin{eqnarray}
  \bs{x}^\dag\bs{B}(\bs{\theta}) \bs{x} &= &
  \sum_{k=1}^{N_{\rm v}} \sum_{l=1}^{N_{\rm v}} x_k^* \beta_k^*(\bs{\theta}) \beta_l(\bs{\theta}) x_l \notag \\
  &=&
  \left[\sum_{m=1}^{\cal D} \left( \sum_{k=1}^{N_{\rm v}} {\cal J}_{mk}(\bs{\theta}) x_k \right)^* \Psi_m(\bs{\theta}) \right]
  \left[\sum_{n=1}^{\cal D} \left( \sum_{l=1}^{N_{\rm v}} {\cal J}_{nl}(\bs{\theta}) x_l \right) \Psi_n^*(\bs{\theta}) \right] \notag \\
  &\leqslant&
  \sum_{m=1}^{\cal D} \left|\sum_{k=1}^{N_{\rm v}} {\cal J}_{mk}(\bs{\theta}) x_k \right|^2
  \sum_{n=1}^{\cal D} |\Psi_n(\bs{\theta})|^2 
  = \bs{x}^\dag\bs{\gamma}(\bs{\theta}) \bs{x},  
\end{eqnarray}
where the Cauchy-Schwarz inequality is used to obtain the third line and 
$ \sum_{n=1}^{\cal D} |\Psi_n(\bs{\theta})|^2=1$ is used
for the last equality.
We thus have shown that 
$\bs{\gamma}(\bs{\theta})\geqslant \bs{B}(\bs{\theta})$,
implying that for any nonzero real vector $\bs{y}$,  
$\bs{y}^T \bs{G}(\bs{\theta}) \bs{y}
=\bs{y}^T {\rm Re}[\bs{\gamma}(\bs{\theta})-\bs{B}(\bs{\theta})] \bs{y}
=\bs{y}^T (\bs{\gamma}(\bs{\theta})-\bs{B}(\bs{\theta})) \bs{y}\geqslant 0$, 
which thus confirms the positive semidefiniteness of $\bs{G}(\bs{\theta})$.
In summary, we have
${\rm Re}\bs{\gamma}(\bs{\theta}) \geqslant \bs{G}(\bs{\theta}) \geqslant 0$ and hence
$0 \leqslant [{\rm Re}\bs{\gamma}(\bs{\theta})]^{-1} 
\leqslant \bs{G}^{-1}(\bs{\theta})$~\cite{yamamoto2019natural}.
Note that 
$\bs{G}(\bs{\theta})$ is induced by $\langle \dd \Psi_\perp| \dd\Psi_\perp \rangle$ 
as in Eqs.~(\ref{eq:ds})-(\ref{eq:apG}) and similarly 
${\rm Re} \bs{\gamma}(\bs{\theta})$ can be induced by
$\langle \dd \Psi| \dd\Psi\rangle$.

\section{Fermionic symmetry operations}\label{app:fermi}

\subsection{Symmetry operation onto a fermionic state}
We briefly review
how the symmetry operator $\hat{g}_m$ acts on a fermion occupation basis state. 
Let $\hat{c}_{i}$ ($\hat{c}_{i}^\dag$) be a fermion annihilation
(creation) operator of 
a single-particle state labeled as $i\,(=1,2,3,\dots)$. 
The fermion operators satisfy the anticommutation relations
$\{\hat{c}_i, \hat{c}_j^\dag\}\equiv
\hat{c}_i \hat{c}_j^\dag +\hat{c}_j^\dag \hat{c}_i=\delta_{ij}$
and 
$\{\hat{c}_i, \hat{c}_j\}=\{\hat{c}_i^\dag, \hat{c}_j^\dag\}=0$. 
Let $| 0 \rangle_{\rm F}$ be the fermion vacuum that
is annihilated by any fermionic annihilation operator 
\begin{equation}
  \hat{c}_i |0 \rangle_{\rm F} = 0 \text{\quad (for any $i$)},
\end{equation}
and is invariant under any symmetry operation
\begin{equation} 
  \hat{g}_{m}|0\rangle_{\rm F} = |0\rangle_{\rm F} \text{\quad (for any $m$)}.
  \label{gvac}
\end{equation}
Then, 
an occupation-basis state $|b\rangle_{\rm F}$ with $N_{\rm F}$ number of fermions
in an occupation basis can be written as
\begin{equation}
  |b\rangle_{\rm F} \equiv
  \underbrace{
    \hat{c}_{k}^\dag \cdots \hat{c}_{j}^\dag  \cdots\hat{c}_{i}^\dag}
  _{N_{\rm F} \text{\ operators}}
  |0\rangle_{\rm F}, 
  \label{basis}
\end{equation}
where $b$ in $|b\rangle_{\rm F}$ denotes a bit string of length $N$,
explicitly defined in Eq.~(\ref{fermion_basis}).
As a convention for the basis states,
one may choose that the indexes are, e.g.,
in the descending order $k > \cdots >j > \cdots > i$ from left to right
in Eq.~(\ref{basis}). 
Note that $N_{\rm F}$ corresponds
to the number of nonzero $b_i$'s in Eq.~(\ref{fermion_basis}),
i.e., $N_{\rm F}=\sum_{i=1}^{N} b_i$.

Suppose that $\hat{g}_m$
transforms the creation operator $\hat{c}_i^\dag$ as
\begin{equation}
  \hat{g}_m \hat{c}_i^\dag \hat{g}_m^{-1} = \hat{c}_{m(i)}^\dag \text{\quad (for any $i$)},
  \label{ftrans}
\end{equation}
or equivalently $\hat{g}_m \hat{c}_i^\dag = \hat{c}_{m(i)}^\dag \hat{g}_m$.
Then, applying the symmetry operation $\hat{g}_m$
on the state $|b\rangle_{\rm F}$ in Eq.~(\ref{basis}) yields
\begin{alignat}{1}
  \hat{g}_{m} |b\rangle_{\rm F} &=
  \hat{g}_{m}
  \left(
  \hat{c}_{k}^\dag
  \cdots
  \hat{c}_{j}^\dag
  \cdots
  \hat{c}_{i}^\dag |0\rangle_{\rm F}\right)\notag \\
  &=
  \left(
  \hat{g}_{m}
  \hat{c}_{k}^\dag
  \hat{g}_{m}^{-1}
  \right)
  \cdots
  \left(
  \hat{g}_{m}
  \hat{c}_{j}^\dag
  \hat{g}_{m}^{-1}
  \right)
  \cdots
  \left(
  \hat{g}_{m}
  \hat{c}_{i}^\dag
  \hat{g}_{m}^{-1}
  \right)
  \hat{g}_m
  |0\rangle_{\rm F} \notag \\
  &=
  \hat{c}_{m(k)}^\dag
  \cdots
  \hat{c}_{m(j)}^\dag
  \cdots
  \hat{c}_{m(i)}^\dag
  |0\rangle_{\rm F},
  \label{basis_after_sym}
\end{alignat}
where Eqs.~(\ref{gvac}) and (\ref{ftrans}) are used
in the third equality.
In general, the indexes
$m(k), \cdots, m(j), \cdots, m(i)$ in Eq.~(\ref{basis_after_sym})
do not match the convention for the basis states 
(i.e., they are not in the descending order).
To associate $\hat{g}_{m}|b\rangle_{\rm F}$ with 
a basis state $|b^\prime \rangle_{\rm F}$,
one has to reorder the fermion creation operators
$\hat{c}_{m(k)}^\dag \cdots \hat{c}_{m(j)}^\dag  \cdots \hat{c}_{m(i)}^\dag$
to match the convention for the occupation-basis states:
\begin{alignat}{1}
  \hat{g}_{m} |b\rangle_{\rm F} &=
  \hat{c}_{m(k)}^\dag
  \cdots
  \hat{c}_{m(j)}^\dag
  \cdots
  \hat{c}_{m(i)}^\dag
  |0\rangle_{\rm F} \notag \\
  &=
  {\rm sgn}({\cal R})\ 
  \hat{c}_{k^\prime}^\dag
  \cdots
  \hat{c}_{j^\prime}^\dag
  \cdots
  \hat{c}_{i^\prime}^\dag
  |0\rangle_{\rm F} \notag \\
  &\equiv
  {\rm sgn}({\cal R})\ |b^\prime\rangle_{\rm F} 
  \label{basis_after_reorder}
\end{alignat}
where $\cal R$ denotes the permutation 
\begin{equation}
  \mathcal{R}\equiv
  \underbrace{
  \left(
  \begin{matrix}
    m(k) & \cdots & m(j) & \cdots &m(i) \\
    k^\prime  & \cdots & j^\prime  & \cdots & i^\prime \\
  \end{matrix}
  \right)
  }_{N_{\rm F} \text{\ columns}}
  \label{reorder}
\end{equation}
with $k^\prime > \cdots >j^\prime > \cdots > i^\prime$,
and ${\rm sgn}({\cal R})=+1$ ($-1$) if
$\cal R$ is an even (odd) permutation.
The sign factor ${\rm sgn}({\cal R})$ arises because 
of the anticommutation relation for fermion creation operators.

\subsection{Symmetry operator as a product of fermionic-{\sc swap} gates}

To implement $\hat{g}_m$ on a quantum circuit, 
a concrete expression of $\hat{g}_m$ is required. 
To this end, we use the 
fermionic-{\sc swap} operator~\cite{Bravyi2002,Essler2005,Verstraete2009,Barthel2009,Wecker2015,Kivlichan2018}
\begin{equation}
  \hat{\cal F}_{ij} \equiv \hat{1}+\left(\hat{c}_{i}^\dag\hat{c}_j + \rm {H.c.}\right)
  -\hat{c}^\dag_{i} \hat{c}_{i}
  -\hat{c}^\dag_{j} \hat{c}_{j}.
  \label{fswap}
\end{equation}
It readily follows from Eq.~(\ref{fswap}) that
$\hat{\cal F}_{ji}
=\hat{\cal F}_{ij}
=\hat{\cal F}_{ij}^\dag
=\hat{\cal F}_{ij}^{-1}$
and $\hat{\cal F}_{ii}=\hat{1}$. 
$\hat{\cal F}_{ij}$ also satisfies the following relations: 
\begin{equation}
  \hat{\cal F}_{ij} |0\rangle_{\rm F} = |0\rangle_{\rm F}
\end{equation}
and
\begin{equation}
  \hat{\cal F}_{ij} \hat{c}_i^\dag \hat{\cal F}_{ij}^{-1} = \hat{c}_{j}^\dag,
  \label{fcf}
\end{equation}
or equivalently $\hat{\cal F}_{ij}\hat{c}_i^\dag = \hat{c}_j^\dag \hat{\cal F}_{ij}$, 
which are analogous to Eqs.~(\ref{gvac}) and (\ref{ftrans}).
Moreover, it can be easily shown that 
$[\hat{\cal F}_{ij},\hat{c}_k^\dag]=0$ 
for $i\not=k$ and $j\not=k$, and
$[\hat{\cal F}_{ij},\hat{\cal F}_{kl}]=0$ 
for $\{i,j\}\not=\{k,l\}$. 
Now one can confirm that
$\hat{\cal F}_{i\delta(i)}$ satisfies Eqs.~(\ref{fswap00})-(\ref{fswap11}).
For example, Eq.~(\ref{fswap11}) can be confirmed as follows:
\begin{alignat}{1}
  &\hat{\cal F}_{j\delta(j)}
  \left(
  \hat{c}_{k}^\dag
  \cdots \hat{c}_{j}^\dag \hat{c}_{\delta(j)}^\dag \cdots
  \hat{c}_i^\dag|0\rangle_{\rm F}
  \right) \notag \\
  =&
  \hat{c}_{k}^\dag
  \cdots
  \left(
  \hat{\cal F}_{j\delta(j)}
  \hat{c}_{j}^\dag
  \hat{\cal F}_{j\delta(j)}^{-1}
  \right)
  \left(
  \hat{\cal F}_{j\delta(j)}
  \hat{c}_{\delta(j)}^\dag
  \hat{\cal F}_{j\delta(j)}^{-1}
  \right)
  \cdots
  \hat{c}_i^\dag
  \hat{\cal F}_{j\delta(j)}
  |0\rangle_{\rm F} \notag
  \\
  =&
  \hat{c}_{k}^\dag
  \cdots \hat{c}_{\delta(j)}^\dag \hat{c}_{j}^\dag \cdots
  \hat{c}_i^\dag|0\rangle_{\rm F}\notag\\
  =&
  -\hat{c}_{k}^\dag
  \cdots \hat{c}_{j}^\dag \hat{c}_{\delta(j)}^\dag \cdots
  \hat{c}_i^\dag|0\rangle_{\rm F},
\end{alignat}
From the above properties of $\hat{\cal F}_{ij}$,
it is now obvious that $\hat{g}_m$ for fermions can be 
obtained by replacing the {\sc swap} operators $\hat{\cal S}_{i\delta(i)}$
in Eq.~(\ref{amida})
with the fermionic-{\sc swap} operators $\hat{\cal F}_{i\delta(i)}$
as in Eq.~(\ref{amida2}).

\section{Two-qubit unitary circuits}~\label{app:two_level_unitaries}

In this appendix, we discuss quantum circuits
and their matrix representations 
for typical two-qubit two-level unitaries
that appear in quantum computations of quantum many-body systems. 

\subsection{Two-qubit two-level unitaries}

\begin{center}
  \begin{figure*}
    \includegraphics[width=2\columnwidth]{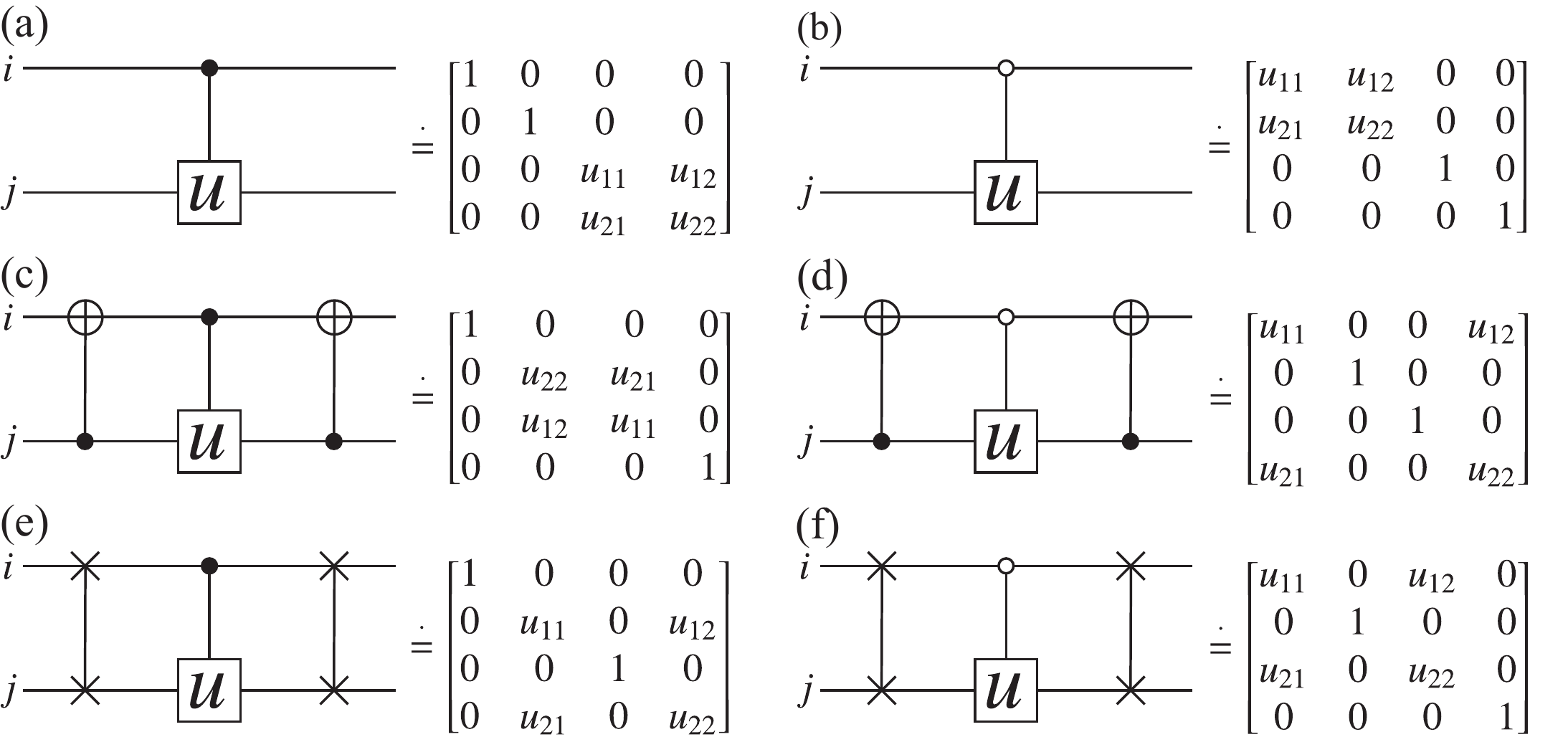}
    \caption{
      {
        List of the two-qubit two-level unitaries.
        The corresponding matrix representations 
        are also shown with the basis states $|0\rangle_i |0\rangle_j$, $|0\rangle_i |1\rangle_j$, 
        $|1\rangle_i |0\rangle_j$, and $|1\rangle_i |1\rangle_j$ for qubits $i$ and $j$. 
      }
      \label{fig.twolevel}
    }
  \end{figure*}
\end{center}

Let $\hat{u}$ be a one-qubit unitary operator such that 
\begin{equation}
  \hat{u}\overset{\cdot}{=}
  \begin{bmatrix}
    u_{11} & u_{12} \\
    u_{21} & u_{22}
  \end{bmatrix}, 
\end{equation}
where the matrix representation is for
the basis states $|0\rangle$ and $|1\rangle$ in the computational basis. 
Assuming the basis states
$|0\rangle_i |0\rangle_j$,
$|0\rangle_i |1\rangle_j$,
$|1\rangle_i |0\rangle_j$, and
$|1\rangle_i |1\rangle_j$ for the $i$th and $j$th qubits
(hereafter we assume $i \not = j$),  
the matrix representation of the controlled-$u$ gate
is then given by a block-diagonal matrix with nontrivial $2\times 2$ entries 
in the lower-right part~\cite{Barenco1995}, 
as shown in Fig.~\ref{fig.twolevel}(a). 
By permuting the basis states with CNOT, SWAP, and $X$ gates,
one can complete the quantum circuits
for the two-qubit two-level unitaries~\cite{NielsenChuang}, 
as shown in Figs.~\ref{fig.twolevel}(b)-\ref{fig.twolevel}(f). 

\subsection{Normal quadratic terms of fermion operators}\label{app:Givens}

Consider the following general quadratic form of fermion operators that conserves the number of fermions (i.e., normal term):  
\begin{alignat}{1}
  &\hat{h}^{\rm (n)}_{ij}=
  \begin{bmatrix}
    \hat{c}_i^\dag & 
    \hat{c}_j^\dag 
  \end{bmatrix}
  \begin{bmatrix}
    a & b \\
    b^* & -a
  \end{bmatrix}
  \begin{bmatrix}
    \hat{c}_i\\ 
    \hat{c}_j
  \end{bmatrix}
  \notag \\
  &=
  {{\rm Re}b} \left(\hat{c}_i^\dag \hat{c}_j + \hat{c}_j^\dag \hat{c}_i \right) 
  +{\imag{\rm Im}b} \left(\hat{c}_i^\dag \hat{c}_j - \hat{c}_j^\dag \hat{c}_i \right) 
  +a\left(\hat{n}_i - \hat{n}_j\right) \label{eq:math} \\
  & = \hat{h}^{\rm (n,1)}_{ij} + \hat{h}^{\rm (n,2)}_{ij} + \hat{h}^{\rm (n,3)}_{ij}, \nonumber  \\
  &\overset{\rm JWT}{=}
  \frac{{\rm Re}b}{2} \left(\hat{X}_i \hat{X}_j + \hat{Y}_i \hat{Y}_j\right) \hat{Z}_{{\rm JW},ij} 
  -\frac{{\rm Im}b}{2} \left(\hat{X}_i \hat{Y}_j - \hat{Y}_i \hat{X}_j\right) \hat{Z}_{{\rm JW},ij}  \notag \\
  &\quad+\frac{a}{2} \left(\hat{Z}_j-\hat{Z}_i\right), 
\end{alignat}
where $a$ and $b$ are real and complex numbers, respectively, such that $a^2+|b|^2=1$, and 
hence the $2\times 2$ matrix is traceless, unitary, and Hermitian.
The $2\times 2$ matrix reduces to Pauli $X$, $Y$, and $Z$ matrices when
$(a,b)=(0,1)$, $(0,-\imag)$, and $(1,0)$, respectively. 
$\hat{h}^{\rm (n,1)}_{ij}$, $\hat{h}^{\rm (n,2)}_{ij}$, and $\hat{h}^{\rm (n,3)}_{ij}$ correspond to each term in Eq.~(\ref{eq:math}) 
and do not commute with each other. 
We shall now consider the exponentiated $\hat{h}^{\rm (n,1)}_{ij}$, $\hat{h}^{\rm (n,2)}_{ij}$, and $\hat{h}^{\rm (n,3)}_{ij}$, 
i.e., ${\rm e}^{-\imag\theta\hat{h}^{\rm (n,1)}_{ij}/2}$, ${\rm e}^{-\imag\theta\hat{h}^{\rm (n,2)}_{ij}/2}$, 
and ${\rm e}^{-\imag\theta\hat{h}^{\rm (n,3)}_{ij}/2}$, and express 
these two-qubit unitary operators in quantum circuits. 

For this purpose, it is useful to notice that 
\begin{equation}
  \left(\hat{c}_i^\dag \hat{c}_j + \hat{c}_j^\dag \hat{c}_i \right)^l =
  \left\{
  \begin{array}{ll}
    \hat{c}_i^\dag \hat{c}_j + \hat{c}_j^\dag \hat{c}_i & \text{for }l\text{ odd}, \\
   \hat{n}_i+\hat{n}_j-2\hat{n}_i\hat{n}_j & \text{for }l\text{ even}, \\
  \end{array}
  \right. \label{ccevenodd}
\end{equation}
\begin{equation}
  \left\{-\imag \left(\hat{c}_i^\dag \hat{c}_j - \hat{c}_j^\dag \hat{c}_i \right)\right\}^l =
  \left\{
  \begin{array}{ll}
    -\imag \left(\hat{c}_i^\dag \hat{c}_j - \hat{c}_j^\dag \hat{c}_i \right) & \text{for }l\text{ odd}, \\
  \hat{n}_i+\hat{n}_j-2\hat{n}_i\hat{n}_j & \text{for }l\text{ even}, \\
  \end{array}
  \right.
\end{equation}
and 
\begin{equation}
  \left(\hat{n}_i - \hat{n}_j \right)^l =
  \left\{
  \begin{array}{ll}
    \hat{n}_i - \hat{n}_j & \text{for }l\text{ odd}, \\
  \hat{n}_i+\hat{n}_j-2\hat{n}_i\hat{n}_j & \text{for }l\text{ even}, \\
  \end{array}
  \right.
\end{equation}
where $l$ is positive integer, or equivalently in the qubit representation that   
\begin{equation}
  \left\{ \frac{1}{2} \left(\hat{X}_i \hat{X}_j + \hat{Y}_i \hat{Y}_j\right) \hat{Z}_{{\rm JW},ij} \right\}^l =
  \left\{
  \begin{array}{ll}
    \frac{1}{2} \left(\hat{X}_i \hat{X}_j + \hat{Y}_i \hat{Y}_j\right) \hat{Z}_{{\rm JW},ij} & \text{for }l\text{ odd}, \\
  \frac{1}{2}\left(1-\hat{Z}_i\hat{Z}_j\right) & \text{for }l\text{ even}, \\
  \end{array}
  \right.
  \label{xevenodd}
\end{equation}
\begin{equation}
  \left\{ \frac{1}{2} \left(\hat{X}_i \hat{Y}_j - \hat{Y}_i \hat{X}_j\right) \hat{Z}_{{\rm JW},ij}\right\}^l =
  \left\{
  \begin{array}{ll}
     \frac{1}{2} \left(\hat{X}_i \hat{Y}_j - \hat{Y}_i \hat{X}_j\right) \hat{Z}_{{\rm JW},ij} & \text{for }l\text{ odd}, \\
  \frac{1}{2}\left(1-\hat{Z}_i\hat{Z}_j\right) & \text{for }l\text{ even}, \\
  \end{array}
  \right.
  \label{yevenodd}
\end{equation}
and
\begin{equation}
  \left\{ \frac{1}{2} \left(\hat{Z}_j - \hat{Z}_i \right) \right\}^l =
  \left\{
  \begin{array}{ll}
     \frac{1}{2} \left(\hat{Z}_j - \hat{Z}_i \right) & \text{for }l\text{ odd}, \\
  \frac{1}{2}\left(1-\hat{Z}_i\hat{Z}_j\right) & \text{for }l\text{ even}. \\
  \end{array}
  \right.
  \label{zevenodd}
\end{equation}
Note that Eqs.~(\ref{xevenodd}) and (\ref{yevenodd}) also hold when the Jordan-Wigner string $\hat{Z}_{{\rm JW},ij}$ 
is absent in these equations. 
Let us now consider an operator $\hat{\cal{A}}$ that satisfies $\hat{\cal{A}}^l=\hat{\cal{A}}$ for $l$ odd and 
$\hat{\cal{A}}^l=\hat{\cal{B}}$ for $l$ even, as in Eqs.~(\ref{ccevenodd})-(\ref{zevenodd}). Then, it is easily shown that 
\begin{equation}
{\rm e}^{-\imag\frac{\theta}{2}\hat{\cal{A}}} = 1 + \left(\cos\frac{\theta}{2}-1\right)\hat{\cal{B}}-\imag\sin\frac{\theta}{2}\hat{\cal{A}}. 
\label{eq:expA}
\end{equation}

Given that the Jordan-Wigner string $\hat{Z}_{{\rm JW},ij}$
commutes with the other operators in Eqs.~(\ref{xevenodd}) and (\ref{yevenodd}),  
and $\hat{Z}_{{\rm JW},ij}^2=1$, 
the matrix representations 
\begin{alignat}{1}
  &\frac{1}{2} \left(\hat{X}_i \hat{X}_j + \hat{Y}_i \hat{Y}_j\right)  \overset{\cdot}{=}
  \begin{bmatrix}
    0 & 0 & 0 & 0 \\
    0 & 0 & 1 & 0 \\
    0 & 1 & 0 & 0 \\
    0 & 0 & 0 & 0
  \end{bmatrix}, \label{xmat}\\
  &\frac{1}{2} \left(\hat{X}_i \hat{Y}_j - \hat{Y}_i \hat{X}_j\right)  \overset{\cdot}{=}
  \begin{bmatrix}
    0 & 0 & 0 & 0 \\
    0 & 0 & \imag & 0 \\
    0 & -\imag & 0 & 0 \\
    0 & 0 & 0 & 0
  \end{bmatrix}, \label{ymat}\\
  &\frac{1}{2} \left(\hat{Z}_j - \hat{Z}_i \right)  \overset{\cdot}{=}
  \begin{bmatrix}
    0 & 0 & 0 & 0 \\
    0 & -1 & 0 & 0 \\
    0 & 0 & 1 & 0 \\
    0 & 0 & 0 & 0
  \end{bmatrix},\label{zmat} 
\end{alignat}
and
  \begin{alignat}{1}
  \frac{1}{2} \left(1-\hat{Z}_i \hat{Z}_j \right)  \overset{\cdot}{=}
  \begin{bmatrix}
    0 & 0 & 0 & 0 \\
    0 & 1 & 0 & 0 \\
    0 & 0 & 1 & 0 \\
    0 & 0 & 0 & 0
  \end{bmatrix} \label{imat}
\end{alignat}
also confirm Eqs.~(\ref{xevenodd}), (\ref{yevenodd}), and (\ref{zevenodd}). 
Here, the basis states
$|0\rangle_i |0\rangle_j$,
$|0\rangle_i |1\rangle_j$,
$|1\rangle_i |0\rangle_j$, and
$|1\rangle_i |1\rangle_j$ for the $i$th and $j$th qubits are assumed. 
Now it is obvious from the matrix representations that 
the operators in Eqs.~(\ref{xmat})-(\ref{imat}) act as
Pauli $X$, $Y$, $Z$, and identity $I$ operators 
on the basis states
$|1\rangle_i |0\rangle_j$ and 
$|0\rangle_i |1\rangle_j$
[note the order of these basis states, see also Fig.~\ref{fig.twolevel}(c)],
and as zero on the basis states
$|0\rangle_i |0\rangle_j$ and $|1\rangle_i |1\rangle_j$.
Therefore, the exponentials of 
the matrices in Eqs.~(\ref{xmat})-(\ref{zmat}) are given by
\begin{alignat}{1}
  & \exp\left[-\imag \frac{\theta}{4} \left(\hat{X}_i \hat{X}_j + \hat{Y}_i \hat{Y}_j\right) \right] \overset{\cdot}{=}
  \begin{bmatrix}
    1 & 0 & 0 & 0 \\
    0 & \cos{\frac{\theta}{2}} & -\imag\sin{\frac{\theta}{2}} & 0 \\
    0 & -\imag\sin{\frac{\theta}{2}} & \cos{\frac{\theta}{2}} & 0 \\
    0 & 0 & 0 & 1
  \end{bmatrix},
  \label{expxmat}\\
  & \exp\left[-\imag \frac{\theta}{4} \left(\hat{X}_i \hat{Y}_j - \hat{Y}_i \hat{X}_j\right) \right] \overset{\cdot}{=}
  \begin{bmatrix}
    1 & 0 & 0 & 0 \\
    0 & \cos{\frac{\theta}{2}} & \sin{\frac{\theta}{2}} & 0 \\
    0 & -\sin{\frac{\theta}{2}} & \cos{\frac{\theta}{2}} & 0 \\
    0 & 0 & 0 & 1 
  \end{bmatrix}, \label{expymat}
\end{alignat} 
and
\begin{alignat}{1}
   \exp\left[-\imag \frac{\theta}{4} \left(\hat{Z}_j - \hat{Z}_i \right) \right] \overset{\cdot}{=}
  \begin{bmatrix}
    1 & 0 & 0 & 0 \\
    0 & \e^{\imag \theta/2} & 0 & 0 \\
    0 & 0 & \e^{-\imag \theta/2} & 0 \\
    0 & 0 & 0 & 1
  \end{bmatrix} \label{expzmat}
\end{alignat} 
in the basis states
$|0\rangle_i |0\rangle_j$,
$|0\rangle_i |1\rangle_j$,
$|1\rangle_i |0\rangle_j$, and
$|1\rangle_i |1\rangle_j$. 
These are also derived directly from Eq.~(\ref{eq:expA}). 

The two-qubit unitaries in 
Eqs.~(\ref{expxmat}), (\ref{expymat}), and (\ref{expzmat}) can be
implemented with the quantum circuit of the form 
in Fig.~\ref{fig.twolevel}(c)
with
$\hat{u}=\hat{R}_{X}(\theta)=\e^{-\imag \theta \hat{X}/2}$,
$\hat{u}=\hat{R}_{Y}(\theta)=\e^{-\imag \theta \hat{Y}/2}$, and
$\hat{u}=\hat{R}_{Z}(\theta)=\e^{-\imag \theta \hat{Z}/2}$, respectively.
Notice that Eq.~(\ref{expxmat}) is identical with the exchange-type gate in Eq.~(\ref{eq:kinetic}), and
Eq.~(\ref{expymat}) is identical with 
the Givens-rotation gate in Eqs.~(\ref{Givens}) and (\ref{eq:mgivens}).
Since the Jordan-Wigner string $\hat{Z}_{{\rm JW},ij}$
can be implemented as the sequences of CZ or fermionic SWAP gates,
as shown in Fig.~\ref{fig:fswap_exp}, the two-qubit unitary operators 
${\rm e}^{-\imag\theta\hat{h}^{\rm (n,1)}_{ij}/2}$, ${\rm e}^{-\imag\theta\hat{h}^{\rm (n,2)}_{ij}/2}$, 
and ${\rm e}^{-\imag\theta\hat{h}^{\rm (n,3)}_{ij}/2}$ can also be implemented in quantum circuits.

\subsection{Anomalous quadratic terms of fermion operators}\label{app:Bogoliubov}

A similar analysis can be made for a quadratic form of fermion operators that does not conserve the number of fermions 
(i.e., anomalous term):  
\begin{alignat}{1}
 &\hat{h}^{\rm (a)}_{ij}=
  \begin{bmatrix}
    \hat{c}_i^\dag & 
    \hat{c}_j 
  \end{bmatrix}
  \begin{bmatrix}
    a & b \\
    b^* & -a
  \end{bmatrix}
  \begin{bmatrix}
    \hat{c}_i\\ 
    \hat{c}_j^\dag
  \end{bmatrix}
  \notag \\
  &=
  {{\rm Re}b} \left(\hat{c}_i^\dag \hat{c}_j^\dag + \hat{c}_j \hat{c}_i \right) 
  +{\imag{\rm Im}b} \left(\hat{c}_i^\dag \hat{c}_j^\dag - \hat{c}_j \hat{c}_i \right) 
  +a\left(\hat{n}_i + \hat{n}_j - 1\right) \label{eq:math2} \\
  & = \hat{h}^{\rm (a,1)}_{ij} + \hat{h}^{\rm (a,2)}_{ij} + \hat{h}^{\rm (a,3)}_{ij}, \nonumber  \\
  &\overset{\rm JWT}{=}
  \frac{{\rm Re}b}{2} \left(\hat{X}_i \hat{X}_j - \hat{Y}_i \hat{Y}_j\right) \hat{Z}_{{\rm JW},ij} 
 +\frac{{\rm Im}b}{2} \left(\hat{X}_i \hat{Y}_j + \hat{Y}_i \hat{X}_j\right) \hat{Z}_{{\rm JW},ij}  \notag \\
  &\quad-\frac{a}{2} \left(\hat{Z}_i+\hat{Z}_j\right), 
\end{alignat}
where $a$ and $b$ are real and complex numbers, respectively, such that $a^2+|b|^2=1$.  
$\hat{h}^{\rm (a,1)}_{ij}$, $\hat{h}^{\rm (a,2)}_{ij}$, and $\hat{h}^{\rm (a,3)}_{ij}$ correspond to each term 
in Eq.~(\ref{eq:math2}) and do not commute with each other. 
We shall now consider the exponentiated $\hat{h}^{\rm (a,1)}_{ij}$, $\hat{h}^{\rm (a,2)}_{ij}$, and $\hat{h}^{\rm (a,3)}_{ij}$, 
i.e., ${\rm e}^{-\imag\theta\hat{h}^{\rm (a,1)}_{ij}/2}$, ${\rm e}^{-\imag\theta\hat{h}^{\rm (a,2)}_{ij}/2}$, 
and ${\rm e}^{-\imag\theta\hat{h}^{\rm (a,3)}_{ij}/2}$, and express 
these two-qubit unitary operators in quantum circuits.

For this purpose, it is useful to notice that 
\begin{equation}
  \left(\hat{c}_i^\dag \hat{c}_j^\dag + \hat{c}_j \hat{c}_i \right)^l =
  \left\{
  \begin{array}{ll}
   \hat{c}_i^\dag \hat{c}_j^\dag + \hat{c}_j \hat{c}_i & \text{for }l\text{ odd}, \\
   2\hat{n}_i\hat{n}_j-(\hat{n}_i+\hat{n}_j)+1 & \text{for }l\text{ even}, \\
  \end{array}
  \right. 
\end{equation}
\begin{equation}
 \left\{\imag \left(\hat{c}_i^\dag \hat{c}_j^\dag - \hat{c}_j \hat{c}_i \right)\right\}^l =
  \left\{
  \begin{array}{ll}
   \imag \left(\hat{c}_i^\dag \hat{c}_j^\dag - \hat{c}_j \hat{c}_i \right) & \text{for }l\text{ odd}, \\
   2\hat{n}_i\hat{n}_j-(\hat{n}_i+\hat{n}_j)+1 & \text{for }l\text{ even}, \\
  \end{array}
  \right. 
\end{equation}
and
\begin{equation}
  \left(\hat{n}_i + \hat{n}_j -1\right)^l =
  \left\{
  \begin{array}{ll}
   \hat{n}_i + \hat{n}_j - 1 & \text{for }l\text{ odd}, \\
   2\hat{n}_i\hat{n}_j-(\hat{n}_i+\hat{n}_j)+1 & \text{for }l\text{ even}, \\
  \end{array}
  \right. 
\end{equation}
where $l$ is positive integer, or equivalently in the qubit representation that   
\begin{equation}
  \left\{ \frac{1}{2} \left(\hat{X}_i \hat{X}_j - \hat{Y}_i \hat{Y}_j\right) \hat{Z}_{{\rm JW},ij} \right\}^l =
  \left\{
  \begin{array}{ll}
   \frac{1}{2} \left(\hat{X}_i \hat{X}_j - \hat{Y}_i \hat{Y}_j\right) \hat{Z}_{{\rm JW},ij} & \text{for }l\text{ odd}, \\
   \frac{1}{2}\left(1+\hat{Z}_i\hat{Z}_j\right) & \text{for }l\text{ even}, \\
  \end{array}
  \right. \label{xevenodd2}
\end{equation}
\begin{equation}
  \left\{ \frac{1}{2} \left(\hat{X}_i \hat{Y}_j + \hat{Y}_i \hat{X}_j\right) \hat{Z}_{{\rm JW},ij}\right\}^l =
  \left\{
  \begin{array}{ll}
   \frac{1}{2} \left(\hat{X}_i \hat{Y}_j + \hat{Y}_i \hat{X}_j\right) \hat{Z}_{{\rm JW},ij} & \text{for }l\text{ odd}, \\
   \frac{1}{2}\left(1+\hat{Z}_i\hat{Z}_j\right) & \text{for }l\text{ even}, \\
  \end{array}
  \right. \label{yevenodd2}
\end{equation}
and
\begin{equation}
  \left\{ \frac{1}{2} \left(\hat{Z}_j + \hat{Z}_i \right) \right\}^l =
  \left\{
  \begin{array}{ll}
   \frac{1}{2} \left(\hat{Z}_j + \hat{Z}_i \right)  & \text{for }l\text{ odd}, \\
   \frac{1}{2}\left(1+\hat{Z}_i\hat{Z}_j\right) & \text{for }l\text{ even}, \\
  \end{array}
  \right. \label{zevenodd2}
\end{equation}
Note that Eqs.~(\ref{xevenodd2}) and (\ref{yevenodd2}) also hold when the Jordan-Wigner string $\hat{Z}_{{\rm JW},ij}$ 
is absent in these equations.

Given that the Jordan-Wigner string $\hat{Z}_{{\rm JW},ij}$
commutes with the other operators in Eqs.~(\ref{xevenodd2}) and (\ref{yevenodd2}), and $\hat{Z}_{{\rm JW},ij}^2=1$,
the matrix representations 
\begin{alignat}{1}
  &\frac{1}{2} \left(\hat{X}_i \hat{X}_j - \hat{Y}_i \hat{Y}_j\right)  \overset{\cdot}{=}
  \begin{bmatrix}
    0 & 0 & 0 & 1 \\
    0 & 0 & 0 & 0 \\
    0 & 0 & 0 & 0 \\
    1 & 0 & 0 & 0
  \end{bmatrix}, \label{xmat2}\\
  &\frac{1}{2} \left(\hat{X}_i \hat{Y}_j + \hat{Y}_i \hat{X}_j\right)  \overset{\cdot}{=}
  \begin{bmatrix}
    0 & 0 & 0 & -\imag \\
    0 & 0 & 0 & 0 \\
    0 & 0 & 0 & 0 \\
    \imag & 0 & 0 & 0
  \end{bmatrix}, \label{ymat2}\\
  &\frac{1}{2} \left(\hat{Z}_j + \hat{Z}_i \right)  \overset{\cdot}{=}
  \begin{bmatrix}
    1 & 0 & 0 & 0 \\
    0 & 0 & 0 & 0 \\
    0 & 0 & 0 & 0 \\
    0 & 0 & 0 & -1
  \end{bmatrix},\label{zmat2} 
\end{alignat}
and
\begin{alignat}{1}
  \frac{1}{2} \left(1+\hat{Z}_i \hat{Z}_j \right)  \overset{\cdot}{=}
  \begin{bmatrix}
    1 & 0 & 0 & 0 \\
    0 & 0 & 0 & 0 \\
    0 & 0 & 0 & 0 \\
    0 & 0 & 0 & 1
  \end{bmatrix}
\end{alignat}
also confirm Eqs.~(\ref{xevenodd2}), (\ref{yevenodd2}), and (\ref{zevenodd2}).
Here, the basis states
$|0\rangle_i |0\rangle_j$,
$|0\rangle_i |1\rangle_j$,
$|1\rangle_i |0\rangle_j$, and
$|1\rangle_i |1\rangle_j$ for the $i$th and $j$th qubits are assumed. 
Now it is obvious from the matrix representation that,
these operators act as Pauli $X$, $Y$, $Z$, and identity $I$ operators
on the basis states
$|0\rangle_i |0\rangle_j$ and 
$|1\rangle_i |1\rangle_j$ 
[see also Fig.~\ref{fig.twolevel}(d)], 
and as zero on the basis states
$|0\rangle_i |1\rangle_j$ and $|1\rangle_i |0\rangle_j$.
Therefore, the exponentials of 
the matrices in Eqs.~(\ref{xmat2})-(\ref{zmat2}) are given by
\begin{alignat}{1}
  & \exp\left[-\imag \frac{\theta}{4} \left(\hat{X}_i \hat{X}_j - \hat{Y}_i \hat{Y}_j\right) \right] \overset{\cdot}{=}
  \begin{bmatrix}
    \cos{\frac{\theta}{2}} & 0 & 0 & -\imag\sin{\frac{\theta}{2}} \\
    0 & 1 & 0 & 0 \\
    0 & 0 & 1 & 0 \\   
    -\imag\sin{\frac{\theta}{2}} & 0 & 0 & \cos{\frac{\theta}{2}} \\
  \end{bmatrix}, \label{expxmat2}\\
  & \exp\left[-\imag \frac{\theta}{4} \left(\hat{X}_i \hat{Y}_j + \hat{Y}_i \hat{X}_j\right) \right] \overset{\cdot}{=}
  \begin{bmatrix}
    \cos{\frac{\theta}{2}} & 0 & 0 & -\sin{\frac{\theta}{2}}  \\
    0 & 1 & 0 & 0 \\
    0 & 0 & 1 & 0 \\
    \sin{\frac{\theta}{2}} & 0 & 0 & \cos{\frac{\theta}{2}}  \\
  \end{bmatrix}, \label{expymat2}
\end{alignat} 
and
\begin{alignat}{1}
   \exp\left[-\imag \frac{\theta}{4} \left(\hat{Z}_j + \hat{Z}_i \right) \right] \overset{\cdot}{=}
  \begin{bmatrix}
    \e^{-\imag \theta/2} & 0 & 0 & 0 \\
    0 & 1 & 0 & 0 \\
    0 & 0 & 1 & 0 \\
    0 & 0 & 0 & \e^{\imag \theta/2}
  \end{bmatrix} \label{expzmat2}
\end{alignat} 
in the basis states
$|0\rangle_i |0\rangle_j$,
$|0\rangle_i |1\rangle_j$,
$|1\rangle_i |0\rangle_j$, and
$|1\rangle_i |1\rangle_j$. 
These are also derived directly from Eq.~(\ref{eq:expA}).

The two-qubit unitaries in 
Eqs.~(\ref{expxmat2}), (\ref{expymat2}), and (\ref{expzmat2}) can be
implemented with the quantum circuit of the form 
in Fig.~\ref{fig.twolevel}(d)
with
$\hat{u}=\hat{R}_{X}(\theta)=\e^{-\imag \theta \hat{X}/2}$,
$\hat{u}=\hat{R}_{Y}(\theta)=\e^{-\imag \theta \hat{Y}/2}$, and
$\hat{u}=\hat{R}_{Z}(\theta)=\e^{-\imag \theta \hat{Z}/2}$, respectively.
Notice that Eq.~(\ref{expymat2}) is identical with 
the Bogoliubov-transformation gate in Eqs.~(\ref{eq:BTG}) and (\ref{eq:BT}).
Since the Jordan-Wigner string $\hat{Z}_{{\rm JW},ij}$
can be implemented as the sequences of CZ or fermionic SWAP gates,
as shown in Fig.~\ref{fig:fswap_exp}, the two-qubit unitary operators 
${\rm e}^{-\imag\theta\hat{h}^{\rm (a,1)}_{ij}/2}$, ${\rm e}^{-\imag\theta\hat{h}^{\rm (a,2)}_{ij}/2}$, 
and ${\rm e}^{-\imag\theta\hat{h}^{\rm (a,3)}_{ij}/2}$ can also be implemented in quantum circuits.

\section{Parallelization of VQE simulations}~\label{app:parallel}

\begin{center}
  \begin{figure*}
    \includegraphics[width=2\columnwidth]{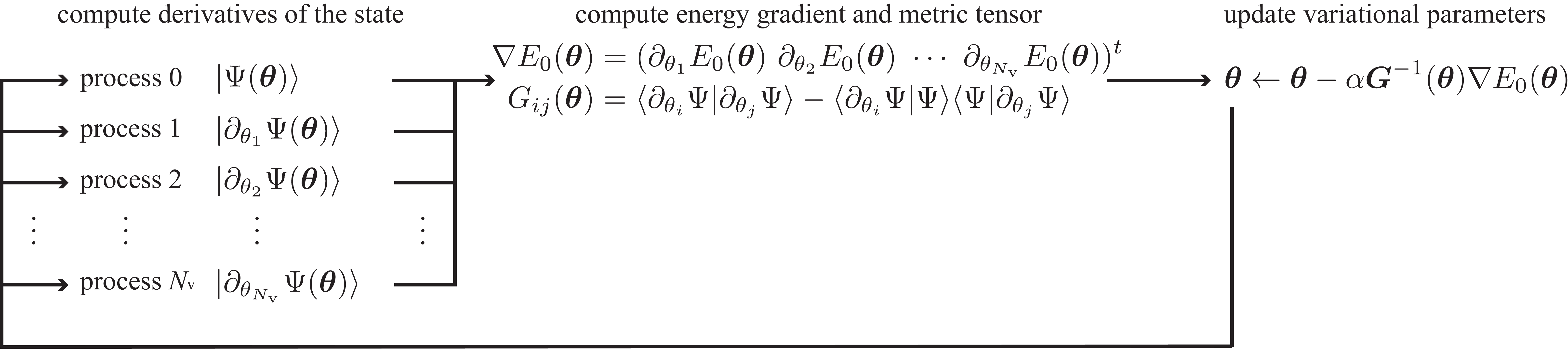}
    \caption{
      Schematic diagram for a parallelization strategy of the VQE simulation with  
      the NGD optimization.
      Here, in this diagram, we assume
      that $d_{\cal U}=1$ without projection operators
      and the number $N_{\rm proc}$ of processes is $N_{\rm proc}=N_{\rm v}+1$, for simplicity. 
    }
    \label{fig.parallel}
  \end{figure*}
\end{center}

Here we briefly describe a simple way of
parallelizing numerical simulations for the VQE method 
by separately and simultaneously evaluating 
derivatives of a variational state $|\Psi(\bs{\theta})\rangle$ with respect to
variational parameters $\bs{\theta}=\{\theta_k\}_{k=1}^{N\nu}$. 
Figure~\ref{fig.parallel} shows a 
schematic diagram of the parallelization strategy. 
In this scheme,
the variational state $|\Psi(\bs{\theta})\rangle$ and its derivatives $|\partial_{\theta_k}\Psi(\bs{\theta})\rangle$
are calculated independently in each process.
All these derivatives 
are then sent to process 0
to compute the
energy gradient $\bs{\nabla} E_0(\bs{\theta})$ and the metric tensor $\bs{G}(\bs{\theta})$.  
Finally, the variational parameters $\bs{\theta}=\{\theta_k\}_{k=1}^{N\nu}$ are updated accordingly
to the NGD scheme in process 0. 
Note that one can simply omit the computation of the metric tensor
$\bs{G}(\bs{\theta})$ when the GD method is used. 
The updated variational parameters are distributed from process 0 
to all other processes (i.e., process~1, process~2, $\cdots$, process~$N_{\rm v}$) for the next iteration. 
Continue this procedure until the convergence is achieved. 
In the schematic diagram shown in Fig.~\ref{fig.parallel}, 
we assume for simplicity
that $d_{\cal U}=1$ without projection operators 
and the number $N_{\rm proc}$ of processes is equal to $N_{\rm v}+1$. 
However, generalization of the parallelization scheme is straightforward 
for the case of $d_{\cal U}>1$ and 
$N_{\rm proc}\not = N_{\rm v}+1$.

\begin{center}
  \begin{figure}
    \includegraphics[width=\columnwidth]{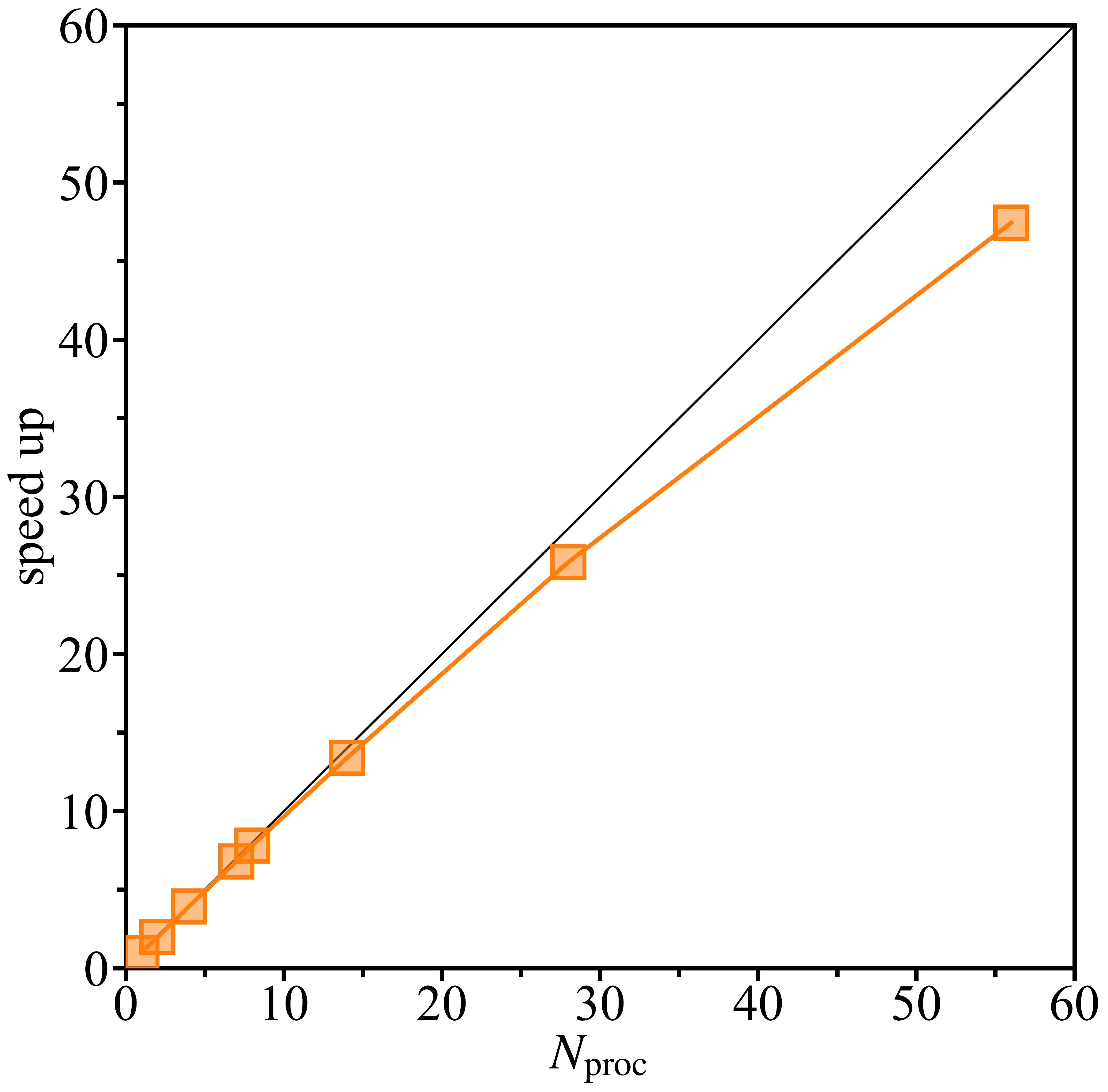}
    \caption{
      Speedup
      as a function of the number $N_{\rm proc}$ of processes 
      for the two-component Fermi-Hubbard model on the $4\times2$ cluster. 
      The benchmark is taken on supercomputer FUGAKU at RIKEN R-CCS. 
    }
    \label{fig.parallel_2}
  \end{figure}
\end{center}

Figure~\ref{fig.parallel_2}
shows the speedup 
as a function of $N_{\rm proc}$ for the case of $N_{\rm v}=56$.
The variational state used here is
the full-projected state $\cal{P}|\psi(\bs{\theta})\rangle$ with 
${N_{S,\rm azimuth}}={N_{\eta,\rm azimuth}}=6$,
${N_{S,\rm polar  }}={N_{\eta,\rm polar  }}=4$, 
$D=2$, and $d_{\cal U}=1$,
and the numerical simulations are performed on supercomputer FUGAKU at RIKEN at R-CCS.
We use 6-thread OpenMP parallelization in each process and MPI between different processes. 
As shown in Fig.~\ref{fig.parallel_2}, an efficient speedup is obtained with increasing $N_{\rm proc}$. 
This simple parallelization strategy is practical for 
small- to medium-size (e.g., $N\leqslant 30$) problems
where a state vector can be stored in the memory of a single node 
and hence communications between different processes do not
occur during
the computation of the derivatives of the state (see Fig.~\ref{fig.parallel}).
For larger problems where
  a state vector must be distributed over the memories on a
  distributed-memory system,
massively parallel quantum-computer simulators for distributed memory systems,
such as that in Ref.~\cite{DeReadt2019}, will be promising.

\section{Normalization factor of a symmetry-projected state in the fidelity calculation}\label{app:fidelity}

Here, we show that the normalization factor due to the 
projection operator in a symmetry-projected state has to be treated with care in the fidelity calculation. 
For simplicity, let us assume the Krylov-subspace dimension $d_{\cal U}=1$. In this case,  
the normalized approximated ground state is given by 
\begin{equation}
  |\Psi_{\cal U}^{(0)}(\bs{\theta})\rangle = \frac{\hat{\cal P} |\psi(\bs{\theta})\rangle}
  {\sqrt{\langle \psi(\bs{\theta}) |\hat{\cal P}|\psi(\bs{\theta}) \rangle }}, 
  \label{eq:nawf}
\end{equation} 
and hence the fidelity reads 
\begin{equation}
  F(\bs{\theta}) = \frac{|\langle \Psi_0|\hat{\cal P} |\psi(\bs{\theta})\rangle |^2}
  {\langle \psi(\bs{\theta}) |\hat{\cal P}|\psi(\bs{\theta}) \rangle }, 
  \label{eq:fidelity}
\end{equation}
where 
\begin{equation}
  \hat{\cal P}^2 = \hat{\cal P}
\end{equation}
is used for the normalization of $|\Psi_{\cal U}^{(0)}(\bs{\theta})\rangle$ in Eq.~(\ref{eq:nawf}). 
By substituting Eq.~(\ref{eq:PPP2}) into Eq.~(\ref{eq:fidelity}), 
we obtain 
\begin{alignat}{1}
  &F(\bs{\theta})
  =
  \frac{2\eta+1}{2}
  \frac{2S+1}{2}
  \frac{d_{\alpha}}{|{\cal G}|}\notag \\
\times&
\frac{\left|
  \sum_{j,l,m}
  w_{j,\eta}
  w_{l,S}
  P_{\eta}
  P_{S}
  \left[\chi^{(\alpha)}\right]^* 
  \langle \Psi_0|
  \e^{-\imag \beta_{j,\eta} \hat{\eta}_y}
  \e^{-\imag \beta_{l,S} \hat{S}_y}
  \hat{g}_m |\psi(\bs{\theta})\rangle 
\right|^2
}
{
  \sum_{j,l,m}
  w_{j,\eta}
  w_{l,S}
  P_{\eta}
  P_{S}
  \left[\chi^{(\alpha)}\right]^* 
  \langle \psi(\bs{\theta})|
  \e^{-\imag \beta_{j,\eta} \hat{\eta}_y}
  \e^{-\imag \beta_{l,S} \hat{S}_y}
  \hat{g}_m |\psi(\bs{\theta})\rangle 
}.
\label{eq:fidelity2}
\end{alignat}
Namely, because of the mismatch of the power exponents 
between the matrix elements of the projection operator $\hat{\cal P}$ 
in the numerator and the denominator on the right-hand side of Eq.~(\ref{eq:fidelity}), 
the normalization factor $(2\eta+1)(2S+1)d_\alpha/4|{\cal G}|$ 
due to the projection operator remains in Eq.~(\ref{eq:fidelity2}).

On the other hand, the expectation value of an observable $\hat{A}$ 
that commutes with the projection operator, $[\hat{A},{\hat {\cal P}}]=0$, 
can be evaluated as 
\begin{equation}
  \langle \Psi_{\cal U}^{(0)}(\bs{\theta})|\hat{A}|\Psi_{\cal U}^{(0)}(\bs{\theta}) \rangle 
  = 
  \frac{\langle \psi(\bs{\theta})| \hat{A} \hat{\cal P} |\psi(\bs{\theta})\rangle }
  {\langle \psi(\bs{\theta}) |\hat{\cal P}|\psi(\bs{\theta}) \rangle }.   
  \label{eq:A}
\end{equation}
Examples of $\hat{A}$ include 
$\hat{A}=\hat{\cal H}$, $\hat{S}^2$, and  $\hat{\eta}^2$. 
Since the power exponents of the matrix elements involving the projection operator $\hat{\cal P}$
in the numerator and the denominator 
on the right-hand side of Eq.~(\ref{eq:A}) are the same, 
the normalization factor due to the projection operator is irrelevant 
for the evaluation of Eq.~(\ref{eq:A}).

\section{Hamiltonian variational ansatz}\label{app:trotter}

In this appendix, the HVA method~\cite{Wecker2015vqe,Wiersema2020} is used to simulate the two-component 
Fermi-Hubbard model described by the Hamiltonian $\hat{\cal{H}}$ in Eq.~(\ref{Ham_Hubbard}) on the $4\times2$ cluster under 
open boundary conditions (see Fig.~\ref{fig.geometry}) at half filling. 
The HVA is a quantum circuit ansatz, based on a principle of the quantum approximate optimization algorithm (QAOA)~\cite{Farhi2014}, 
which is constructed by discretizing a quantum adiabatic 
process~\cite{Barends2015,Barends2016,Ho2019,Mbeng2019,Mbeng2019B,Wauters2020,Shirakawa2021}.    

\begin{center}
  \begin{figure}
    \includegraphics[width=0.35\textwidth]{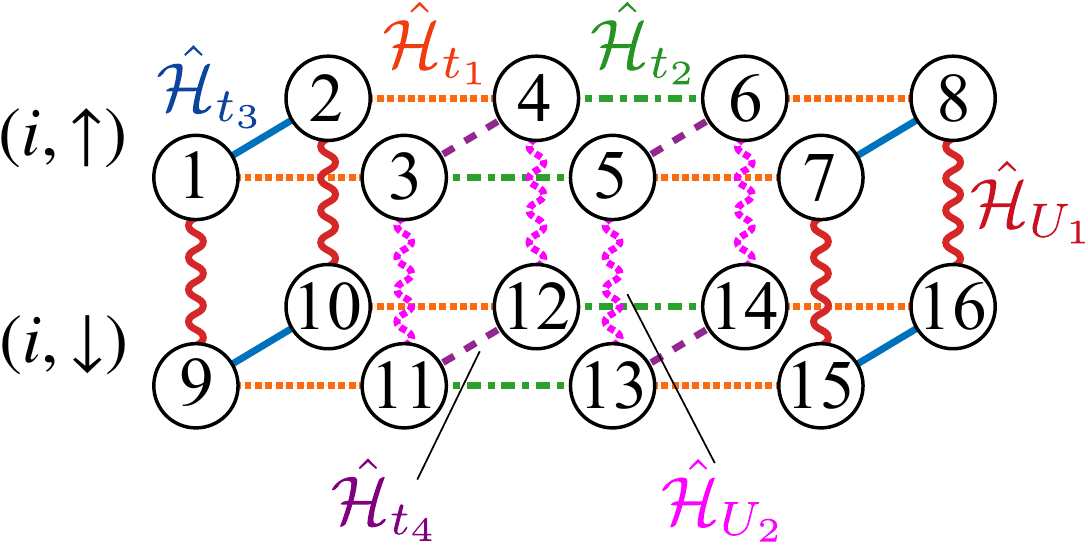}
    \caption{
      In the HVA, the Fermi-Hubbard Hamiltonian $\hat{\cal H}$ on the $4\times 2$ cluster 
      is divided into six parts $\hat{\cal H}_{t_1}$, $\hat{\cal H}_{t_2}$, 
      $\hat{\cal H}_{t_3}$, $\hat{\cal H}_{t_4}$, $\hat{\cal H}_{U_1}$, and $\hat{\cal H}_{U_2}$, 
      where $\hat{\cal H}_{t_1}$, $\hat{\cal H}_{t_2}$, $\hat{\cal H}_{t_3}$, and $\hat{\cal H}_{t_4}$ are 
      the hopping ($t$) terms between sites connected by different types of bonds (indicated by 
      orange dotted, green dash-dotted, blue solid, and purple dashed lines,
      respectively), and $\hat{\cal H}_{U_1}$ and $\hat{\cal H}_{U_2}$ are 
      the interaction ($U_{\rm H}$) terms on nonequivalent sites (indicated by
      red solid curvy and magenta dashed curvy lines, respectively). 
      Circles represent qubits that are numbered from 1 to 8 for single-particle states at site $i\,(=1,2,\dots,8)$ with spin up and 
      from 9 to16 for single-particle states at site $i\,(=1,2,\dots,8)$ with spin down. 
    }\label{fig:lattice_hva}
  \end{figure}
\end{center}

\begin{center}
  \begin{figure*}
    \includegraphics[width=1\textwidth]{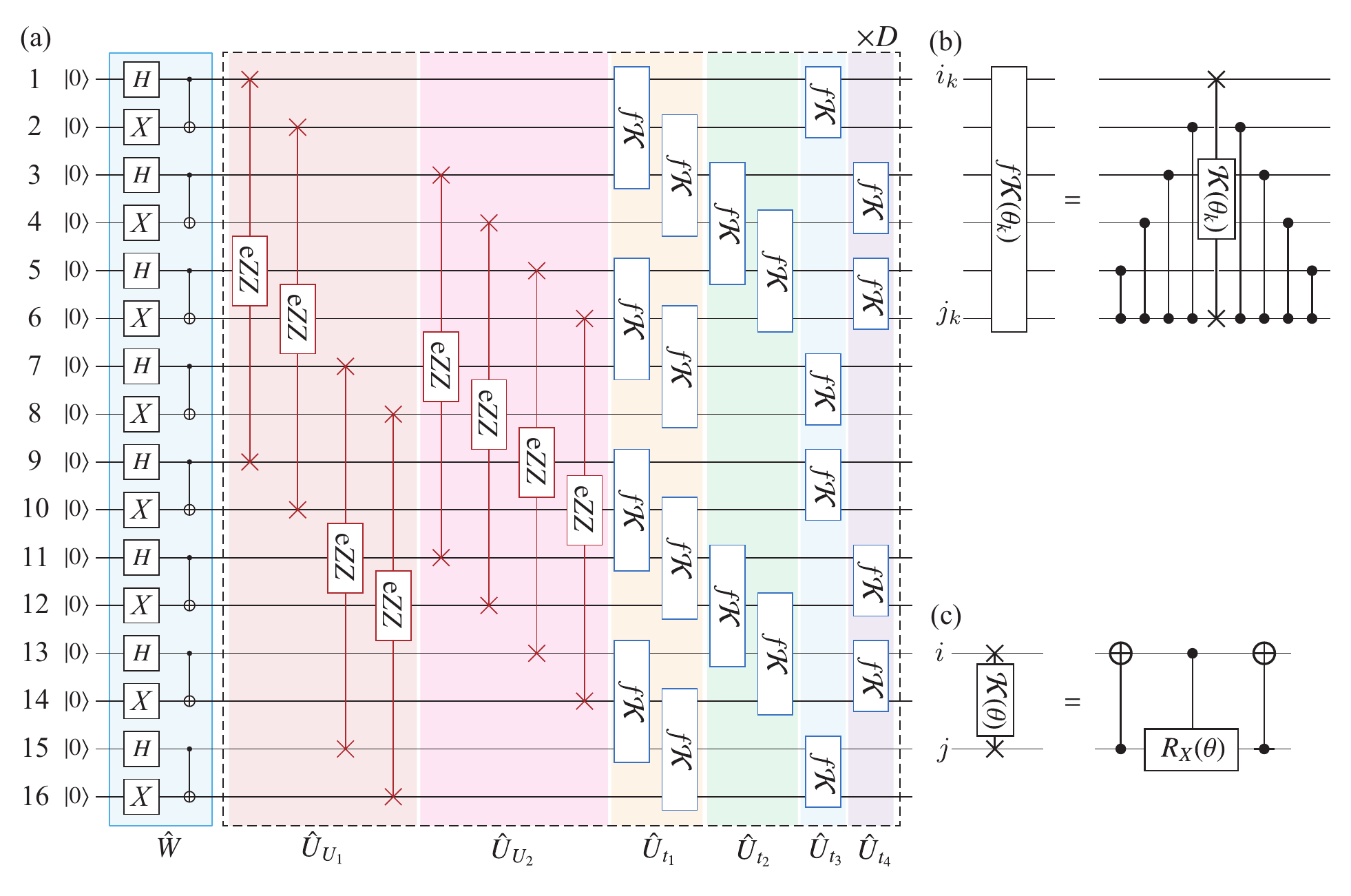}
    \caption{
      (a) A quantum circuit for preparing the 
      HVA state $|\psi_{\rm HVA}(\bs{\theta})\rangle=\prod_{l=D}^{1}\hat{U}_{{\rm HVA},l}(\bs{\theta}_l)
      \hat{W}|0\rangle^{\otimes N}$ with 
      $\hat{U}_{{\rm HVA},l}(\bs{\theta}_{l})=
      \hat{U}_{t_4}(\theta_{6,l})
      \hat{U}_{t_3}(\theta_{5,l})
      \hat{U}_{t_2}(\theta_{4,l})
      \hat{U}_{t_1}(\theta_{3,l})
      \hat{U}_{U_2}(\theta_{2,l})
      \hat{U}_{U_1}(\theta_{1,l})$.
      $f{\cal K}$ denotes a fermionic exchange gate $\widehat{f{\cal K}}_{i j}(\theta)$
      acting on qubits $i$ and $j$ 
      with $\theta = -t\theta_{k,l}$~\cite{notepara},  
      defined in Eq.~(\ref{eq:def_T}).  
      eZZ denotes a gate $\exp(-\imag \hat{Z}_{i} \hat{Z}_{j} \theta/2)$ 
      acting on qubits $i$ and $j$ with $\theta=U_{\rm H}\theta_{k,l}/4$~\cite{notepara}, 
      defined in Eq.~(\ref{eZZ}) and Fig.~\ref{fig.circuit}(c). 
      The qubit numbers indicated in the left most side correspond to the numbering of qubits 
      in Fig.~\ref{fig:lattice_hva} for the two-component Fermi-Hubbard model on the $4\times2$ cluster. 
      (b) A decomposition of a fermionic exchange gate
      ${f{\cal K}}(\theta)$ into a exchange gate ${\cal K}(\theta)$ sandwiched with CZ gates, 
      as in Eq.~(\ref{CZTCZ}).
      (c) A decomposition of the exchange gate $\mathcal{K}(\theta)$ defined in Eq.~(\ref{eq:kinetic}). 
      Here, $\hat{R}_{X}(\theta)=\e^{-\imag \theta \hat{X}_i/2}$ acting on qubit $i$.
    } \label{fig.circuit_hav}
  \end{figure*}
\end{center}

The HVA state $|\psi_{\rm HVA}(\bs{\theta})\rangle$ adopted in this study is given by 
\begin{equation}
  |\psi_{\rm HVA}(\bs{\theta})\rangle
  =\prod_{l=1}^{D}\hat{U}_{{\rm HVA},l}(\bs{\theta}_l)
    \hat{W}|0\rangle^{\otimes N},
\end{equation}
where $\bs{\theta}=\{\bs{\theta}_l\}_{l=1}^D$ with $\bs{\theta}_l=\{\theta_{k,l}\}_{k=1}^6$ is a set of variational parameters 
and 
\begin{alignat}{1}
  \hat{U}_{{\rm HVA},l}(\bs{\theta}_{l})
  &=
  \hat{U}_{t_4}(\theta_{6,l})
  \hat{U}_{t_3}(\theta_{5,l})
  \hat{U}_{t_2}(\theta_{4,l})
  \hat{U}_{t_1}(\theta_{3,l})\notag \\
  &\times
  \hat{U}_{U_2}(\theta_{2,l})
  \hat{U}_{U_1}(\theta_{1,l})
\end{alignat}
with $\hat{U}_{t_\alpha}(\theta)={\rm e}^{-\imag\theta\hat{\cal{H}}_{t_\alpha}/2}$ and
$\hat{U}_{U_\beta}(\theta)={\rm e}^{-\imag\theta\hat{\cal{H}}_{U_\beta}/2}$ being unitary operators generated by
hopping ($t$) and interaction ($U_{\rm H}$) terms, respectively (see Fig.~\ref{fig:lattice_hva}). 
For example, one of the former unitary operators is defined as 
\begin{equation}
\hat{U}_{t_3}(\theta) = {\rm e}^{-\imag\theta\hat{h}_{1,2}^{\uparrow}/2} {\rm e}^{-\imag\theta\hat{h}_{7,8}^{\uparrow}/2}
{\rm e}^{-\imag\theta\hat{h}_{1,2}^{\downarrow}/2} {\rm e}^{-\imag\theta\hat{h}_{7,8}^{\downarrow}/2}
\end{equation}
with $\hat{h}_{i,j}^\sigma = -t(c_{i\sigma}^\dag c_{j\sigma} + c_{j\sigma}^\dag c_{i\sigma})$. 
With the Jordan-Wigner transformation, 
\begin{alignat}{1}
  {\rm e}^{-\imag\theta\hat{h}_{i,j}^{\sigma}/2}&
  \overset{\rm JWT}{=}
  \exp\left[
    \imag\frac{t\theta}{4} \left(
    \hat{X}_{i_\sigma} \hat{X}_{j_\sigma} +
    \hat{Y}_{i_\sigma} \hat{Y}_{i_\sigma} 
    \right)\hat{Z}_{{\rm JW},i_\sigma,i_\sigma} \right]
  \label{XXYY}\\
  &=
  \left[\prod_{i_\sigma \lessgtr k \lessgtr j_\sigma } \widehat{{\rm CZ}}_{j_\sigma k}\right]
  \hat{\cal K}_{i_\sigma j_\sigma}\left(-t\theta \right)
  \left[\prod_{i_\sigma \lessgtr k \lessgtr j_\sigma} \widehat{{\rm CZ}}_{j_\sigma k}\right]
  \label{CZTCZ} \\
  &\equiv \widehat{f{\cal K}}_{i_\sigma j_\sigma}(-t\theta), \label{eq:def_T} 
\end{alignat}
where the 
${\rm CZ}$ gates in Eq.~(\ref{CZTCZ})
account for the Jordan-Wigner string in Eq.~(\ref{XXYY}),
as shown for the more general case in Eq.~(\ref{eq:fswap_exp}), and 
\begin{equation}
  \hat{\cal K}_{i j}(\theta)=
  \exp\left[
    {-}
    \imag\frac{\theta}{4} \left(
    \hat{X}_{i} \hat{X}_{j} +
    \hat{Y}_{i} \hat{Y}_{j} 
    \right) \right]
  \label{eq:kinetic}
\end{equation}
is the exchange-type gate as in Eq.~(\ref{expxmat}) [also see Fig.~\ref{fig.circuit_hav}(b)]. 
$\hat{U}_{U_\beta}(\theta)$ operators can be implemented in a quantum circuit by a product of the eZZ gates 
defined in Eq.~(\ref{eZZ}) and Fig.~\ref{fig.circuit}(c).

Therefore, as shown in Fig.~\ref{fig.circuit_hav}(a), there are six independent
variational parameters $\{\theta_{i,l}\}_{i=1}^6$ for each layer $l$ and hence $6D$ parameters in total for the 
HVA state $|\psi_{\rm HVA}(\bs{\theta})\rangle$ composed of $D$ layers. 
These variational parameters are optimized by the NGD method so as to 
minimize the expectation value of energy. 
Notice that the initial state $\hat{W}|0\rangle^{\otimes N}$ in $|\psi_{\rm HVA}(\bs{\theta})\rangle$ is the ground state of 
$\hat{\cal{H}}_{t_3}+\hat{\cal{H}}_{t_4}$
and thus $|\psi_{\rm HVA}(\bs{\theta})\rangle$ indeed follows a discretized version of a 
quantum adiabatic process in which an optimal unitary evolution path, i.e., $\bs{\theta}=\{\bs{\theta}_l\}_{l=1}^D$, 
is determined variationally. 
Notice also that as opposed to the parametrization adopted in Sec.~\ref{varstate}, 
here the same variational parameters are set for the parametrized gates that represent equivalent hopping bonds 
and on-site interaction sites. Apart from this difference, the quantum circuit considered here and that in 
Sec.~\ref{varstate} (also see Fig.~\ref{fig.circuit}) are essentially the same if 
$\hat{U}_{t_\alpha}(\theta)={\rm e}^{-\imag\hat{\cal{H}}_{t_\alpha}\theta/2}$ in $|\psi_{\rm HVA}(\bs{\theta})\rangle$ is replaced 
with the parametrized exponentiated fermionic-{\sc swap} operator in Eq.~(\ref{efswap}) and Fig.~\ref{fig.circuit}(b).

\begin{center}
  \begin{figure*}
    \includegraphics[width=2\columnwidth]{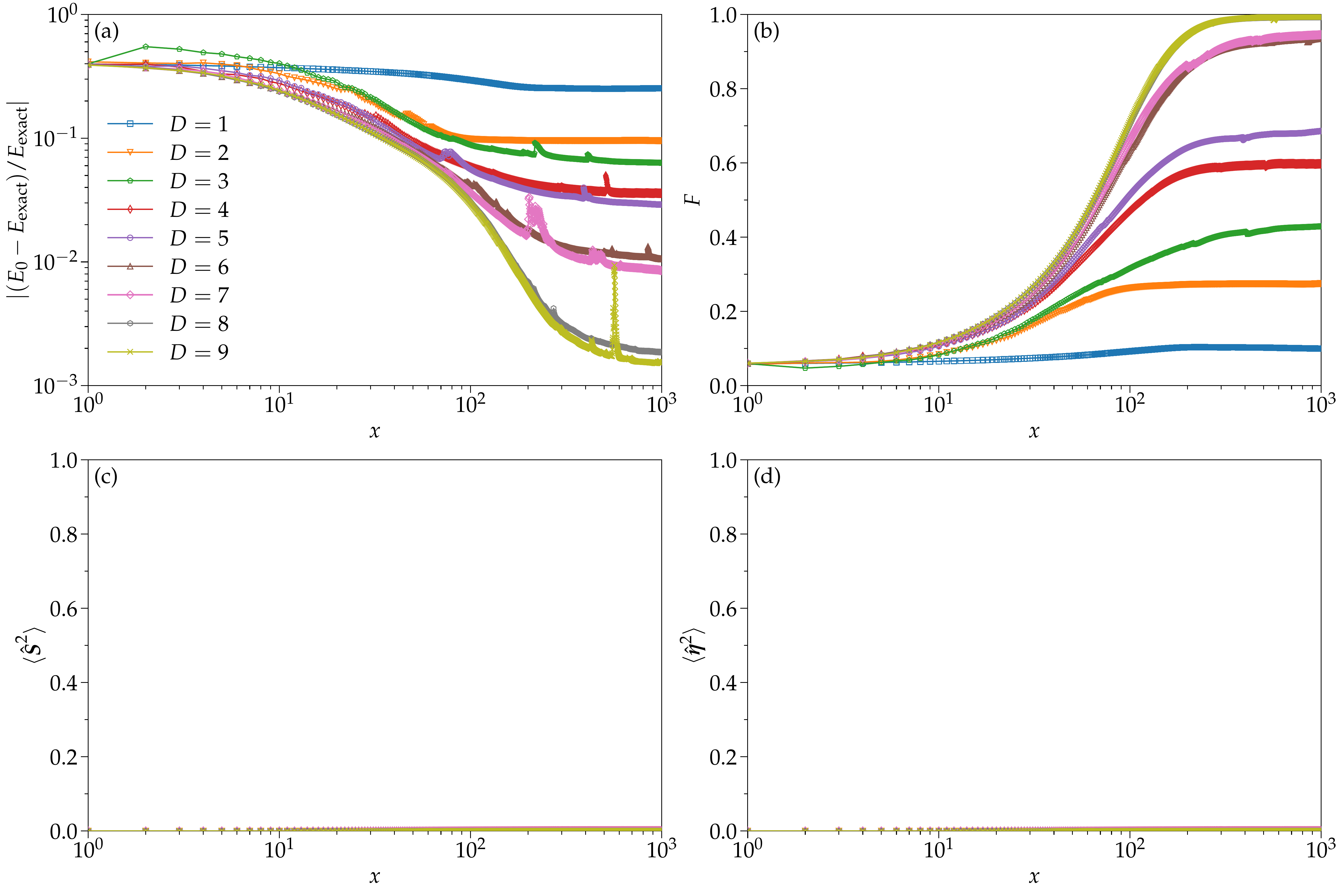}
    \caption{
      (a) The ground-state energy $E_0(\bs{\theta}^{(x)})$,
      (b) the ground-state fidelity $F(\bs{\theta}^{(x)})$,
      (c) the expectation value of total spin squared $ \langle \hat{S}^2 \rangle_{\bs{\theta}^{(x)}}$, and
      (d) the expectation value of total $\eta$ squared $\langle \hat{\eta}^2 \rangle_{\bs{\theta}^{(x)}}$
      as a function of the optimization iteration $x$ 
      obtained by the HVA for different $D$. 
      Each result with a different value of $D$ is obtained by 
      averaging over 64 independent calculations started with 64 different sets of random initial parameters $\bs{\theta}^{(1)}$. 
      The spiky behavior is occasionally observed in the ground-state energy $E_0(\bs{\theta}^{(x)})$ 
      during the parameter optimization iteration. This is due to the same reason described in the caption of Fig.~\ref{fig.energyd1n2}.
    }  \label{fig.energyhva}
  \end{figure*}
\end{center}

In the numerical simulations, we set the learning rate $\tau=0.005/t$ in the NGD optimization 
because we have found that the optimization tends to be less stable if we use the same learning rate 
$\tau=0.025/t$ as in the numerical simulations for the Krylov-extended SAVQE method 
shown in Sec.~\ref{sec:numerical}.
Figure~\ref{fig.energyhva} shows
the ground-state energy, the ground-state fidelity, and the expectation values of total spin squared and total $\eta$ squared
as a function of the optimization iteration $x$.
The results are averaged over 64 independent calculations 
with different sets of initial parameters, each of which is
randomly distributed in $[-0.05,0.05]$.
It is found that the ground-state energy and fidelity are improved systematically
with increasing $D$. As expected, the total spin squared $\langle \hat{S}^2 \rangle_{\bs{\theta}^{(x)}}$ 
and the total eta squared $\langle \hat{\eta}^2 \rangle_{\bs{\theta}^{(x)}}$ are always zero, independently of the optimization 
iteration $x$.

\bibliography{biball}

\begin{thebibliography}{132}
\bibitem [{not()}]{notepara}%
  \BibitemOpen
  \href@noop {} {}\bibinfo {note} {Since $\theta_{k,l}$ are variational
  parameters to be optimized, the prefactors such as $-t$ and $U_{\rm H}/4$ are
  not important in practice.}\BibitemShut {Stop}%
\end{thebibliography}
\end{document}